\documentclass[prd,12pt,reprint, twocolumn,noshowpacs,nofootinbib,amsmath,amssymb,superscriptaddress,preprintnumbers]{revtex4-1}
\usepackage{graphicx,color}
\usepackage{tikz-feynman} 
\usepackage{caption}
\usepackage{amssymb,amsmath,mathrsfs}
\usepackage{cancel}
\usepackage{subfig}
\usepackage{hyperref}
\usepackage{verbatim,slashed}
\usepackage{xspace}
\usepackage{afterpage}
\usepackage[normalem]{ulem}

\newcommand{\beq}{\begin{equation}}
\newcommand{\eeq}{\end{equation}}
\newcommand{\be}{\begin{eqnarray}}
\newcommand{\ee}{\end{eqnarray}}

\newcommand{\bi}{\begin{itemize}}
\newcommand{\ei}{\end{itemize}}
\newcommand{\benum}{\begin{enumerate}}
\newcommand{\eenum}{\end{enumerate}}

\def\lsim{\mathrel{\rlap{\lower4pt\hbox{\hskip1pt$\sim$}}
    \raise1pt\hbox{$<$}}}
\def\gsim{\mathrel{\rlap{\lower4pt\hbox{\hskip1pt$\sim$}}
    \raise1pt\hbox{$>$}}}

\begin{document}
\title{Electroweak Axion Portal to  Dark Matter}
\author{Stephanie Allen}
\affiliation{Harvey Mudd College, 301 Platt Blvd., Claremont, CA 91711, USA}
\author{Albany Blackburn}
\affiliation{Harvey Mudd College, 301 Platt Blvd., Claremont, CA 91711, USA}
\affiliation{Berkeley Center for Theoretical Physics, University of California, Berkeley, CA 94720, USA}
\author{Oswaldo Cardenas}
\affiliation{Harvey Mudd College, 301 Platt Blvd., Claremont, CA 91711, USA}
\author{Zoe Messenger}
\affiliation{Harvey Mudd College, 301 Platt Blvd., Claremont, CA 91711, USA}
\author{Ngan H. Nguyen }
\affiliation{Harvey Mudd College, 301 Platt Blvd., Claremont, CA 91711, USA}
\affiliation{Johns Hopkins University, 3400 N Charles St., Baltimore, MD 21218, USA}
\author{Brian Shuve}
\affiliation{Harvey Mudd College, 301 Platt Blvd., Claremont, CA 91711, USA}

\date{\today}

\begin{abstract}
Axion-like particles (ALPs) are good candidates for mediators to the dark sector. We explore scenarios in which an ALP mediates interactions between dark matter and electroweak gauge bosons. These models  yield testable electromagnetic signals in astrophysical, cosmological, and terrestrial probes. We find promising prospects for both indirect detection and accelerator tests, with interesting parameter space already constrained by current experiments. Our work provides concrete benchmarks for future tests of the electroweak ALP portal.

\end{abstract}
\maketitle

\section{Introduction}\label{sec:intro}

Axion-like particles (ALPs) are among the best-motivated candidates
for new particles beyond the Standard Model (SM). They arise as pseudo-Goldstone bosons whenever there exist spontaneously broken (approximate) global symmetries that are anomalous under
SM gauge interactions, also known as Peccei-Quinn (PQ) symmetries \cite{Peccei:1977hh,Peccei:1977ur,Weinberg:1977ma,Wilczek:1977pj}. ALPs are generic features of
ultraviolet (UV) theories such as string theory \cite{Witten:1984dg,Conlon:2006tq,Svrcek:2006yi,Arvanitaki:2009fg} and supersymmetry \cite{Frere:1983ag,Nelson:1993nf,Bagger:1994hh}. Although axions were originally hypothesized to explain the absence of $CP$-violation in quantum chromodynamics (QCD) \cite{Peccei:1977hh,Peccei:1977ur,Weinberg:1977ma,Wilczek:1977pj},
ALPs have also been proposed to play a role in explaining
the smallness of the electroweak scale relative to the Planck scale \cite{Graham:2015cka}.

While the PQ symmetry is spontaneously broken at some energy scale $f$, ALPs have masses that are proportional to the scale of explicit breaking, and are therefore naturally much lighter than $f$. The approximate shift symmetry enjoyed by ALPs requires that they couple to other fields via derivatives, and the strengths of these couplings are suppressed by  powers of $\frac{1}{f}$. Consequently, ALPs are ideal mediators between dark matter (DM) and SM particles through what has been dubbed the ``axion portal'' \cite{Nomura:2008ru,Freytsis:2010ne,Dolan:2014ska,Kaneta:2017wfh,Kamada:2017tsq,Hochberg:2018rjs,Gola:2021abm,Fitzpatrick:2023xks,Dror:2023fyd,Armando:2023zwz}. 

If SM fermions are also charged under the PQ symmetry, the ALP couples derivatively to the corresponding fermion current, and the phenomenology resembles that of a spin-0 particle with Yukawa-like couplings \cite{Dolan:2014ska}. By contrast, when only beyond-SM fermions are charged under both the PQ symmetry and SM gauge interactions, the predominant ALP coupling to the SM is to pairs of SM gauge bosons. Recent studies have comprehensively assessed the status of axion-portal DM with predominant couplings to gluons \cite{Fitzpatrick:2023xks,Dror:2023fyd}.

We consider instead the scenario in which the ALP interacts with electroweak gauge bosons. Such a limit is interesting because it allows direct annihilation of DM into photon pairs via the ALP mediator, and consequently there exist strong constraints from indirect detection searches for photon lines \cite{Dolan:2017osp}. These bounds on DM annihilation are complemented by a vast array of direct experimental searches for ALPs, as well as constraints on other particles that we have reinterpreted to apply to ALPs \cite{Bjorken:1988as,Blumlein:1990ay,L3:1994shn,L3:1995nbq,L3:1997exg,E949:2005qiy,DELPHI:2008uka,KTeV:2008nqz,PrimEx:2010fvg,CMS:2011bsw,ATLAS:2011ab,ATLAS:2012fgo,Jaeckel:2012yz,CDF:2013lma,NA62:2014ybm,ATLAS:2015rsn,Jaeckel:2015jla,Knapen:2016moh,Izaguirre:2016dfi,Bauer:2017ris,Dolan:2017osp,ATLAS:2017jag,CMS:2018erd,KOTO:2018dsc,Aloni:2019ruo,ATLAS:2020hii,Belle-II:2020jti,Gori:2020xvq,NA64:2020qwq,NA62:2020xlg,BaBar:2021ich,CMS:2021juv,CMS:2021klu,ATLAS:2021uiz,BESIII:2022rzz,Jerhot:2022chi,Mitridate:2023tbj,ATLAS:2023zfc,CMS:2023jgd}. We derive the DM-motivated parameter space for the electroweak axion portal scenario, showing the interplay between indirect detection searches for DM and direct probes of the ALP mediator. A similarly motivated study, albeit one with very different phenomenology, was recently performed for the case where the ALP couples to leptons in addition to photons \cite{Armando:2023zwz}.

Since the relative sizes of the DM-ALP and SM-ALP couplings are \emph{a priori} undetermined, we propose simple benchmarks with SM gauge anomaly coefficients that are either ``small'' or ``large'' and that parametrize different well-motivated limits of the parameter space. We find that current constraints already set meaningful bounds on parameters predicting the DM abundance through thermal freeze-out, and that improvements of one to two orders of magnitude in sensitivity to ALP couplings would test a substantial fraction of the remaining parameter space. Our benchmarks therefore provide insight into gaps in current coverage of ALP-mediated DM and motivation for future searches. We additionally consider the parameter space for freeze-in, identifying the relevant processes driving the abundance in each parameter regime. ALP-portal freeze-in DM has also been recently considered in Refs.~\cite{Bharucha:2022lty,Fitzpatrick:2023xks,Dror:2023fyd,Ghosh:2023tyz}.

We first present the ALP effective field theory and benchmarks in Sec.~\ref{sec:model}. We then study the predictions for thermal DM in Sec.~\ref{sec:DM_abundance}, presenting interplay with indirect detection constraints in Sec.~\ref{sec:cosmo_constraints}, and direct searches for ALPs in Sec.~\ref{sec:acc}.  Finally, we consider the motivated parameter space of and detection prospects for freeze-in DM in Sec.~\ref{sec:freeze_in}.

\section{Mediator Effective Field Theory}\label{sec:model}

We consider an effective field theory (EFT) in which DM, $\chi$, is a Dirac fermion with unit charge under an anomalous
PQ symmetry. If the DM is sufficiently weakly coupled to the scalar(s) responsible for breaking the PQ symmetry, 
only two states remain in the EFT:~DM and the ALP, $a$. We take the ALP mass to be a free parameter:~its mass could originate
either from explicit PQ breaking or from a confining gauge interaction in the hidden sector. In any case, $M_a$ is technically natural. The validity of the EFT requires that $M_\chi$ and $M_a$ be well below the radial-mode masses of the symmetry-breaking sector, which are comparable to the ALP decay constant, $f$. The ALP-DM Lagrangian is
\be\label{eq:DM-ALP}
\mathcal{L}_{\rm DM-ALP} &=& -\frac{1}{2f}\,\partial_\mu a\,\bar\chi\gamma^\mu\gamma^5\chi.
\ee
It is common in the literature to integrate this term by parts in the action and apply the fermion equations of motion, which results in an $i\frac{M_\chi a}{f}\bar\chi\gamma^5\chi$ interaction. However, this is only valid up to corrections of $\mathcal{O}(\frac{1}{f^2})$ at matrix-element level. Thus, if we wish to correctly calculate rates of processes like $\bar\chi\chi\to aa$ whose matrix elements scale like $\frac{1}{f^2}$, we cannot apply the equations of motion in this simplified manner\footnote{More correctly, ``applying the equations of motion'' is the result of a field redefinition. When this field redefinition is done correctly to $\mathcal{O}(\frac{1}{f^2})$, one obtains both the pseudoscalar coupling $i\frac{M_\chi a}{f}\bar\chi\gamma^5\chi$ and an additional contact term $\frac{M_\chi a^2}{2f^2}\bar\chi\chi$. This is equivalent to using Eq.~\eqref{eq:DM-ALP} for processes with matrix elements of $\mathcal{O}(\frac{1}{f^2})$.} and should instead  use Eq.~\eqref{eq:DM-ALP}.

The ALP  can also couple to SM fields. We consider the scenario in which SM fields are neutral under the PQ symmetry; in this case, the ALP does not  couple directly to SM fermions. The ALP can, however, couple to SM gauge bosons via intermediate states with masses $\sim f$ carrying both PQ and SM charges. At low energies, the Lagrangian for the ALP-SM couplings is:
\begin{align}
\label{eq:gauge-alp}
\mathcal{L}_{\rm gauge-ALP} =& -\frac{C_{aB}}{4}\,aB_{\mu\nu}\tilde{B}^{\mu\nu}-\frac{C_{aW}}{4}\,aW^b_{\mu\nu}\tilde{W}^{b\mu\nu}\nonumber\\
&{}-\frac{C_{ag}}{4}\,aG_{\mu\nu}^b\tilde{G}^{b\mu\nu},
\end{align}
where $B$, $W$, and $G$ are respectively the hypercharge, $\mathrm{SU}(2)_L$, and QCD gauge fields, and $\tilde{V}^{\mu\nu}= \frac{1}{2}\epsilon^{\mu\nu\rho\sigma}V_{\rho\sigma}$ is the dual gauge boson field strength. The constants $C_{aV}$ depend on the charges of the heavy states to SM gauge bosons $V$ and are proportional to $\frac{\alpha_V}{\pi f}$. 

In our study, we focus on the couplings between ALPs and electroweak gauge bosons. After electroweak symmetry breaking, the physically relevant couplings are those to the photon, $W^\pm$, and $Z$ bosons. The Lagrangian is
\begin{align}
    \mathcal{L}_{\rm gauge-ALP} =& -\frac{g_{a\gamma\gamma}}{4}\,a F_{\mu\nu}\tilde{F}^{\mu\nu} - \frac{g_{aZZ}}{4}\,a Z_{\mu\nu}\tilde{Z}^{\mu\nu}\nonumber\\
&{}-\frac{g_{a\gamma Z}}{2} \,aF_{\mu\nu}\tilde{Z}^{\mu\nu}-\frac{g_{aWW}}{2}\,a W_{\mu\nu}^+\tilde{W}^{-\mu\nu},
\end{align}
where
\be
g_{a\gamma\gamma} &=& C_{aB}\cos^2\theta_{\rm W} + C_{aW}\sin^2\theta_{\rm W},\\
g_{aZZ} &=& C_{aB}\sin^2\theta_{\rm W} + C_{aW}\cos^2\theta_{\rm W},\\
g_{a\gamma Z}&=& \left(C_{aW}-C_{aB}\right)\sin\theta_{\rm W}\cos\theta_{\rm W},\\
g_{aWW} &=& C_{aW},
\ee
and $\theta_{\rm W}$ is the weak mixing angle.

Additional couplings to SM fermions are generated by renormalization group (RG) evolution from the cutoff of the EFT ($\lesssim 4\pi f$) down to phenomenologically relevant scales \cite{Bauer:2020jbp,Bauer:2021mvw}. These additional couplings are UV sensitive and model dependent, and we generally do not include them in our calculations. We find that the electroweak ALP portal predicts relatively low EFT cutoffs, which makes RG-induced couplings less relevant than they otherwise might be. However, in our analysis, we highlight where these RG-induced couplings could have a significant impact on the phenomenology. 

\subsection{Specific Model Benchmarks}\label{sec:benchmarks}
The undetermined coefficients $C_{aV}$ in Eq.~\eqref{eq:gauge-alp} do not allow a direct correspondence between the DM and gauge couplings of the ALP. However, it is possible to relate the couplings in specific UV completions. For example, consider a Dirac fermion $\Psi$ that is a hypercharge singlet but vectorlike under  $\mathrm{SU}(2)_L$ with representation $R$. $\Psi$ has an axial charge under the PQ symmetry such that it acquires its mass through the mechanism that spontaneously breaks $\mathrm{U}(1)_{\rm PQ}$. In this case, $\Psi$ acquires a coupling to the ALP of (upon applying equations of motion\footnote{This is valid here because, in our analysis, only terms in the matrix element that are linear in the gauge boson coupling are relevant.})
\be
\mathcal{L}_{\Psi a} &=& \frac{iM_\Psi a}{f} \bar\Psi \gamma^5 \Psi.
\ee
We can then calculate the ALP coupling to $\mathrm{SU}(2)_L$ gauge bosons explicitly, obtaining
\be\label{eq:Caw_minimal}
C_{aW} &=& \frac{\alpha_W T(R)}{\pi f},
\ee
where $T(R)$ is the index of the representation $R$ defined by $\mathrm{Tr}(T^a_R T^b_R) \equiv T(R) \delta^{ab}$, and $T_R^a$ are generators for the representation $R$. The coupling in Eq.~\eqref{eq:Caw_minimal}  can also be obtained using anomalous field redefinitions on the corresponding $\theta$-parameter.
Note that $T(\square)=\frac{1}{2}$  for the fundamental representation, and this is the minimum that a field $\Psi$ can contribute to the ALP coupling to $\mathrm{SU}(2)_L$ gauge bosons.

Similarly, we can consider a benchmark where $\Psi$ is instead an $\mathrm{SU}(2)_L$ singlet with hypercharge $Q_Y$. In this case,
\be
C_{aB} &=& \frac{\alpha_Y Q_Y^2}{\pi f}.
\ee

This motivates us to consider the following benchmarks in our study:
\be
(1) &:& C_{aW} = 0, \,C_{aB}= \frac{\alpha_Y Q_Y^2}{\pi f},\\
(2) &:& C_{aW} =  \frac{\alpha_W T(R)}{\pi f},\,C_{aB} = 0
\ee
for various choices of $R$ and $Q_Y$.

We are interested in a ``small coupling'' and ``large coupling'' for each of the hypercharge and $\mathrm{SU}(2)$ couplings. We choose $Q_Y=1$ and $T(R)=1/2$ for the small-coupling benchmarks, and $Q_Y=10$ and $T(R)=110$ for the large-coupling benchmarks.

\section{Thermal Dark Matter Abundance} \label{sec:DM_abundance}

The ALP mediates production and annihilation of DM to/from SM states. Which of these two processes is most relevant depends on the DM-ALP mass hierarchy. In this section, we focus on thermal DM scenarios and we return to freeze-in in Sec.~\ref{sec:freeze_in}. There are two regimes relevant for our thermal DM study:\\

\noindent {\bf Secluded Scenario.} In this scenario, $M_a < M_\chi$ and the dominant annihilation mode is typically $\chi\bar\chi\to aa$. The relic abundance is determined entirely by the ALP-DM coupling which scales like $\frac{1}{f}$. However, the ALP-SM couplings also scale like $\frac{1}{f}$ and the two are therefore related in a manner that is parametrically different from renormalizable portals. Thus, the electroweak ALP portal  provides relatively concrete predictions  of phenomenological signatures even in the secluded scenario.\\

\noindent {\bf Annihilation to SM.} In this scenario, $M_a \ge M_\chi$ and DM annihilation to ALPs is kinematically forbidden at zero temperature.  The dominant annihilation mode is instead $\chi\bar\chi\to \gamma\gamma$ (or other SM gauge boson pairs). This regime faces strong constraints from gamma-ray line searches, except when DM annihilates resonantly through the ALP in the early universe but not in the present \cite{Dolan:2017osp}. \\

\noindent {\bf Study Methodology:}~We determine the viable parameter space consistent with the observed DM abundance in two ways:
\begin{itemize}
\item We  implement our model into the \texttt{CalcHEP} format\footnote{The CalcHEP model files are available at \url{https://github.com/bshuve/ALP_CalcHEP}.} \cite{Belyaev:2012qa} using \texttt{FeynRules} \cite{Alloul:2013bka}, and we then use \texttt{micrOMEGAs 5.2.13} \cite{Belanger:2018ccd} to solve for the DM abundance.

\item We derive and solve a thermally averaged Boltzmann equation for $\chi$ that includes all annihilation and decay processes (see Appendices \ref{app:BE} and \ref{app:CS} for more details on our Boltzmann equation and rates, respectively). In thermal scenarios, we can assume that the ALPs are always in equilibrium. 
\end{itemize}
In our Boltzmann equation, the only ALP-SM interaction is with the photon, whereas in \texttt{micrOMEGAs} we include all SM gauge bosons. We have checked that the predictions for the DM abundance of the two approaches agree to within 20\% for most parameters within the regime of mutual validity, $M_a\lesssim M_Z$. We find stronger disagreement very close to resonance, $(M_a-2M_\chi)/M_a < 0.01$; in this limit, our calculation of the thermally averaged cross section agrees with the prediction in the narrow-width approximation. For most parameter points, we use \texttt{micrOMEGAs} because it includes all gauge interactions. However, very close to resonance, we use our own solution to the Boltzmann equations, with the on-shell cross section modified by  appropriate powers of the photon branching fraction when the ALP can decay into multiple gauge bosons.

Because of  subtleties in the treatment of annihilation near resonance, we now provide more details on the relevant rates in that region of parameter space.

\subsection{DM Annihilation Near Resonance}\label{sec:resonance}
When $M_a\approx 2M_\chi$, there is a resonant enhancement of the DM annihilation cross section. We must evaluate the thermally averaged cross section,
\begin{align}
    \langle \sigma_{\bar\chi\chi\to\gamma\gamma}v\rangle = &\frac{g_{a\gamma\gamma}^2 }{1024\pi M_\chi^2 f^2 T \,K_2(M_\chi/T)^2}\nonumber\\
&\int_{4M_\chi^2}^\infty\,\frac{s^{7/2}\sqrt{1-4M_\chi^2/s}}{(s-M_a^2)^2+M_a^2\Gamma_a^2} \, K_1(\sqrt{s}/T)\,ds,
\end{align}
in this limit, where $K_1$ and $K_2$ are modified spherical Bessel functions of the second kind and $s$ is the Mandelstam variable. The correct treatment of resonance depends on whether DM is heavier or lighter than $M_a/2$; in the latter case, the parametric dependence of the annihilation cross section is also sensitive to whether the ALP predominantly decays into DM or SM gauge bosons. To simplify our analysis, in this section we restrict our consideration to $M_\chi \lesssim 45$ GeV, in which case the only open annihilation modes to the SM in the resonance region are to pairs of photons.

Those who are interested only in our quantitative results can skip to the summary of our findings at the end of this section. We now provide details of our calculation of resonant annihilation in various limits.

Above but close to resonance, $M_a > 2M_\chi$, the ALP can decay into both DM and photons.  In the narrow-width approximation, the ALP propagator can be approximated with a Dirac delta function,
\be
\frac{1}{(s-M_a^2)^2+M_a^2\Gamma_a^2}\to\frac{\pi}{M_a\Gamma_a}\delta(s-M_a^2),
\ee
allowing a straightforward evaluation of the thermally averaged cross section:

\begin{equation}
    \langle \sigma_{\bar\chi\chi\to\gamma\gamma}v\rangle = \frac{g_{a\gamma\gamma}^2 M_a^6\sqrt{1-4M_\chi^2/M_a^2}K_1(M_a/T)}{1024f^2 M_\chi^2 \Gamma_a T\,K_2(M_\chi/T)^2} .
\end{equation}

Because we generally expect the ALP coupling to DM to be larger than its coupling to photons, the decay width is dominated by the decay to DM even for $M_a$ quite close to $2M_\chi$. In this case, we can approximate $\Gamma_a\approx \Gamma(a\to\chi\bar\chi)$, where 
\be
\Gamma(a\to\bar\chi\chi) &\approx& \frac{M_aM_\chi^2}{8\pi f^2}\sqrt{1-\frac{4M_\chi^2}{M_a^2}}.
\ee
Substituting this into the thermally averaged cross section gives,
\begin{equation}
    \label{eq:aboveresonance-DM}
\langle \sigma_{\bar\chi\chi\to\gamma\gamma}v\rangle \approx \frac{\pi g_{a\gamma\gamma}^2 M_a^5 \,K_1(M_a/T)}{128 M_\chi^4 T \,K_2(M_\chi/T)^2}.
\end{equation}
Note that this rate depends only on the ALP coupling to SM fields and not on the DM coupling.  It does, however, implicitly assume that the DM coupling is much larger than $g_{a\gamma\gamma}$ for the decay to DM to dominate. The parametric dependence of the cross section in Eq.~\eqref{eq:aboveresonance-DM} agrees with that found in Ref.~\cite{Dolan:2017osp}. In particular, for $M_a \approx 2M_\chi$ and freeze-out occurring at $M_\chi/T\approx20$, this  yields
\begin{equation}
    \langle\sigma v\rangle \approx (6\times10^{-26}\,\,\mathrm{cm}^3/\mathrm{s})\left(\frac{g_{a\gamma\gamma}}{1.2\times10^{-5}\,\,\mathrm{GeV}^{-1}}\right)^2,
\end{equation}
such that $g_{a\gamma\gamma}\approx1.2\times10^{-5}\,\,\mathrm{GeV}^{-1}$ will give a thermal cross section. This prediction is independent of $M_\chi$.

The above estimate, however, does not apply  very near to resonance. $\Gamma(a\to\bar\chi\chi)$ goes to zero in the limit $M_a\to 2M_\chi$ while the decay width to photons,
\be
\Gamma(a\to\gamma\gamma) &=& \frac{g_{a\gamma\gamma}^2M_a^3}{64\pi},
\ee
does not. Therefore, sufficiently close to resonance, the ALP decay to photons dominates over the decay to DM. When this happens (while continuing to use the delta-function approximation of the ALP propagator),  the thermally averaged cross section depends only on the coupling to DM:
\be\label{eq:aboveresonance-SM}
\langle \sigma_{\bar\chi\chi\to\gamma\gamma}v\rangle &\approx& \frac{\pi M_a^3\, K_1(M_a/T)\sqrt{1-4M_\chi^2/M_a^2}}{16 M_\chi^2 f^2 T\, K_2(M_\chi/T)^2}.
\ee
Note that the cross section appears to vanish in the $M_a\to 2M_\chi$ limit! 

The reason this appears to go to zero on resonance is that the delta function only has support at $s=M_a^2$, but the phase space also vanishes in this limit.   We must therefore be more careful with the thermal average integral, treating $\Gamma_a$ as small but non-vanishing when we take $M_a\to 2M_\chi$. To illustrate our approach, we evaluate the integral exactly on resonance, $M_a=2M_\chi$. If the width is narrow, we can approximate $s\approx M_a^2=4M_\chi^2$ everywhere \emph{except} in two places: in the denominator, and in the phase space factor in the numerator:

\be
\langle \sigma_{\bar\chi\chi\to\gamma\gamma}v\rangle_{\rm res} &=& \frac{g_{a\gamma\gamma}^2 M_a^5 \,K_1(M_a/T) }{256\pi  f^2 T\, K_2(\frac{M_a}{2T})^2}\nonumber\\
&& \int_{M_a^2}^\infty\,\frac{\sqrt{1-M_a^2/s}}{(s-M_a^2)^2+M_a^2\Gamma_a^2}\,ds.
\ee

The integral can be evaluated analytically assuming $\Gamma_a\ll M_a$, in which case:
\be
\langle \sigma_{\bar\chi\chi\to\gamma\gamma}v\rangle_{\rm res} &\approx& \frac{g_{a\gamma\gamma}^2 M_a^5 \,K_1(M_a/T) }{256  f^2 T \,K_2(\frac{M_a}{2T})^2}\,\frac{1}{\sqrt{2M_a^3\Gamma_a}}.
\ee
As expected, we find the thermally averaged cross section is non-zero even exactly on resonance.

The non-analytic dependence on the width gives rise to interesting scaling behavior. For example, substituting the ALP decay width to SM photons gives
\be\label{eq:thermalaverage-onresonance}
\langle \sigma_{\bar\chi\chi\to\gamma\gamma}v\rangle_{\rm res} &\approx&\sqrt{\frac{\pi}{2}} \frac{g_{a\gamma\gamma} M_a^2\, K_1(M_a/T) }{32  f^2 T \,K_2(\frac{M_a}{2T})^2},
\ee
with a linear dependence on the coupling to photons. Defining $x\equiv \frac{M_\chi}{T} = \frac{M_a}{2T}$, which is typically $\sim15-20$ for thermal freeze-out, we have
\be
\langle \sigma_{\bar\chi\chi\to\gamma\gamma}v\rangle_{\rm res} &\approx&\sqrt{\frac{\pi}{2}} \frac{g_{a\gamma\gamma} M_a \,x K_1(2x) }{16  f^2  \,K_2(x)^2}.
\ee

The linear dependence of the cross section on the coupling has interesting implications. For example, our benchmarks in Sec.~\ref{sec:benchmarks} all have $g_{a\gamma\gamma}\sim\frac{1}{f}$. If we want Eq.~\eqref{eq:thermalaverage-onresonance} to give a fixed thermal cross section to achieve the correct relic abundance, we predict a  $\frac{1}{f}\propto \frac{1}{M_a^{1/3}}$ scaling for the thermal benchmark.

When $M_a < 2M_\chi$, the resonance condition cannot be exactly satisfied. If $0 < (4M_\chi^2-M_a^2)\lesssim M_a\Gamma_a$, the annihilation still occurs close enough to resonance that the above calculation  applies in an approximate manner. Away from resonance, the cross section plummets rapidly for fixed couplings and the results are the same as those for annihilation through an off-shell ALP.

For $M_a\gtrsim90$ GeV, additional decay modes open to $\gamma Z$, $ZZ$, and $WW$ depending on the ALP mass. This leads to a rescaling of the resonant annihilation cross section. Above resonance, with the ALP predominantly decaying to DM, $\Gamma_a$ is unchanged by an increase in the effective coupling to the SM while the numerator increases due to more DM annihilation modes. Consequently, $\langle \sigma_{\bar\chi\chi}v\rangle$ increases by $\mathrm{BF}(a\to\gamma\gamma)^{-1}$ relative to the photon-only scenario. Above, but sufficiently close to resonance that the ALP decays predominantly to SM states, $\langle \sigma_{\bar\chi\chi}v\rangle$ is independent of the SM coupling and is therefore independent of  $\mathrm{BF}(a\to\gamma\gamma)$. Finally, exactly on resonance we have that the cross section scales as $\mathrm{BF}(a\to\gamma\gamma)^{-1/2}$ due to the enhancement from additional SM annihilation modes being partially cancelled by the factor of $\Gamma_a^{-1/2}$ in the denominator.  \\

\noindent {\bf Summary of findings:}
\begin{itemize}
    \item When $M_a\lesssim 2M_\chi$ but very close to resonance, the correct thermally averaged cross section is given by Eq.~\eqref{eq:thermalaverage-onresonance};
    \item When $M_a>2M_\chi$, we  determine for the given ALP mass whether the decay is predominantly to DM or to photons. If it decays predominantly to DM, then Eq.~\eqref{eq:aboveresonance-DM} applies. Otherwise, Eq.~\eqref{eq:aboveresonance-SM} applies.
\end{itemize}

\section{Indirect Detection Constraints on Dark Matter}\label{sec:cosmo_constraints}

\subsection{Overview}
Given that no definitive electromagnetic signals of DM have been discovered to date, indirect detection sets powerful constraints on DM coupled to photons. In this section, we apply photon-line searches and CMB ionization bounds to the electroweak ALP portal, comparing constraints to parameters predicted by the DM thermal relic abundance. 

The process  $\bar\chi\chi\to\gamma\gamma$ in the DM halo today leads to a monochromatic photon-line feature with $E_\gamma = M_\chi$. This is an $s$-wave scattering process at low momentum, and consequently it is unsuppressed at late times. In contrast, the process $\bar\chi\chi \to aa$, when kinematically allowed, occurs in the $p$-wave; even in the secluded scenario when this process determines the DM abundance, the direct annihilation to photons provides the strongest bounds today. 

In the limit $v\to0$, the thermally averaged diphoton cross section is
\begin{equation}
\langle\sigma_{\bar\chi\chi\to\gamma\gamma}v\rangle_{v\to0} = \frac{g_{a\gamma\gamma}^2 M_\chi^6}{4\pi f^2}\,\frac{1}{(4M_\chi^2-M_a^2)^2+M_a^2\Gamma_a^2}.    
\end{equation}
We now discuss constraints on this annihilation rate from astrophysical and cosmological observables.\\

\noindent\textbf{Photon-line searches:} We apply constraints on $\langle\sigma_{\bar\chi\chi\to\gamma\gamma} v\rangle$ from galactic photon-line searches with Fermi-LAT \cite{Albert:2014hwa,Foster:2022nva}, MAGIC \cite{MAGIC:2022acl}, H.E.S.S \cite{HESS:2018cbt}, EGRET, and COMPTEL \cite{Pullen:2006sy,Boddy:2015efa}\footnote{Current searches have low sensitivity to gamma rays in the MeV-GeV energy range. These can be improved with future telescopes such as e-ASTROGAM \cite{e-ASTROGAM:2016bph}.}. Because  constraints from gamma-ray lines reach sub-thermal cross sections, indirect detection bounds are relevant even when $\bar\chi\chi\to\gamma\gamma$ is a subdominant DM annihilation mode. These constraints are sensitive to the DM profile in the Milky Way, especially in the galactic center region. For concreteness, we present bounds assuming a Navarro-Frenk-White (NFW) profile \cite{Navarro:1995iw}.  Due to uncertain systematic uncertainties affecting modelling of the photon continuum in the EGRET dataset, we present both conservative \cite{Boddy:2015efa}  and optimistic limits \cite{Pullen:2006sy} from EGRET. Ref.~\cite{Boddy:2015efa} also includes constraints from COMPTEL.

There are, in principle, constraints on continuum gamma-ray production, either from DM annihilation to ALP pairs or, at high masses, to $W$ and $Z$ bosons. Since these bounds are weaker than those from photon lines (see, \emph{e.g.,} \cite{Cirelli:2010xx}) we do not include them in our analysis.\\

\noindent \textbf{Cosmic Microwave Background:} DM annihilation to photons during recombination leads to distortions in the cosmic microwave background (CMB) that are incompatible with Planck data \cite{Planck:2018vyg}. We apply the constraints from Ref.~\cite{Slatyer:2015jla} to our model. \\

Along with the above constraints, we impose two model-dependent consistency conditions. For the first, we require that there exists a perturbative UV completion of the EFT such that DM annihilation occurs within the regime of validity of the EFT. Since the radial mode of PQ-symmetry breaking has a mass that is at most $\sim4\pi f$, and the typical momentum flowing through the ALP is $2M_\chi$ during late-time annihilation, this leads to the consistency bound of $M_\chi \le 2\pi f$. For the ALP to be the dominant mediator between DM and the SM, we additionally require $M_a\le 4\pi f$.

\begin{figure*}[t]
          \includegraphics[width=3.25in]{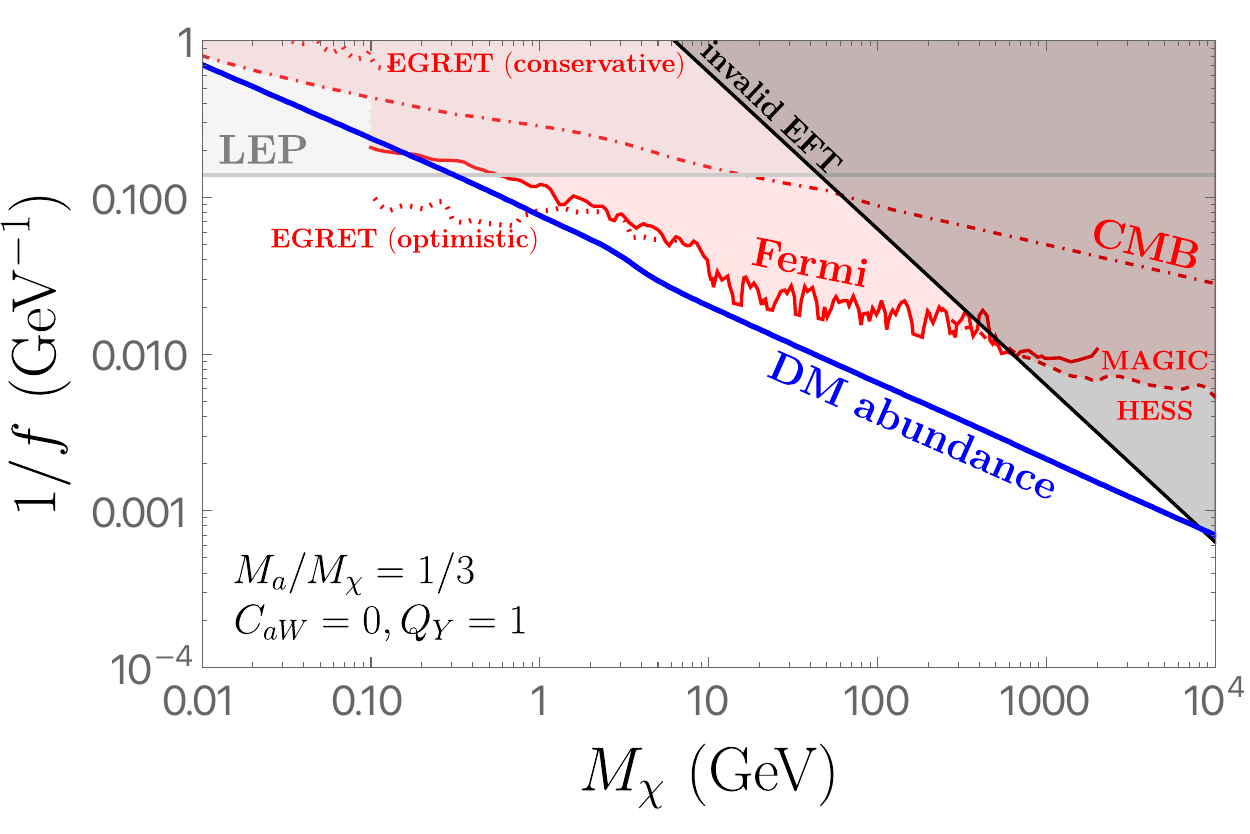}
           \quad 
           \includegraphics[width=3.25in]{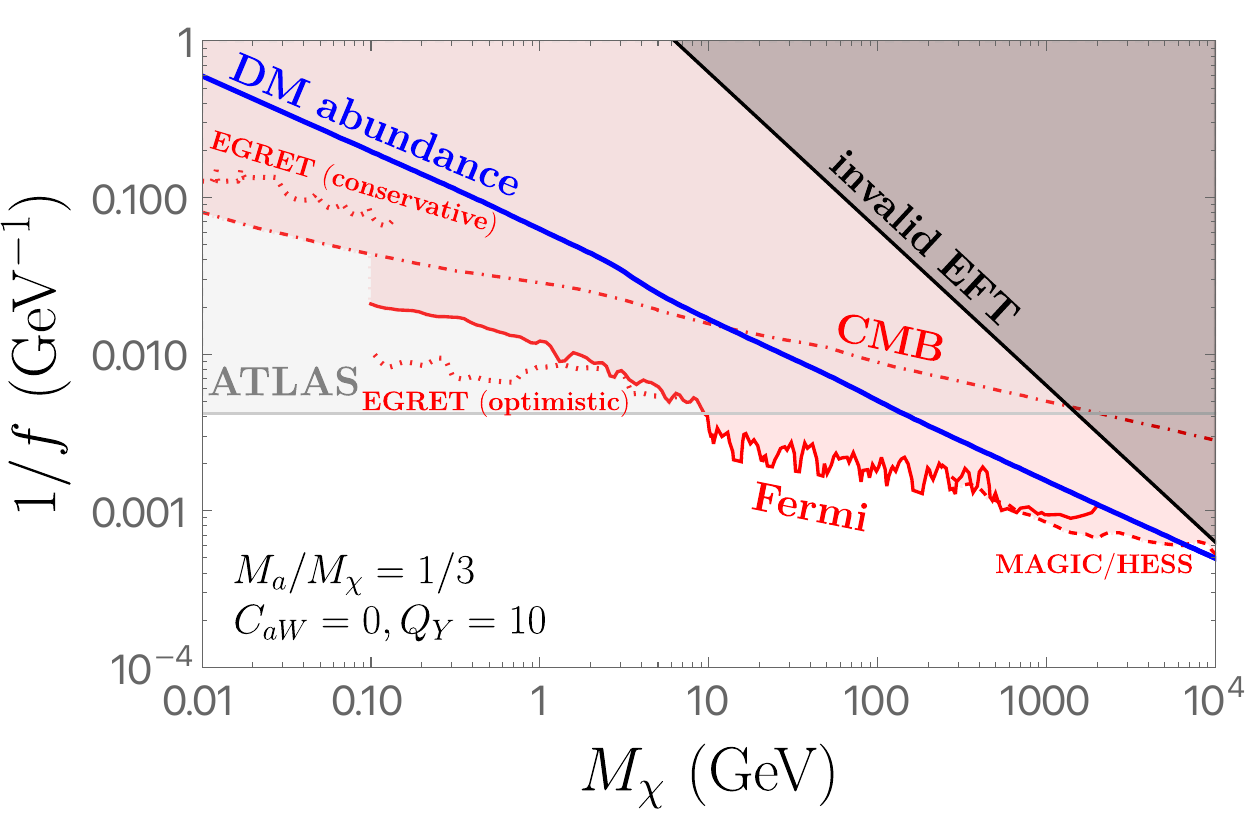}\\
                 \includegraphics[width=3.25in]{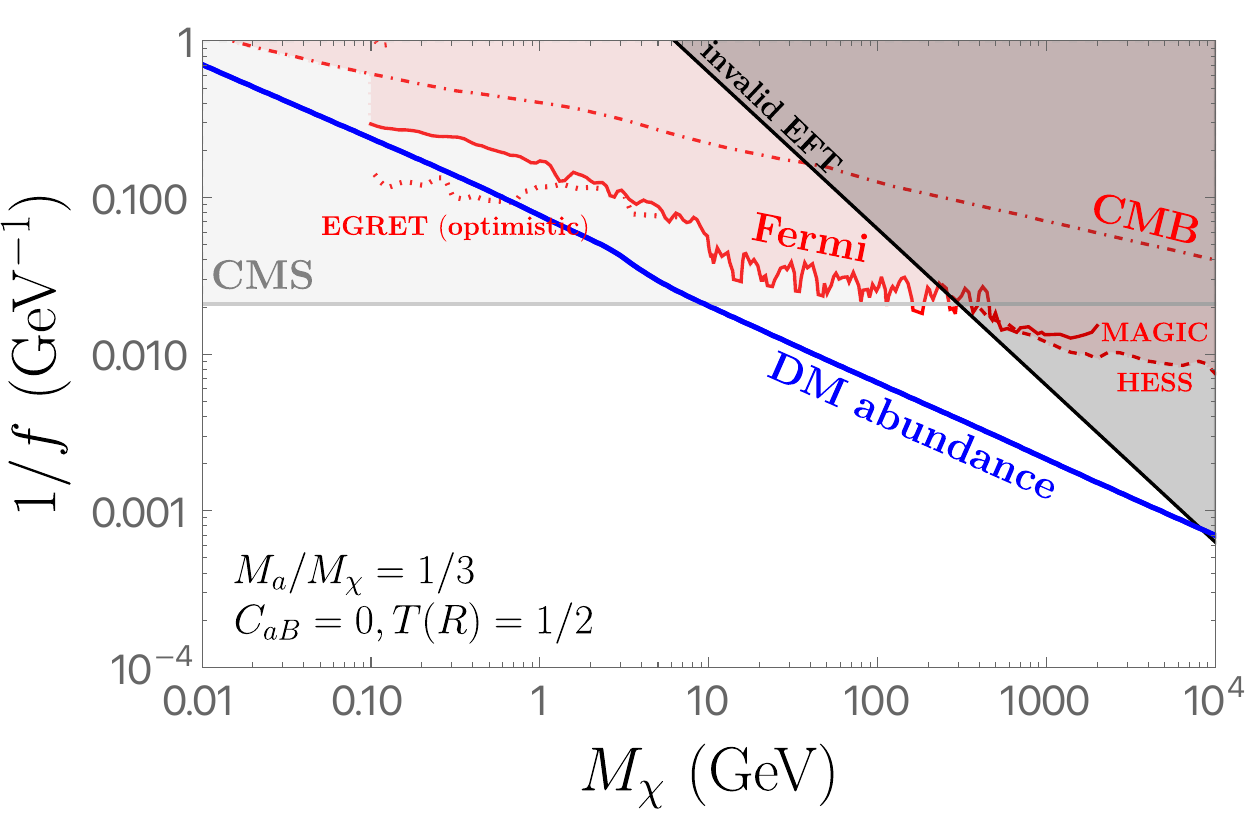}
           \quad 
        \includegraphics[width=3.25in]{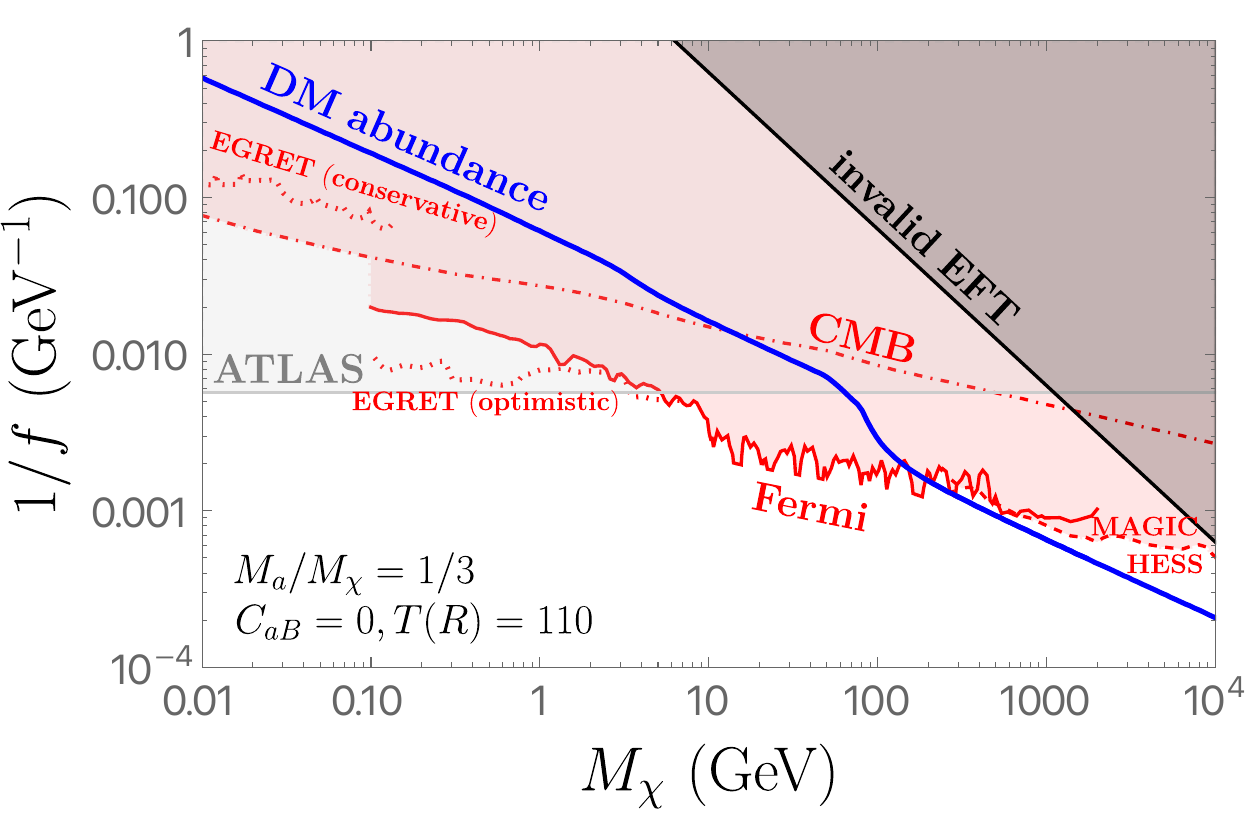}
\caption{
DM thermal-relic parameters along with indirect-detection constraints for the secluded scenario ($M_a/M_\chi=1/3$). The top row shows results for the hypercharge-coupled scenario with (left) $Q_Y=1$; (right) $Q_Y=10$; the bottom row shows results for the $\mathrm{SU}(2)_L$-coupled scenario with (left) $T(R)=1/2$ and (right) $T(R)=110$. Constraints are from: (solid) Fermi-LAT; (dashed) MAGIC and H.E.S.S.; (dot-dashed) CMB; (dotted) EGRET. The region shaded in dark gray is inconsistent with our EFT framework, while the light gray shaded region indicates model-dependent collider constraints on the charged states in the minimal UV completion of the model.}
\label{fig:secluded}
\end{figure*}

The second consistency condition comes from the fact that we have not observed direct production of the charged fermions, $\Psi$, which mediate interactions between the ALP and SM gauge bosons. Since the EFT  depends only on the sum of the anomaly coefficients of all such particles, it is not possible to directly impose such a bound. For the purposes of comparison, we derive approximate constraints assuming the existence of a single such fermion with a mass equal to $4\pi f$. 

For the hypercharge-coupled case and $Q_Y=1$, LEP constrains new charged fermion masses to be above 90 GeV in minimal models \cite{Egana-Ugrinovic:2018roi}. For $Q_Y=10$, the new charged states form bound states that annihilate to photon pairs, and they can be constrained from an ATLAS search for diphoton resonances \cite{ATLAS:2021uiz,Altakach:2022hgn}; this leads to a stronger constraint than searches for heavy stable charged particles. Using Ref.~\cite{Altakach:2022hgn}, we estimate a bound of 3 TeV on heavy fermion masses, corresponding to a bound-state mass of 6 TeV.

When the ALP couples predominantly to $\mathrm{SU}(2)_L$ gauge bosons, the fundamental representation has fermions of charge $\pm\frac{1}{2}$ for which a CMS search for fractionally charged particles constrains masses to be above 600 GeV \cite{CMS:2024eyx}. The value $T(R)=110$ corresponds to an eleven-plet with maximum charge of 5, and so the  ATLAS diphoton search constrains the mass to be above approximately 2.2 TeV \cite{ATLAS:2021uiz,Altakach:2022hgn}.

We reiterate that each of these constraints on new charged states is highly model dependent, and in many cases  the same ALP-gauge coupling can be approximately achieved with drastically changed bounds on the charged state(s).

Finally, we note that direct-detection constraints on DM are expected to be negligible for DM through the electroweak axion portal \cite{Frandsen:2012db}.

\subsection{Secluded Scenario Results}

In Fig.~\ref{fig:secluded}, we present our results for the secluded scenario, $M_a<M_\chi$. When $M_a\ll M_\chi$, we find that the results are independent of the precise $M_a/M_\chi$ ratio, and so we choose $M_a/M_\chi=1/3$ as a benchmark for our study. In this limit, the DM annihilation cross section to ALPs scales like $\frac{M_\chi^2}{f^4}$, leading to a prediction of $\frac{1}{f}\sim\frac{1}{\sqrt{M_\chi}}$ scaling for the thermal abundance. This scaling can be seen in our numerical results for the DM abundance curve, although there are deviations in this scaling for masses in the 1-10 GeV range due to rapid changes in $g_*$ with temperature during freeze-out for these masses (this effect also shows up in other scenarios). Additionally, when large couplings exist to $W$ and $Z$ bosons (for example, with $\mathrm{SU}(2)_L$ coupling and $T(R)=110$), the annihilation rate to $W$ and $Z$ bosons can be large enough to induce deviations in this scaling in the vicinity of the diboson mass threshold.

\begin{figure*}[t]
          \includegraphics[width=3.25in]{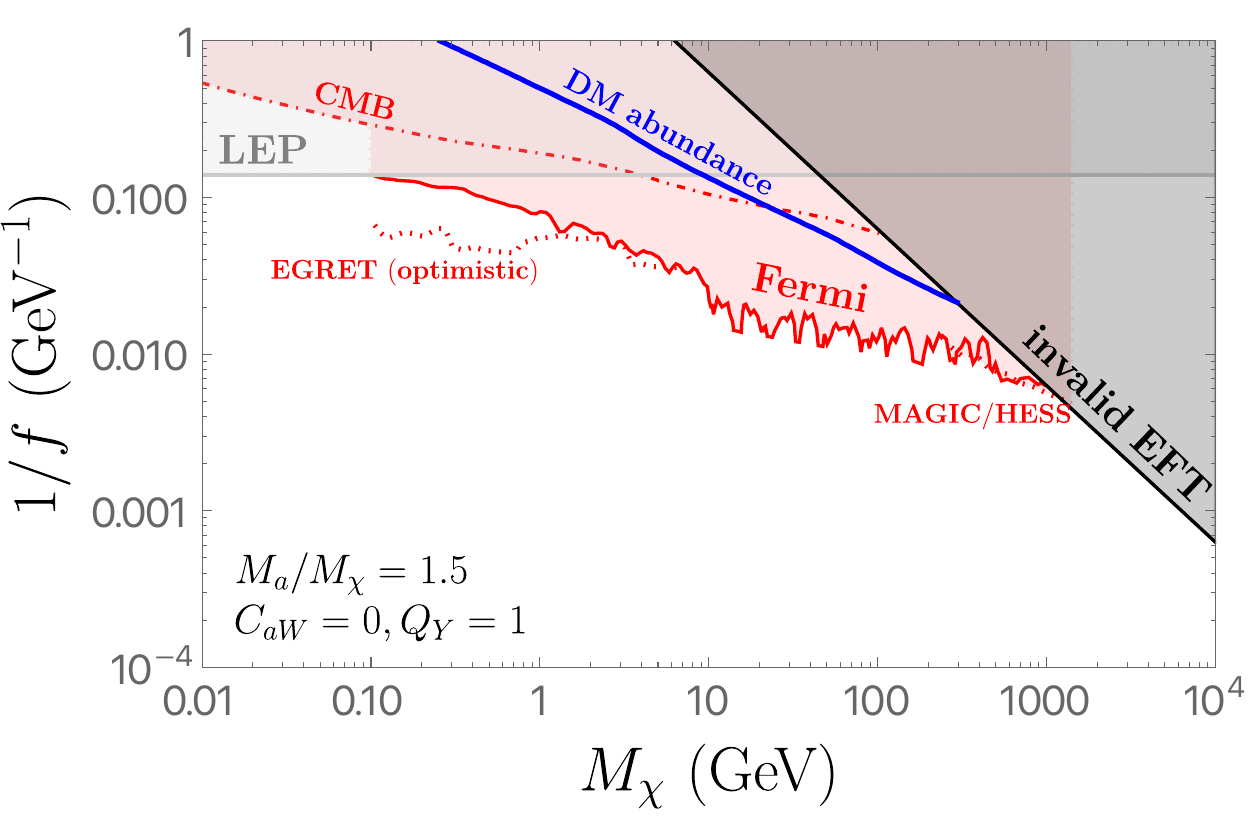}
           \quad 
                  \includegraphics[width=3.25in]{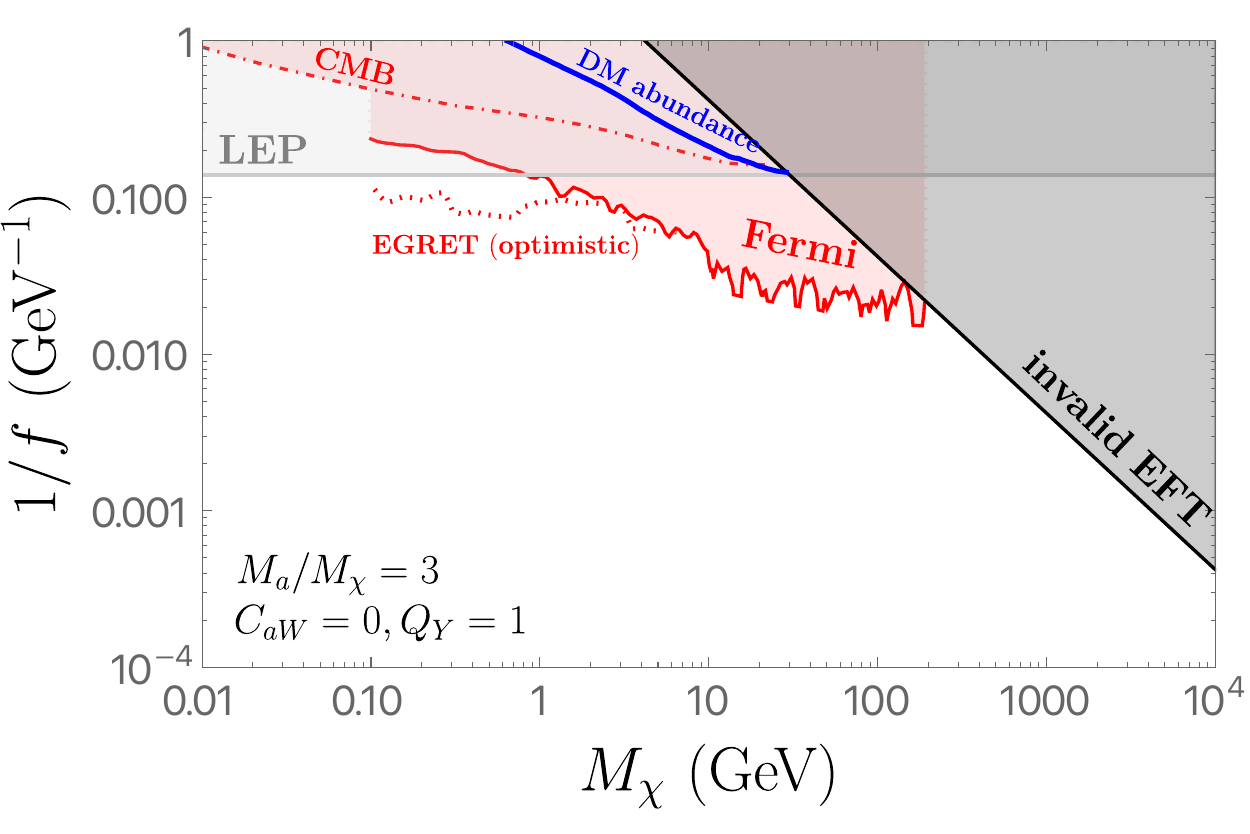}
\caption{DM thermal-relic parameters along with indirect-detection constraints for $M_a > M_\chi$ far from resonance. In the left (right) figure, we show results for DM annihilating below (above) resonance, with $M_a/M_\chi=1.5$ ($M_a/M_\chi=3$). All curves and shaded regions are the same as in Fig.~\ref{fig:secluded}; we do not include MAGIC or H.E.S.S. results for the right plot because they lie entirely in the region of invalid EFT. }\label{fig:ann-to-SM}
\end{figure*}

In the ``small'' gauge coupling limit ($Q_Y=1$ or $T(R)=1/2$), indirect detection constraints are not yet powerful enough to constrain much of the DM parameter space. Improvements to photon-line constraints could meaningfully probe much of the remaining parameter space. By contrast, for ``large'' gauge couplings ($Q_Y=10$ or $T(R)=110$), the predicted DM coupling drops slightly but the constraints get much stronger. Consequently, the DM parameter space is almost completely excluded, except for a small corner in the vicinity of $M_\chi=10$ TeV. We also see that the results are similar for the hypercharge- and $\mathrm{SU}(2)_L$-coupled scenarios. For the rest of this section, we only show results for the hypercharge-coupled scenario although the outcome is similar for both.

For nearly the whole parameter space, $f$ cannot be much above the weak scale and deviations from the predicted EFT behavior are possible. This requires a dedicated analysis that is beyond the scope of our work, but our results show that care must be taken in extrapolating results of the electroweak axion portal to energies at a TeV or above. 

\begin{figure*}[t]

                            \includegraphics[width=3.25in]{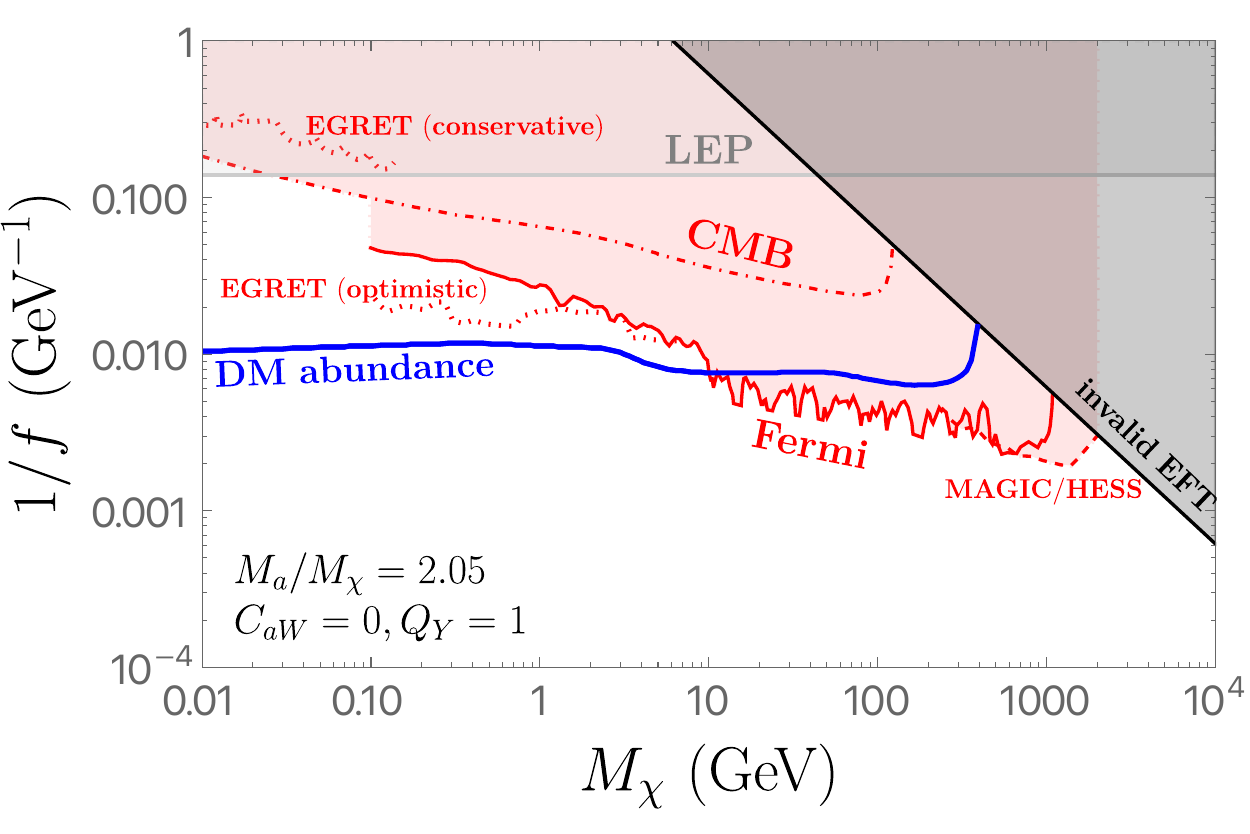}
           \quad 
                  \includegraphics[width=3.25in]{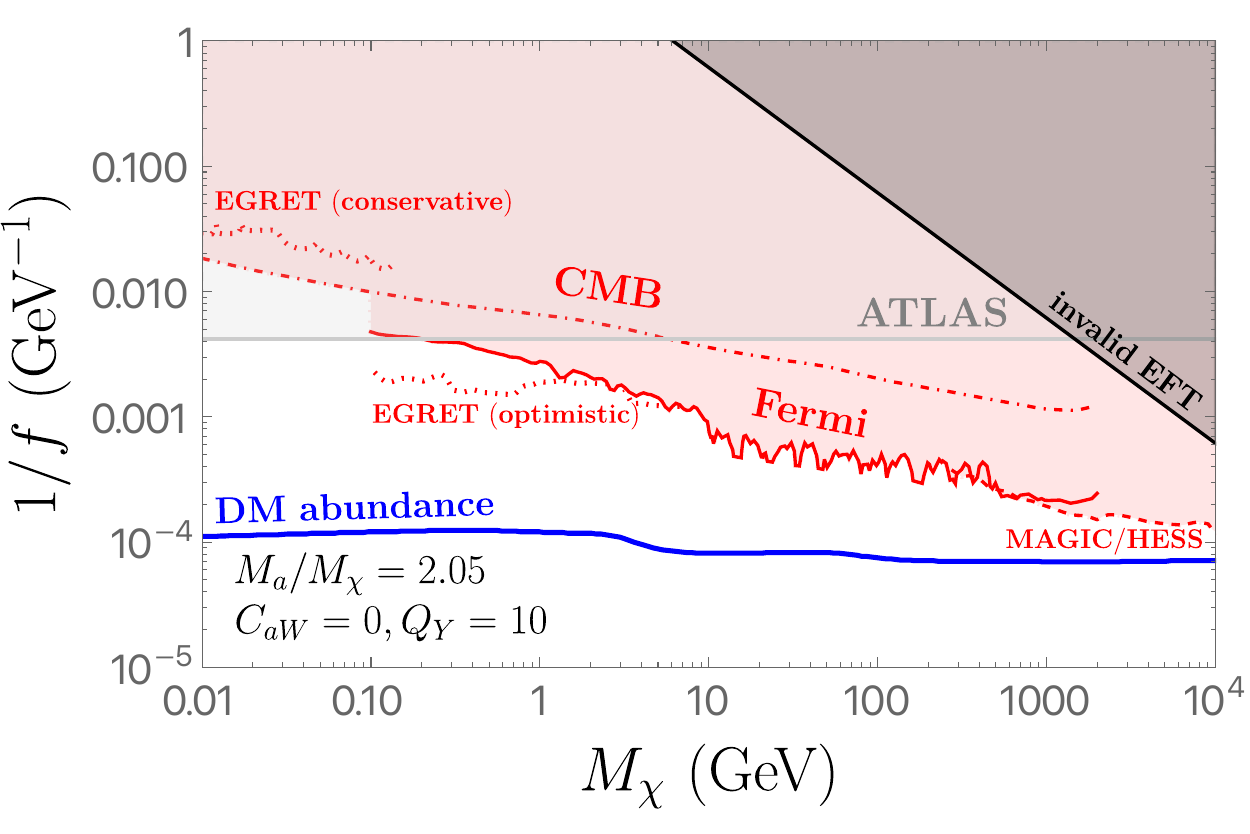}\\
                            \includegraphics[width=3.25in]{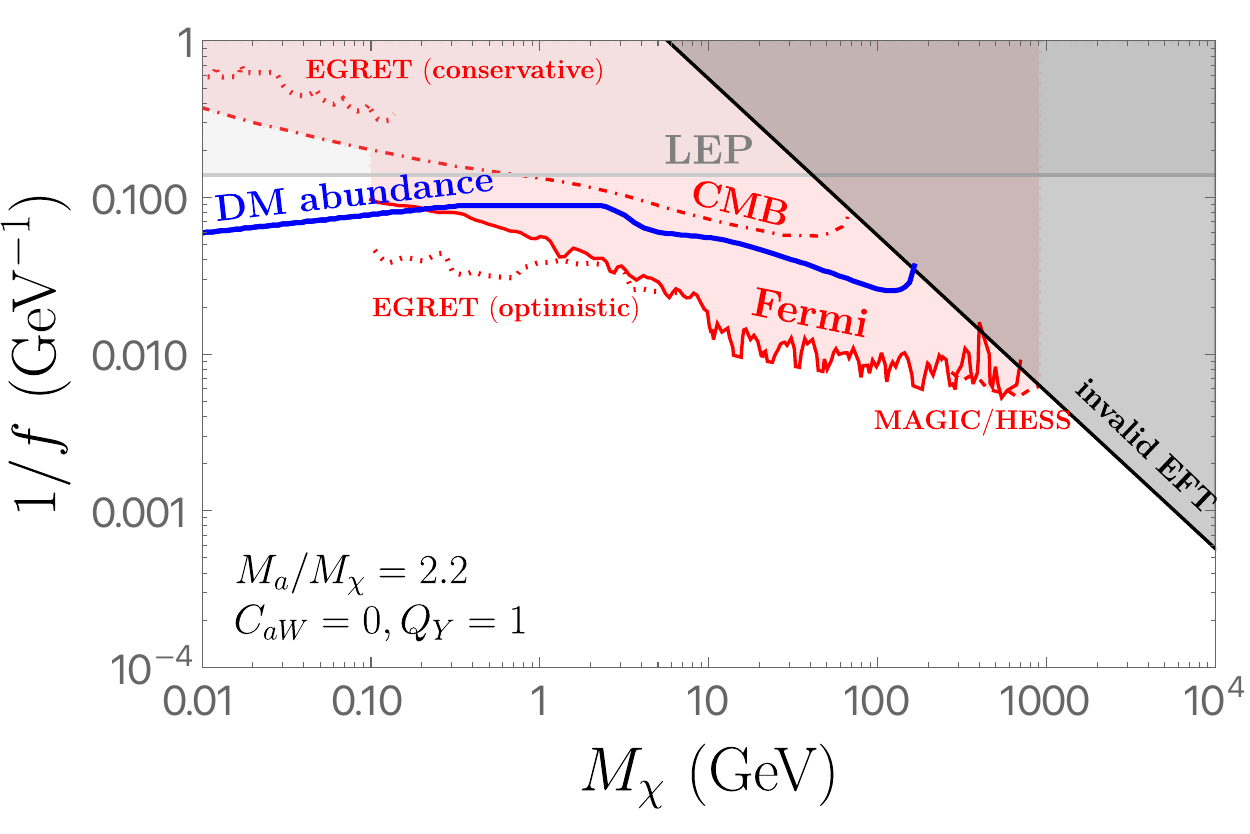}
           \quad 
                  \includegraphics[width=3.25in]{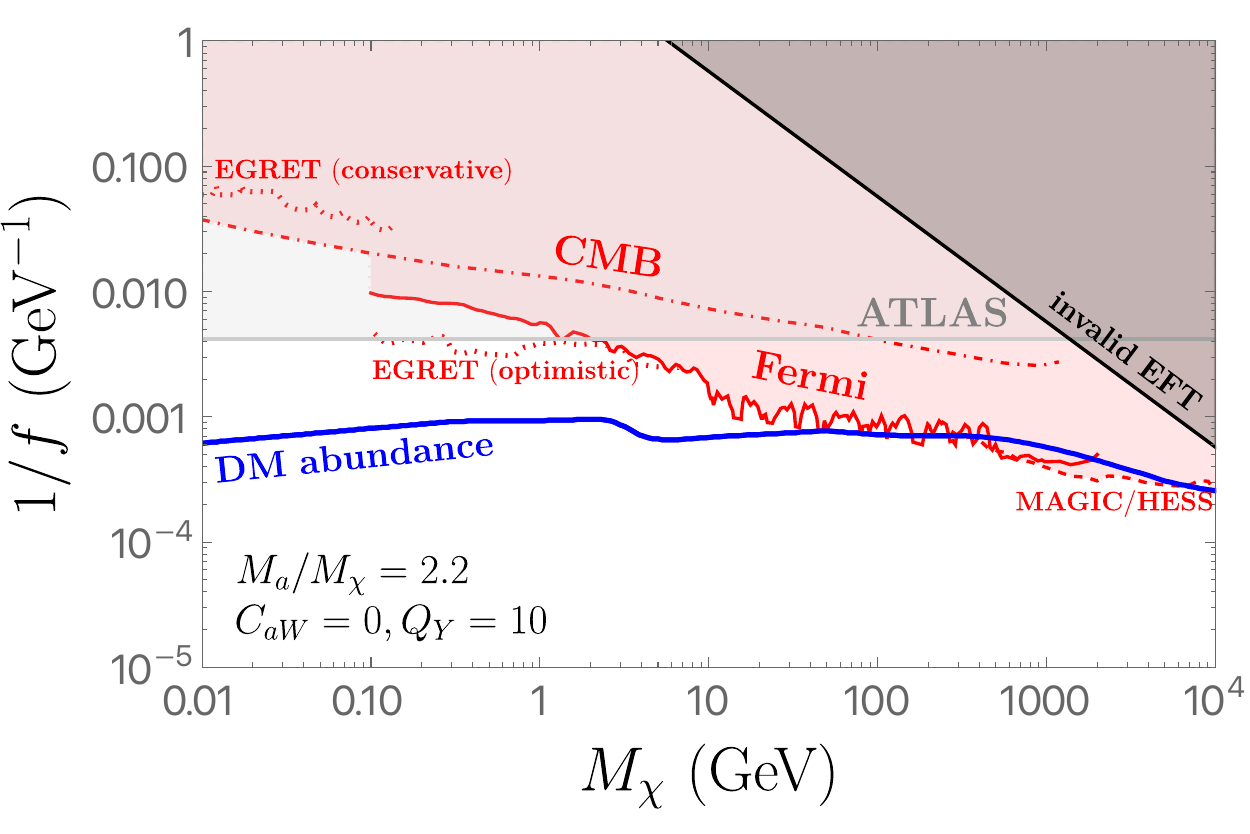}
\caption{DM thermal-relic parameters along with indirect-detection constraints in the resonance region. In the top $(M_a/M_\chi=2.05)$ and bottom $(M_a/M_\chi=2.2)$ rows, we show results for DM annihilation near resonance; the left (right) plots assume $Q_Y=1$ ($Q_Y=10$). All curves and shaded regions are the same as in Fig.~\ref{fig:secluded}}\label{fig:ann-to-SM2}
\end{figure*}

\subsection{Annihilation to SM Far From Resonance}
In the $M_a\ge M_\chi$ regime, DM annihilation could proceed below, on, or above resonance. We defer to the next section an exploration of the resonance region, and so we now choose mass ratios far below ($M_\chi/M_a=1.5$) and above ($M_\chi/M_a=3$) resonance. We show the results of this study in  Fig.~\ref{fig:ann-to-SM}.

In this limit, the s-wave $\bar\chi\chi\rightarrow\gamma\gamma$ process occurs at comparable rates during DM freeze-out and at later times. Therefore, the relative relationship between the DM abundance curve and indirect detection constraints is largely independent of couplings. As expected, we see that indirect detection searches consistently rule out the thermal relic parameters.

\subsection{Annihilation on Resonance}
Perhaps the most interesting scenario is within the resonance region, $M_a/M_\chi\approx 2$. When the DM-ALP mass splitting is such that there is a resonant enhancement of annihilation to photons in the early universe but \emph{not} at later times, it is possible to achieve the observed relic abundance consistent with indirect detection bounds; see Fig.~\ref{fig:ann-to-SM2}. For the parameter  space we show, DM annihilation occurs close to resonance but in the regime where ALP decays to DM still dominate. In this case,  Eq.~\eqref{eq:aboveresonance-DM} holds and $\langle\sigma v\rangle\sim g_{a\gamma\gamma}^2\sim\frac{\alpha_Y^2 Q_Y^4}{\pi^2 f^2}$. Since the thermal relic cross section is a constant, this gives a thermal scaling relation of $\frac{1}{f}\sim \frac{1}{Q_Y^2}$, consistent with the DM abundance curves in Fig.~\ref{fig:ann-to-SM2}. 

The smaller predicted DM-ALP coupling  lessens the constraints from indirect detection. Therefore, when annihilating near resonance, we see the opposite of our findings from the secluded scenario:~in that case, larger ALP-gauge coupling led to strong indirect detection constraints, whereas here the larger gauge coupling  reduces  tension between indirect-detection bounds and the DM abundance.

For most DM masses, the thermal relic curve is approximately independent of $M_\chi$, again in agreement with the prediction of Eq.~\eqref{eq:aboveresonance-DM}. This behavior breaks down for parameters near the boundary of EFT validity, in which case the ALP coupling is relatively strong and the narrow-width approximation no longer applies.

\section{Accelerator and Collider Searches for ALPs}\label{sec:acc}

In recent years, there have been a plethora of new searches for ALPs in accelerators and colliders, motivated by the ubiquity of ALPs in new physics models. Our earlier results showed that the electroweak axion portal gives rise to concrete predictions for the ALP mass and coupling consistent to achieve the correct DM abundance, and we can therefore compare the DM-motivated parameter space to constraints from direct searches for ALPs.

Search strategies for ALPs depend on whether or not the ALP decays predominantly to SM gauge bosons or DM. In the former case, the ALP invariant mass can be reconstructed from its decay products, whereas in the latter case its existence must be inferred through missing momentum. Except just above resonance, we find that the ALPs decay visibly for most of the viable parameter space that is not already ruled out by indirect detection, and we therefore focus on this scenario. Previous studies have been done  to investigate the phenomenology of invisibly decaying ALPs \cite{Izaguirre:2016dfi,Dolan:2017osp,Darme:2020sjf}, with connections to DM explored in the resonant annihilation regime \cite{Dolan:2017osp}.

We find that terrestrial ALP searches already meaningfully constrain ALP portal DM, and improvements in sensitivity to the coupling by an additional order of magnitude or two would allow for a discovery or exclusion of the  majority of the parameter space. Just as for indirect detection, much of the ``large'' gauge coupling parameter space is already ruled out, while the ``small'' gauge coupling parameter space is more open.

We first provide an overview of the main categories of ALP searches, and then present our results in Sec.~\ref{sec:acc-results}.

\begin{figure*}[t]
          \includegraphics[width=3.25in]{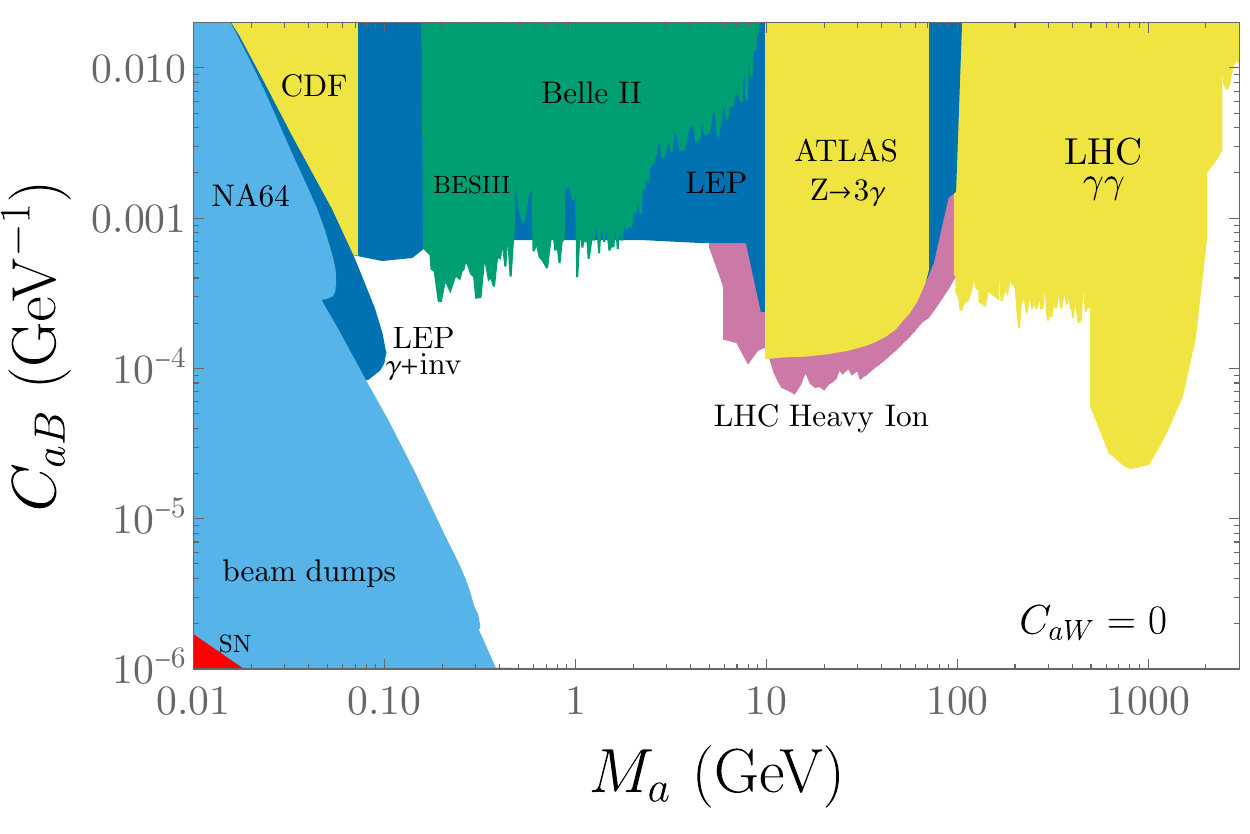}
          \quad
          \includegraphics[width=3.25in]{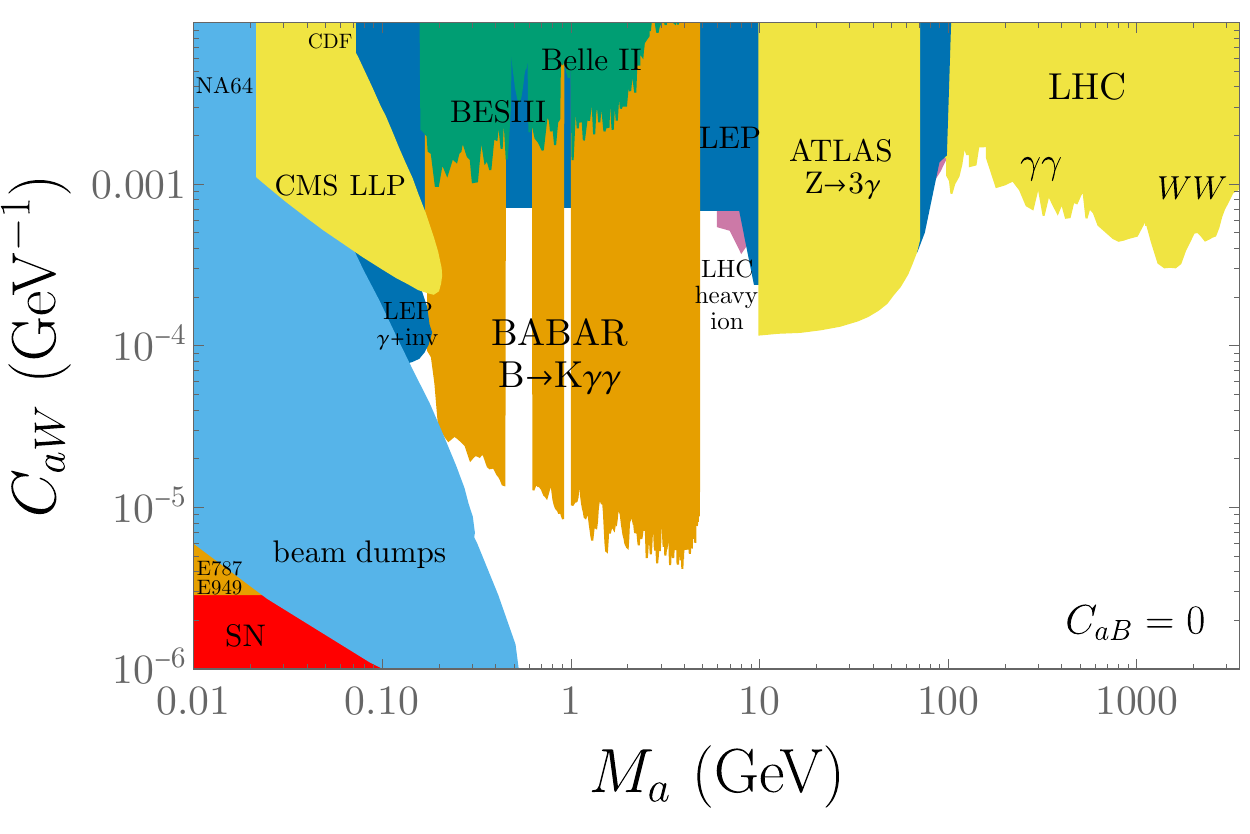}          
\caption{
Bounds from accelerator and collider experiments on (left) the hypercharge-coupled scenario; (right) the $\mathrm{SU}(2)_L$-coupled scenario. Bounds from hadron colliders are in yellow; heavy ion collisions in purple; high-energy electron-positron colliders in dark blue; low-energy electron-positron colliders in green; fixed-target and beam-dump experiments in cyan; and flavor experiments in orange. There is also a sliver of parameter space excluded from supernovae constraints depicted in red. Details of each constraint are found in the text.
}
\label{fig:acc_1}
\end{figure*}

\subsection{Overview of Searches}

Searches for ALPs fall into one of four broad categories:
\begin{itemize}
    \item {\bf Hadron colliders:}~The large center of mass energy of hadron colliders means that these experiment provide the dominant constraints on high-mass ALPs. However, the large quantity of $Z$ and Higgs bosons produced at hadron colliders also provides sensitivity to rare but spectacular decays into lower-mass ALPs. Heavy-ion collisions at the LHC further provide a unique window into ALPs in the 10-100 GeV mass range.

    \item {\bf Electron-positron colliders:}~ALPs can be directly produced in $e^+e^-$ collisions through a virtual $\gamma/Z$ or an on-shell $Z$ boson. Precision studies of rare $Z$ boson decay modes at LEP provide excellent sensitivity to ALPs produced in their decays.

    \item {\bf Fixed-target and beam-dump experiments:}~These experiments provide limited kinematic reach but extremely high luminosities, which provide sensitivity to very feebly coupled ALPs at lower masses.

    \item {\bf Flavor experiments:}~When ALPs couple to $W$ bosons, they can be emitted in flavor-changing neutral current (FCNC) decays \cite{Izaguirre:2016dfi}. The rare nature of these decays in the SM means that flavor experiments have excellent sensitivity to ALPs with masses below the $B$-meson mass.
    
\end{itemize}
We show the  constraints on the hypercharge- and $\mathrm{SU}(2)_L$-coupled scenarios in Fig.~\ref{fig:acc_1}.

We now provide more details on each of the searches. When needed, signal simulations are done using \texttt{MadGraph5\_aMC@NLO} \cite{Alwall:2014hca} with a  \texttt{UFO} \cite{Darme:2023jdn} model file that we created using \texttt{FeynRules} \cite{Alloul:2013bka}.\\

\noindent {\bf Hadron colliders:}~There are  dedicated searches by ATLAS \cite{ATLAS:2023zfc} and CMS/TOTEM \cite{CMS:2023jgd} for photon-coupled ALPs in proton-proton collisions at the LHC. In these searches, the ALP is created in light-by-light scattering with one or both of the protons remaining intact and being tagged in the forward region. These searches collectively span the ALP mass range from 150 GeV to 2 TeV. Because the  results provided by the experimental collaborations are already expressed in terms of the ALP-photon coupling, we can directly map these constraints to our coupling benchmarks.

ALPs are also constrained by inclusive diboson searches. ATLAS has performed an inclusive search for $\gamma\gamma$ resonances \cite{ATLAS:2021uiz}, reporting their results in terms of the fiducial cross section times diphoton branching fraction. We simulate inclusive production of the ALP at parton level, for which the dominant channels are vector-boson-fusion (VBF), $pp\to a+jj$, and associated production, $pp\to a+V$ for some gauge boson $V$. We then decay the ALP into photons and use our simulated events to calculate the fiducial cross section, which ranges from 60-80\% of the total cross section. Finally, we translate the ATLAS bounds onto the ALP coupling parameter space for masses 160 GeV to 3 TeV. This sets the strongest bounds on the lower part of the mass range. This is true even for the $\mathrm{SU}(2)_L$-coupled scenario because the cross section bounds are stronger than other low-mass diboson searches even after the smaller $\gamma\gamma$ branching fraction is taken into account. These limits extend those derived from earlier LHC searches \cite{CMS:2011bsw,ATLAS:2011ab,ATLAS:2012fgo,Jaeckel:2012yz}.

When the ALP couples to $\mathrm{SU}(2)_L$ gauge bosons, the coupling to photons is suppressed by $\tan^2\theta_{\rm W}$ compared to a hypercharge coupling. This dramatically weakens the bounds from light-by-light scattering. This motivates performing additional dedicated ALP searches in $W^+W^-$, $ZZ$, and $\gamma Z$ final states, which have much larger branching fractions than $\gamma\gamma$ in this case. To our knowledge, no such dedicated searches have yet been performed. ATLAS \cite{ATLAS:2017jag} and CMS \cite{CMS:2021klu} have set constraints on VBF production of a narrow spin-0 $WW$ resonance, however, and we can directly apply these bounds to the ALP scenario for masses 300 GeV to 4.5 TeV. Since the ALP decay width scales like $M_a^3$, the width-to-mass ratio increases monotonically with $M_a$ and so we truncate our plots when $\Gamma_a/M_a$ exceeds 10\%, which is approximately the mass resolution of these searches. The $WW$ bound sets the most powerful constraint for the $\mathrm{SU}(2)_L$-coupled scenario for $M_a\gtrsim1.2$ TeV.

We additionally check an ATLAS search for $\gamma Z$ resonances \cite{ATLAS:2017zdf} where the $Z$ decays leptonically. This search is inclusive with respect to production mode but requires that the leptons firing the trigger be part of the reconstructed resonance mass. We include both VBF and associated production in calculating the ALP constraint, but in the case of associated production we conservatively require that the associated gauge boson decay hadronically to avoid the possibility that the event would be triggered on leptons from the associated gauge boson. This removes at most 15\% of the associated production events. We then map the cross section bounds to the ALP coupling. We find that the sensitivity of this search is poorer than in other channels.

Hadron colliders can also probe ALPs much lighter than the weak scale. Light-by-light scattering in ultraperipheral heavy-ion collisions is a sensitive probe of lower-mass ALPs (5-100 GeV) due to the much larger electromagnetic field sourced by ions of large atomic number \cite{Knapen:2016moh}. We apply constraints from ALP searches in heavy-ion collisions at CMS \cite{CMS:2018erd} and ATLAS \cite{ATLAS:2020hii}.

Additionally, the large rate of $Z$ boson production at hadron colliders allows them to have substantial sensitivity to $Z\to \gamma a, a\to\gamma\gamma$. Depending on the ALP mass, the two photons in its decay can either be resolved separately or merged into a single electromagnetic shower. Bounds on the ALP coupling have been derived from an ATLAS search for $Z\to 3\gamma$ \cite{ATLAS:2015rsn,Bauer:2017ris}, which has sensitivity above 10 GeV, and a CDF search for $Z\to2 \gamma$, where the ALP photons are merged for ALP masses below about 75 MeV \cite{CDF:2013lma,Bauer:2017ris}.

Finally, CMS has recently performed a search for long-lived particles decaying in the muon spectrometer \cite{CMS:2021juv}. The muon spectrometer is   used as a sampling calorimeter, giving the search sensitivity to diphoton decays \cite{Mitridate:2023tbj}. Ref.~\cite{Mitridate:2023tbj} did a detailed reinterpretation of the search and presented results for an ALP coupled to $\mathrm{SU}(2)_L$ bosons for masses below 200 MeV, and we include these bounds. They did not do a similar study for ALPs coupled to hypercharge, although the sensitivity is expected to be reduced because of the shorter lifetime associated with the larger diphoton coupling, along with the smaller production cross section from electroweak boson decays. As this search affects only a narrow sliver of parameter space, we leave a reinterpretation of this search to the hypercharge-coupled case to future work.\\

\noindent {\bf Electron-positron colliders:}~Electron-positron colliders currently set strong constraints on ALPs below the $Z$ mass. We apply bounds previously derived from LEP on $Z\to3\gamma$, and on $Z\to \gamma\gamma$ where the two photons from ALP decay have been merged \cite{L3:1994shn,L3:1995nbq,Jaeckel:2015jla}. At smaller ALP masses, it can be long-lived and decay to photon pairs outside the detector. In this regime, LEP constraints on $Z$ boson decays to a photon and an invisible particle put bounds on ALP parameters \cite{L3:1997exg,DELPHI:2008uka,Dolan:2017osp,Mitridate:2023tbj}.

Searches for ALPs decaying to photons have also been performed at low-energy $e^+e^-$ colliders. Belle II has a search for ALPs below 10 GeV in associated production with a photon, $e^+e^-\to\gamma^*\to\gamma a, a\to\gamma\gamma$ \cite{Belle-II:2020jti}. BESIII has two searches in radiative $J/\psi$ decays, $J/\psi\to\gamma a,a\to\gamma\gamma$, which are sensitive to ALP masses below 3 GeV \cite{BESIII:2022rzz,BESIII:2024hdv}.\\

\noindent {\bf Fixed-target and beam-dump experiments:}~The kinematic reach of fixed-target and beam-dump experiment is limited to sub-GeV ALPs, but these experiments have excellent sensitivity to smaller couplings and/or longer lifetimes than other experiments. The strongest constraints from electron beam dumps come from the E137 experiment at SLAC, which places bounds on the photon coupling \cite{Bjorken:1988as,Dolan:2017osp}. The NA64 electron fixed-target experiment probed ALP production via the Primakoff process and subsequent displaced decay of the ALP, either  reconstructed as a photon pair inside one of the downstream calorimeters, or reconstructed as missing energy if the decay occurs outside the detector \cite{NA64:2020qwq}.

For proton beam dumps, the strongest constraints come from the NuCal experiment \cite{Blumlein:1990ay}, and the bounds have recently been updated using improved modelling of ALP production and decay for the hypercharge and $\mathrm{SU}(2)_L$ coupling scenarios \cite{Jerhot:2022chi}. There is a  sliver of parameter space at low mass/coupling that is not excluded  by NuCal. This is covered by supernova bounds; see Sec.~\ref{sec:FI_results} for a more detailed discussion of astrophysical and cosmological bounds on ALPs.

The PrimEx experiment used a photon beam to study the $\pi^0$ decay width via the Primakoff effect \cite{PrimEx:2010fvg}, but the diphoton spectrum can also be used to look for ALP decays into photons \cite{Aloni:2019ruo}. We find that PrimEx constraints are subdominant to other bounds for our benchmarks.\\

\noindent {\bf Flavor experiments:}~When ALPs couple to $\mathrm{SU}(2)_L$ gauge bosons, they can be emitted in FCNC decays \cite{Izaguirre:2016dfi}. The most powerful flavor constraint is set by a BABAR search for $B^\pm \to K^\pm a, a\to\gamma\gamma$ \cite{BaBar:2021ich}. At lower ALP mass/coupling, there are constraints from E787 and E949 on a $K^\pm$ decaying to a $\pi^\pm$ plus an invisible particle  \cite{E787:2004ovg,BNL-E949:2009dza,Ertas:2020xcc} (where the ALP is sufficiently long-lived that it escapes the detector). Additional constraints on invisible ALP decays from NA62 and KOTO, and diphoton ALP decays from E949 and KTeV, are subdominant to the other bounds \cite{Izaguirre:2016dfi,E949:2005qiy,KTeV:2008nqz,NA62:2014ybm,KOTO:2018dsc,Gori:2020xvq,NA62:2020xlg}.

While we have ignored RG-induced couplings because they are typically subdominant, flavor observables are one area where they can have a significant effect on bounds, especially due to the RG-induced ALP-muon coupling. For example, constraints on the decay $B\to K^{(*)}a,a\to\mu^+\mu^-$ \cite{LHCb:2015nkv,LHCb:2016awg} are several orders of magnitude stronger than $B^\pm\to K^\pm a,a\to\gamma\gamma$ and can therefore be competitive with the BABAR ALP search \cite{Izaguirre:2016dfi,Bauer:2021mvw}. We do not include model-dependent dimuon constraints and do not anticipate that they dominate for EFT cutoffs at the 100 GeV to 1 TeV scale, but the interplay of these constraints should be checked in any specific UV completion of the ALP EFT.

\begin{figure*}[t]
          \includegraphics[width=3.25in]{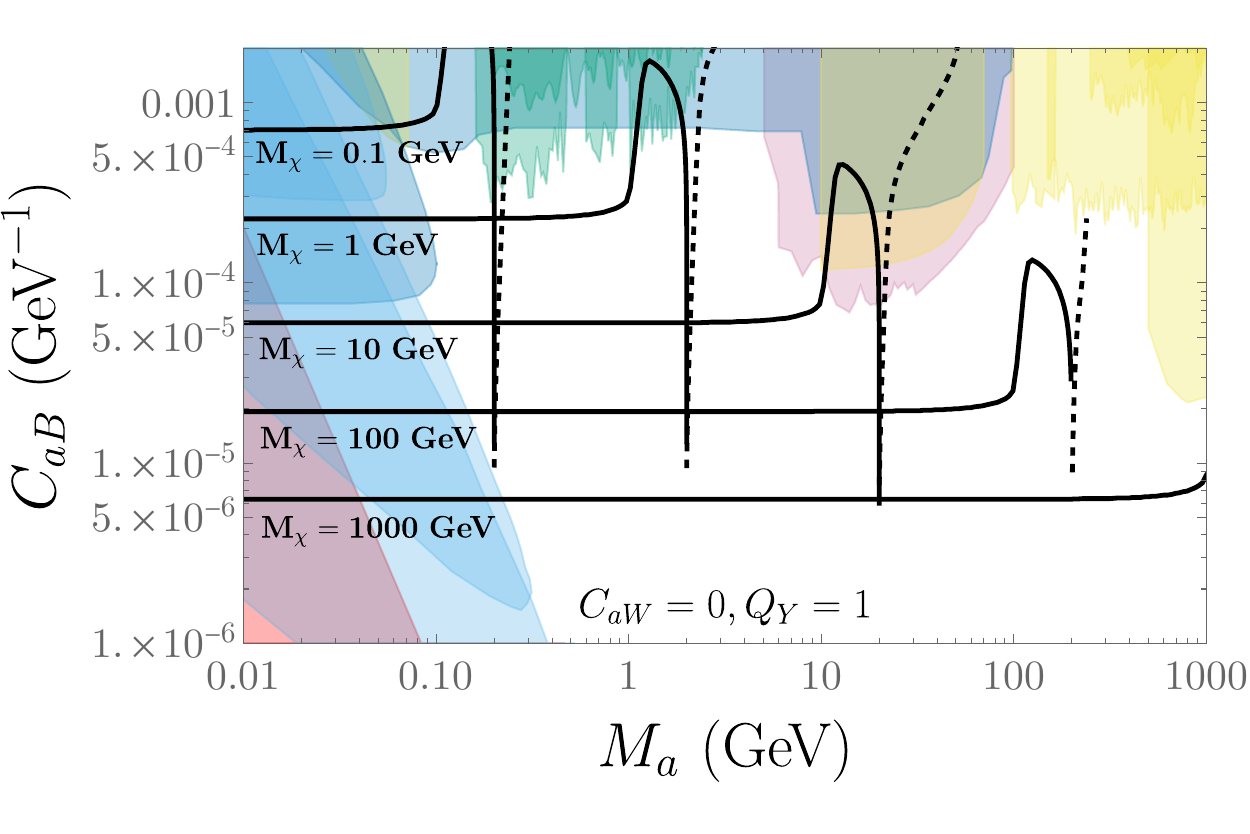}
          \quad
          \includegraphics[width=3.25in]{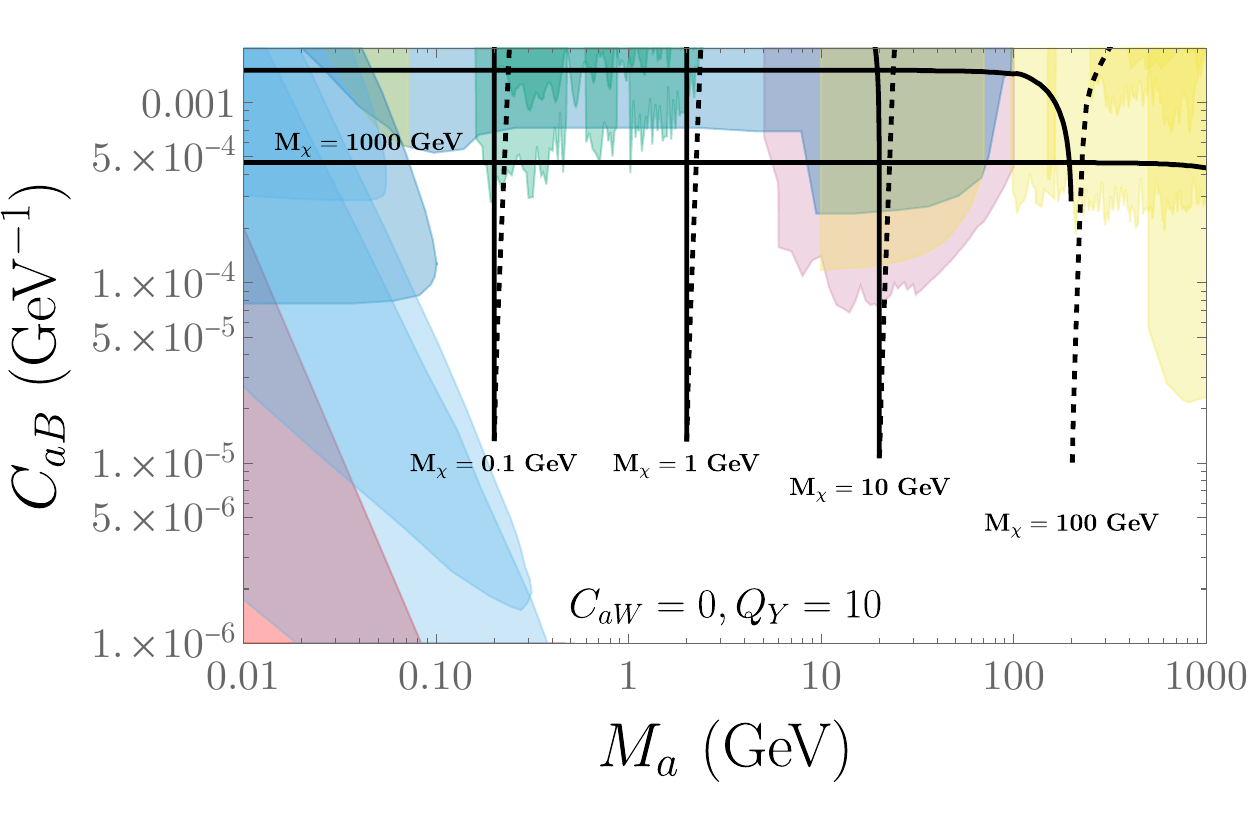}
\caption{
Accelerator and collider bounds on the hypercharge-coupled scenario including thermal-relic contours for various DM masses. These contours assume (left) $Q_Y=1$; (right) $Q_Y=10$. The dashed line indicates the DM parameters for which $\mathrm{BF}(a\to\bar\chi\chi)>50\%$. DM curves are truncated when our treatment of the ALP as a propagating particle is no longer valid, $\Gamma_a/M_a>0.5$. There is also a sliver of parameter space excluded from supernovae constraints depicted in red.
}
\label{fig:acc_Y}
\end{figure*}

\begin{figure*}[t]
          \includegraphics[width=3.25in]{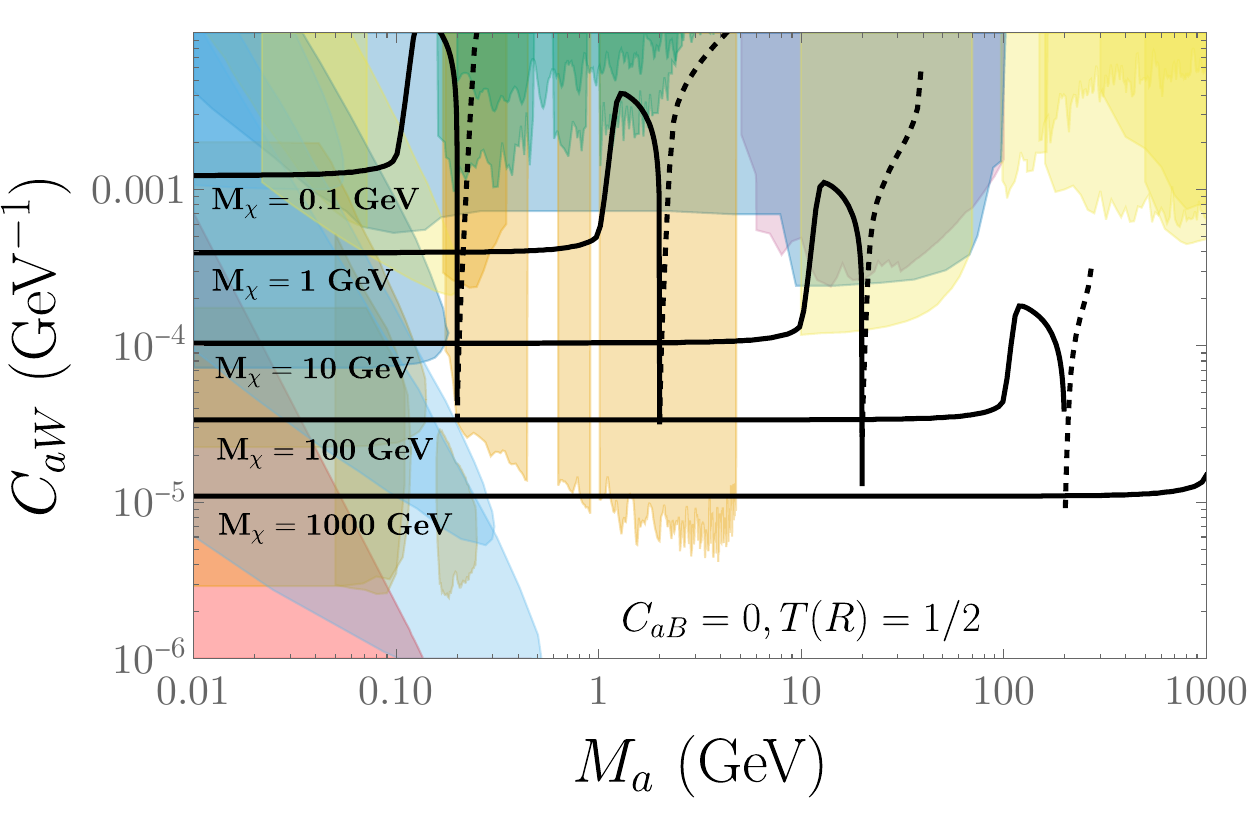}
          \quad
          \includegraphics[width=3.25in]{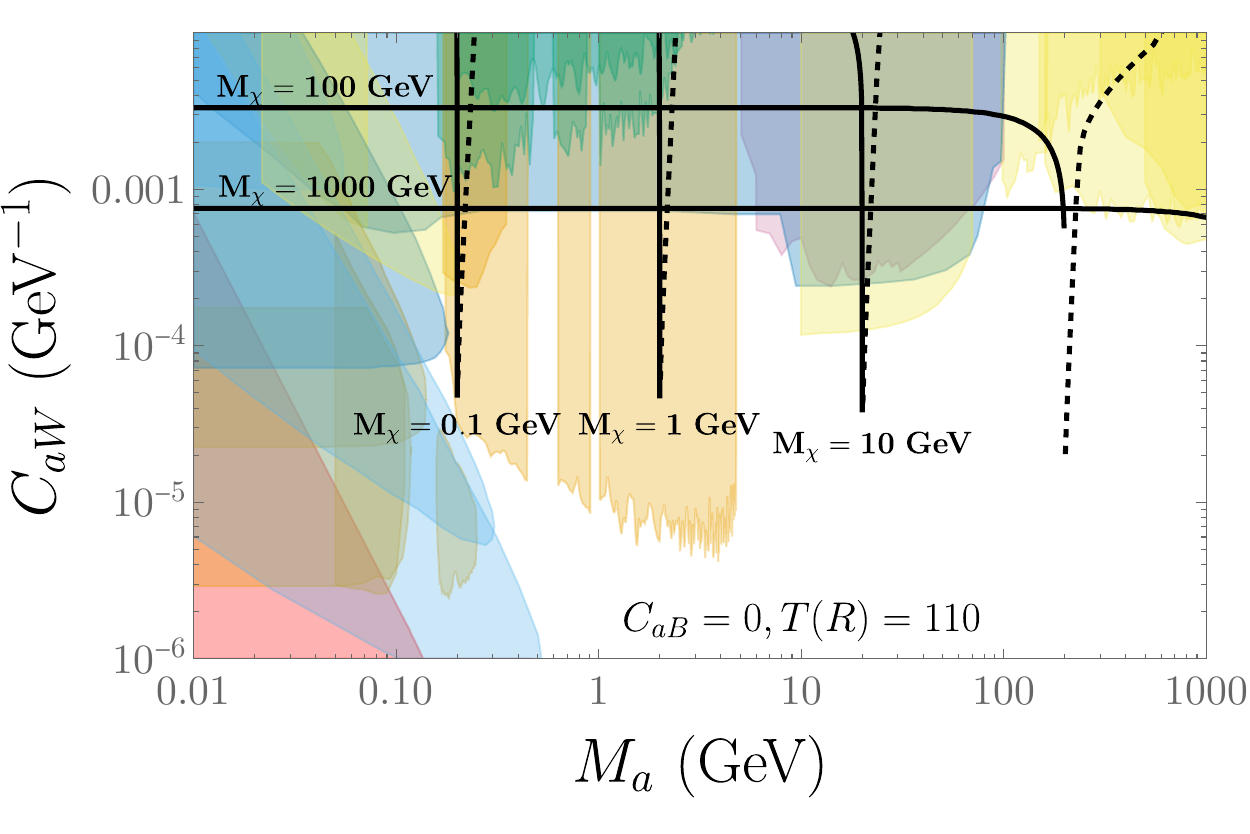}
\caption{
Accelerator and collider bounds on the $\mathrm{SU}(2)_L$-coupled scenario including thermal-relic contours for various DM masses. These contours assume (left) $T(R)=1/2$; (right) $T(R)=110$. The dashed line indicates the DM parameters for which $\mathrm{BF}(a\to\bar\chi\chi)>50\%$. DM curves are truncated when our treatment of the ALP as a propagating particle is no longer valid, $\Gamma_a/M_a>0.5$. There is also a sliver of parameter space excluded from supernovae constraints depicted in red.
}
\label{fig:acc_SU2}
\end{figure*}

\subsection{Results}\label{sec:acc-results}

We first present in Fig.~\ref{fig:acc_1} the constraints on visibly decaying ALPs for each of the gauge-coupling scenarios, independent of any DM considerations. In many parts of the parameter space, the constraints are comparable between the two scenarios, but there are several notable exceptions. First, the gap between 50 MeV and 5 GeV that is present for the hypercharge-coupled ALP is largely filled in the $\mathrm{SU}(2)_L$-coupled scenario because of powerful flavor constraints. There are a few gaps in the sensitivity of the BABAR analysis in the vicinity of the SM diphoton resonances ($\pi^0,\eta,\eta'$), but otherwise much of the low-mass parameter space is already ruled out. At high masses, dedicated LHC searches for ALPs decaying into photon pairs more powerfully constrain the hypercharge scenario than the $\mathrm{SU}(2)_L$ scenario; for the latter, the constraints at high mass predominantly come from $WW$ searches, which suffer larger QCD backgrounds than the $\gamma\gamma$ searches.

We next consider the parameters motivated by the thermal DM abundance. In Figs.~\ref{fig:acc_Y} and \ref{fig:acc_SU2}, we show the coupling needed to obtain the observed DM abundance for fixed DM masses as a function of the ALP mass. We truncate the DM relic curves when our treatment of the ALP as a propagating particle is no longer valid, $\Gamma_a/M_a>0.5$.

In Fig.~\ref{fig:acc_Y}, we show results for the hypercharge-coupled scenario with $Q_Y=1$ and 10, corresponding to ``small'' and ``large'' gauge couplings. For the ``small'' gauge coupling limit, which is largely unconstrained by indirect detection searches, we find that most of the DM-motivated parameter space is beyond current constraints. However, improvements in coupling constraints of one to two orders of magnitude at future experiments would allow either the discovery or exclusion of large portions of the electroweak axion portal parameter space. By contrast, for ``large'' gauge coupling, most of the parameter space is excluded by both indirect detection and accelerator searches except in the immediate vicinity of resonance. Even the resonantly enhanced parameter space could be tested with the next generation of experiments, although for much of this parameter space the ALP decay into DM is kinematically allowed and the ALP no longer has a substantial visible decay fraction. We use dashed lines to indicate the parts of the DM contours  when $\mathrm{BF}(a\to\bar\chi\chi)>50\%$.

In Fig.~\ref{fig:acc_SU2}, we show results for the $\mathrm{SU}(2)_L$-coupled scenario. Once again, we have ``small'' and ``large'' gauge coupling limits of $T(R) = 1/2$ and 110, respectively. The DM relic curves are qualitatively similar to those for hypercharge coupling, except that much of the viable parameter space is ruled out for ALP masses below 5 GeV due to strong flavor constraints. This is true even in the immediate vicinity of resonances, provided the decay to visible states still dominates the ALP decay width.

Finally, we comment on the coupling peak (or, more accurately, dip) of the resonant annihilation region. As argued in Sec.~\ref{sec:resonance}, the coupling needed to obtain a thermal abundance is independent of $M_a$ for annihilation slightly above resonance, but scales like $M_a^{-1/3}$ for annihilation exactly on resonance. Consequently, we find at lower ALP masses that the smallest allowed coupling near resonance is associated with annihilation just above resonance and the peaks are of similar height, modulated only by the fact that $g_*$ changes steeply for $T\sim0.1-10$ GeV and therefore the couplings predicted by freeze-out vary slightly due to this effect. At larger ALP masses, annihilation exactly on resonance becomes more important due to the $M_a^{-1/3}$ scaling of the coupling on resonance, and consequently deeper troughs are predicted at larger $M_a$.

\section{Freeze-In Dark Matter}\label{sec:freeze_in}

For sufficiently small couplings, DM never reaches thermal equilibrium and its abundance is established through freeze-in. In this case, it is also possible that the ALP is out of equilibrium for some or most of the relevant cosmic history during which DM is produced. However, since ALPs are produced at $\mathcal{O}(g_{aVV}^2)\sim\frac{1}{f^2}$ and DM is produced at $\mathcal{O}(\frac{g_{aVV}^2}{f^2})\sim\frac{1}{f^4}$, the ALP thermalizes in a large share of the  DM freeze-in parameter space.

Because ALPs are pseudo-Goldstone bosons, their couplings arise from higher-dimensional operators. The additional degree of momentum-dependence in ALP-mediated reaction densities results in larger scattering rates relative to Hubble expansion at earlier times, and this leads to UV freeze-in \cite{Elahi:2014fsa}. Predictions of UV freeze-in models are sensitive to the reheat temperature, $T_{\rm RH}$, although as we will see there are also freeze-in modes that occur predominantly in the infrared (IR). 

We assume an instantaneous reheating such that the DM and ALP abundances are zero at $T_{\rm RH}$. When freeze-in proceeds through an EFT operator of dimension-7 or lower, as is the case for the axion portal, the instantaneous approximation largely agrees with results from more careful implementations of reheating \cite{Garcia:2017tuj}. However, the approximation breaks down for masses close to or exceeding the reheat temperature, and so we only show numerical results for $M_\chi,M_a\le T_{\rm RH}$.

Our findings qualitatively agree with the regimes identified in previous studies of ALP-portal freeze-in that made different coupling assumptions \cite{Bharucha:2022lty,Ghosh:2023tyz,Fitzpatrick:2023xks,Dror:2023fyd}. We restrict ourselves to the hypercharge-coupled scenario since we expect results with  $\mathrm{SU}(2)_L$ coupling to be similar. We find inconsistencies with the results from the freeze-in module of \texttt{micrOMEGAs}, however, and so we use our own Boltzmann equations to study freeze-in. 

\subsection{Calculation of DM Abundance}

We solve the same Boltzmann equations as before, although we relax the assumption that the ALP and/or DM begin in thermal equilibrium. This means we track the ALP abundance in addition to the DM abundance. DM can be directly produced in $\gamma\gamma\to\bar\chi\chi$ collisions via an intermediate virtual ALP, and from within the hidden sector through ALP decays ($a\to\bar\chi\chi$) and scattering ($aa\to\bar\chi\chi$). ALPs are produced through inverse decays ($\gamma\gamma\to a$) and scattering processes (such as $e^+e^-\to\gamma a$ and $e^-\gamma\to e^- a$) involving SM gauge bosons. The latter are suppressed by $\alpha_{\rm EM}$ relative to inverse decays, but their contribution to the ALP abundance is enhanced by $T_{\rm RH}/M_a$ which can dominate for higher reheat temperatures.

The ALP production rate through $2\to2$ scattering processes depends on which charged species are present in the plasma and is therefore sensitive to various SM mass thresholds. A full treatment of all the mass-dependent effects of every species would be  involved and likely have negligible impact on the results when the reheat temperature is far from any mass thresholds due to the UV-dominance of scattering. We adopt a simplified treatment where we include only scattering involving charged SM fermions, both leptons and quarks ($\bar ff\to \gamma a$, $f\gamma\to fa$, and $\bar f\gamma\to\bar f a$). We neglect fermion masses in these rates  except for when they act as regulators for $t$-channel logarithmic divergences, and we only include scattering involving a fermion $f$ for temperatures $T\ge M_f$\footnote{We use the kaon mass for the strange quark threshold and the pion mass for the up and down quark thresholds.}. We have checked that using the massless rates for the reaction densities agrees  well with the full rates for $T\gtrsim M_f$;  the rates are in any case Boltzmann-suppressed for smaller temperatures. This treatment allows us to efficiently obtain results over a wide parameter space to accuracy in the couplings to better than $\mathcal{O}(1)$ precision.

Our Boltzmann equations assume that the only ALP-SM coupling is to photons:~this is perfectly fine for reheat temperatures below the electroweak scale. For reheat temperatures well above the electroweak scale, the correct treatment in the hypercharge-coupled scenario replaces a photon with a hypercharge boson, and so we can simply replace $g_{a\gamma\gamma}\to C_{aB}$. In this limit, we additionally include $a\to\gamma Z$ and $a\to ZZ$ decays when calculating the ALP width. 

We provide more details of the Boltzmann equations in Appendix \ref{app:BE}.

\subsection{Regimes of Freeze-In}

The dominant mode of DM production depends on the relative values of $M_a$, $M_\chi$, and $T_{\rm RH}$, as well as on the relative strength of the ALP coupling to SM gauge bosons vs.~DM. We choose the same coupling benchmarks as outlined in Sec.~\ref{sec:benchmarks}. We adopt the following approach:~we fix $T_{\rm RH}$ and $M_\chi$, and then study how the freeze-in coupling depends on $M_a$. We then repeat for different values of $T_{\rm RH}$ and $M_\chi$. 

Starting from $M_a\ll M_\chi$ and then gradually increasing $M_a$, we find the following regimes for freeze-in:\\

\noindent {\bf $\mathbf{M_a}$ below DM decay threshold:}~In this limit, $M_a<2M_\chi$, the $a\to\bar\chi\chi$ decay is kinematically forbidden and the ALP always decays visibly. DM is produced via scattering at $\mathcal{O}(\frac{1}{f^4})$, and to obtain an appreciable DM abundance the coupling is large enough to thermalize $a$ (whose production rate is $\sim\frac{1}{f^2}$). The dominant DM production mechanism is typically $aa\to\bar\chi\chi$ and occurs most rapidly just after reheating. If $T_{\rm RH}\gg M_a, M_\chi$, then this DM  production cross section scales like $\frac{M_\chi^2}{f^4}$ and is independent of $M_a$. Consequently, we see an essentially flat dependence of the freeze-in coupling with increasing $M_a$ in this regime, while $C_{aB}\sim\frac{1}{f}$ varies with DM mass like $M_\chi^{-1/2}$.\\

\noindent {\bf $\mathbf{M_a}$ just above DM decay threshold:}~As $M_a$ increases above $2M_\chi$, the DM production channel $a\to\bar\chi\chi$ opens up. For our benchmarks, the ALP coupling to SM gauge bosons is smaller than that to DM, so the ALP does not need to be much above threshold before $\mathrm{BF}(a\to\bar\chi\chi)$ becomes nearly 100\%. Since ALPs are produced at $\mathcal{O}(\frac{1}{f^2})$ and every ALP decays into a $\chi$ and a $\bar\chi$, the DM abundance is now also established at $\mathcal{O}(\frac{1}{f^2})$. Consequently, the only viable parameter space has both DM and the ALP out of equilibrium, and the coupling needed to achieve the correct DM abundance plummets by many orders of magnitude. The fact that the ALP decays invisibly in this regime also affects certain terrestrial and cosmological constraints on ALPs.

The predicted freeze-in coupling is largely independent of $M_a$ in this limit because the ALP production rate is also largely independent of $M_a$, with a minor dependence on $M_a$ arising from the fact that the inverse decay rate $\gamma\gamma\to a$ depends on $M_a$ but is typically subdominant to $2\to2$ scattering (along with a mild logarithmic dependence on $M_a$ for scattering processes with a $t$-channel singularity). Since $Y_\chi^{\rm after\,\, decay} = Y_a^{\rm before\,\, decay}$, the DM abundance remains fixed for fixed coupling independent of $M_a$.\\

\noindent {\bf The ALP is visible again:}~As $M_a$ continues to increase, its visible branching fraction also increases until it dominates the decay width. To see why this is the case, note that $\mathrm{BF}(a\to\bar\chi\chi) \sim \frac{M_\chi^2}{M_a^2}$ and decreases with larger ALP masses. We now have $Y_\chi^{\rm after\,\,decay} = \mathrm{BF}(a\to\bar\chi\chi)Y_a^{\rm before\,\, decay}$, and consequently a larger ALP-photon coupling is needed for larger ALP masses to produce more ALPs in order to compensate for the diminished DM branching fraction. According to this argument, we predict a scaling $C_{aB}\sim M_a$. Somewhat counterintuitively, the ALP also decays visibly in experiments, even though its decay to DM is kinematically allowed. The value of $M_a$ at which we enter this regime depends on the relative coupling strength to DM vs.~photons:~a smaller coupling to photons implies larger values of $M_a$ for the transition to visible decays.\\

\noindent {\bf The ALP reaches equilibrium:}~The scaling in the previous regime continues until the ALP reaches equilibrium at reheating. At this point, increasing the SM coupling to ALPs no longer increases the ALP abundance since $Y_a$ saturates at $Y_a^{\rm eq}$. The result is a small range of ALP masses over which the freeze-in coupling can vary by many orders of magnitude since its effect on the ALP abundance (and consequently the DM abundance) has little effect in the approach to equilibrium. This transition region can appear as a sharp kink or a mild wiggle depending on the relative values of various parameters.

\begin{figure*}[t]
          \includegraphics[width=3.25in]{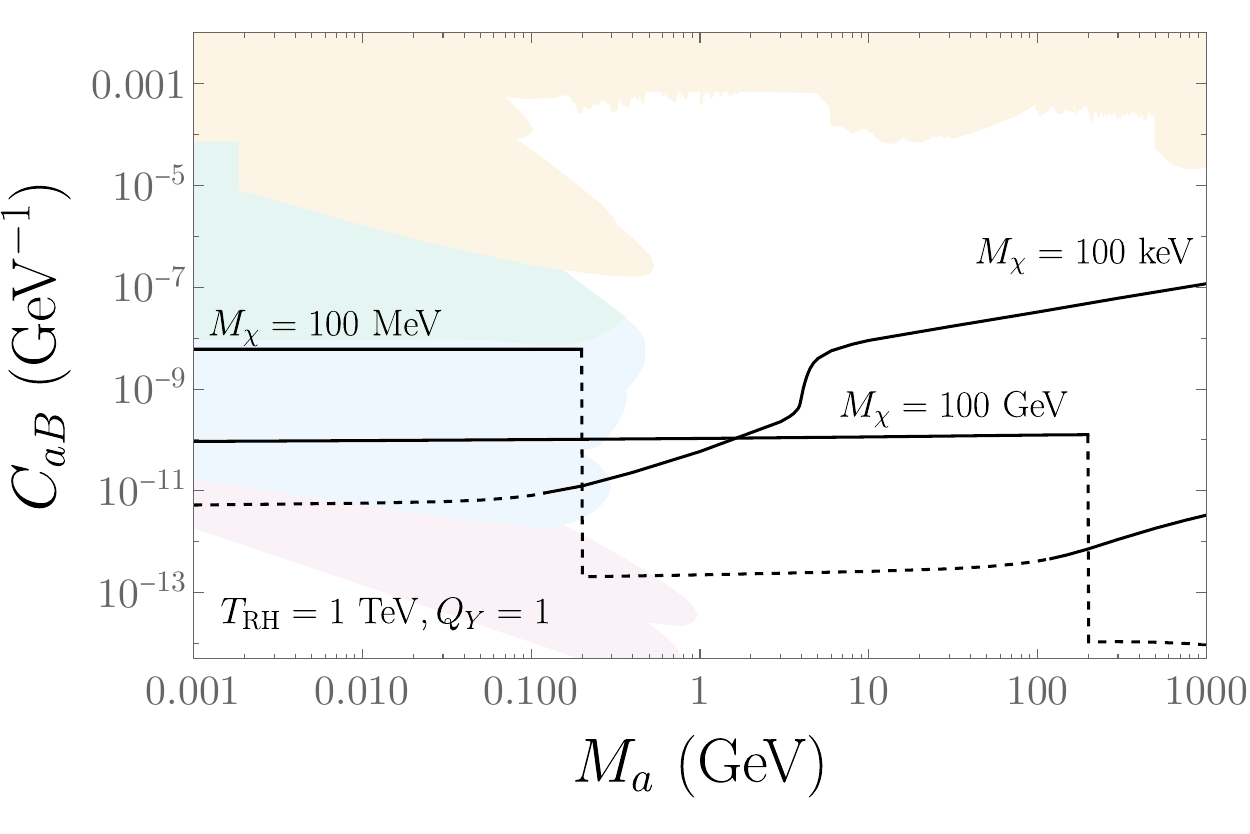}
          \quad
          \includegraphics[width=3.25in]{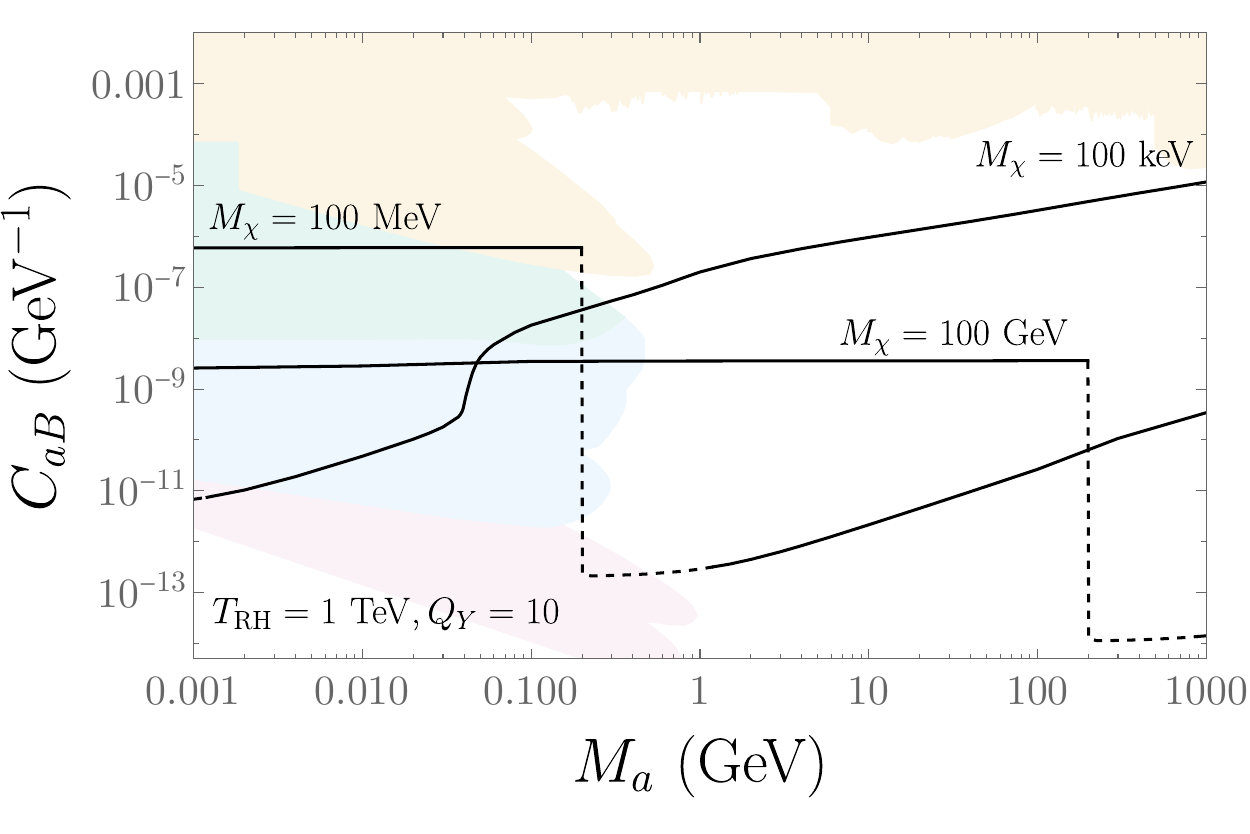}\\
        \includegraphics[width=3.25in]{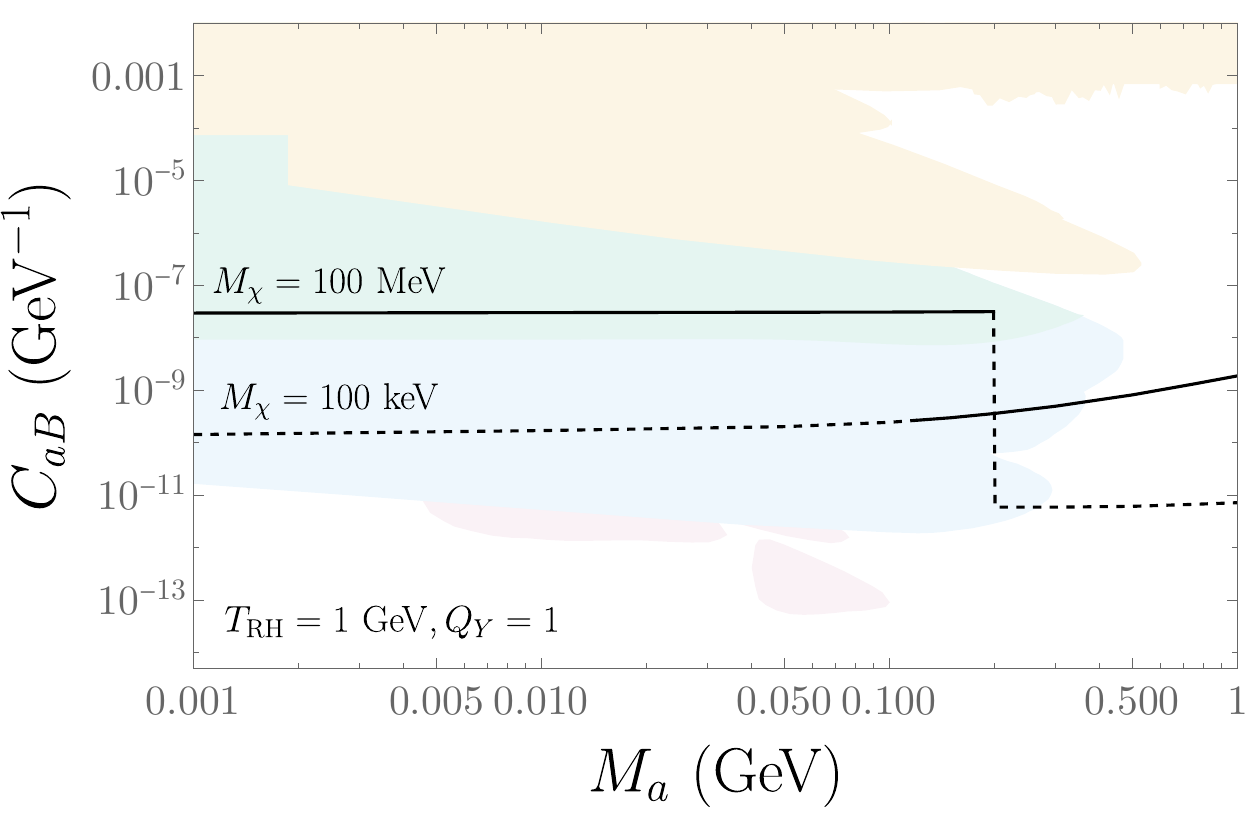}
          \quad
          \includegraphics[width=3.25in]{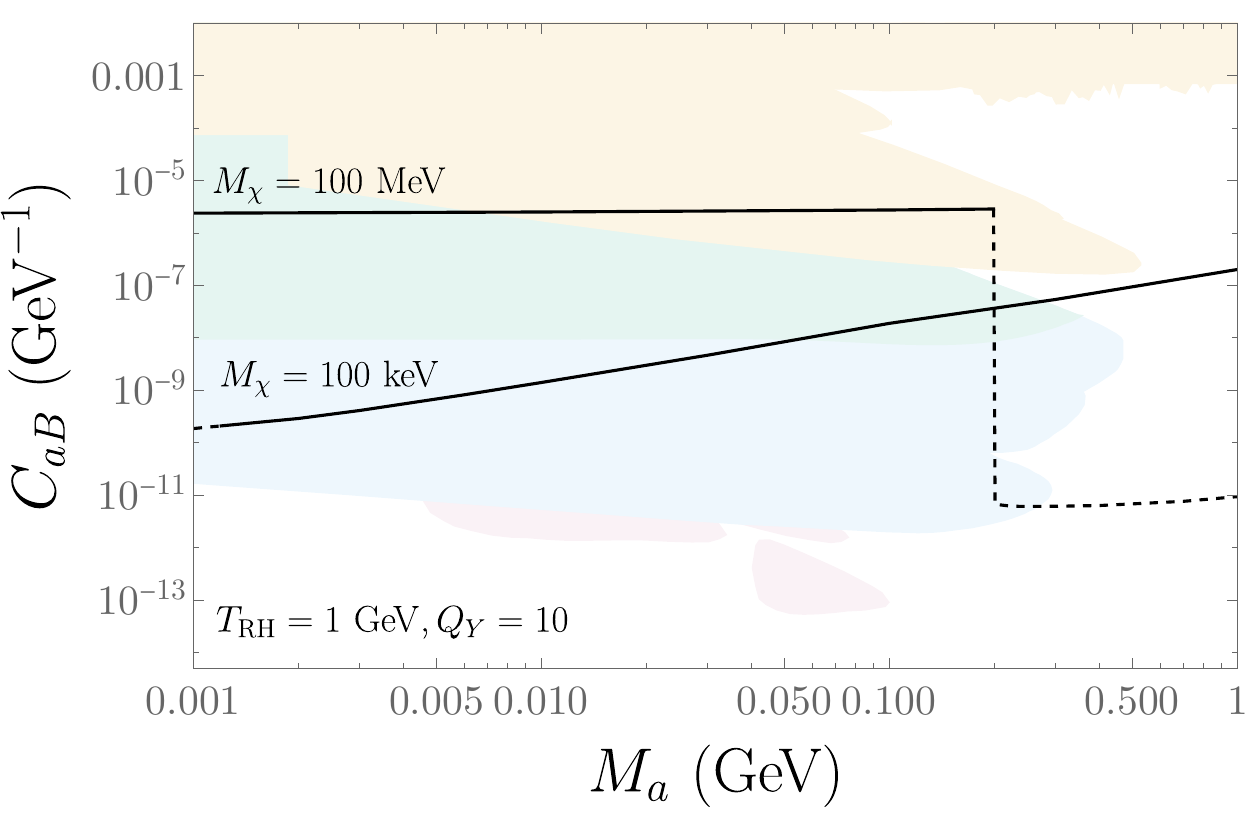}\\
        \includegraphics[width=3.25in]{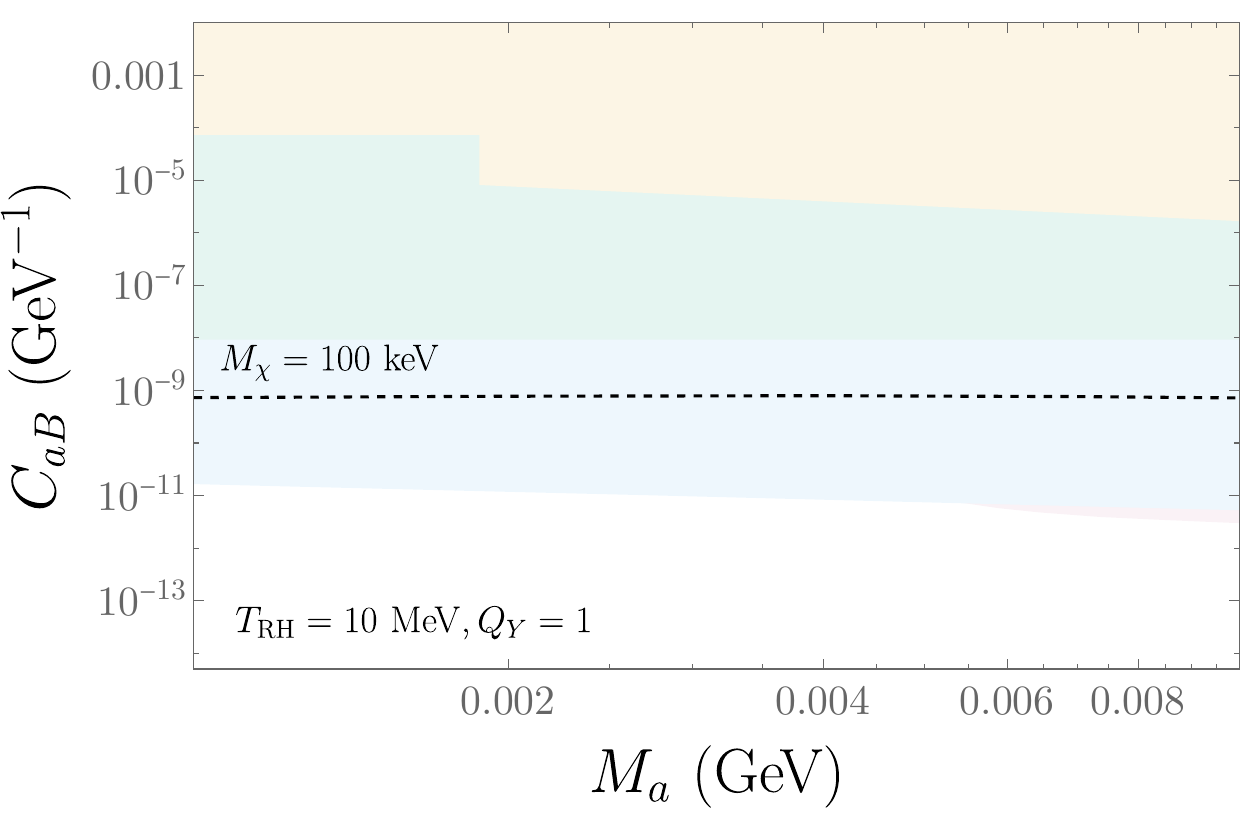}
          \quad
          \includegraphics[width=3.25in]{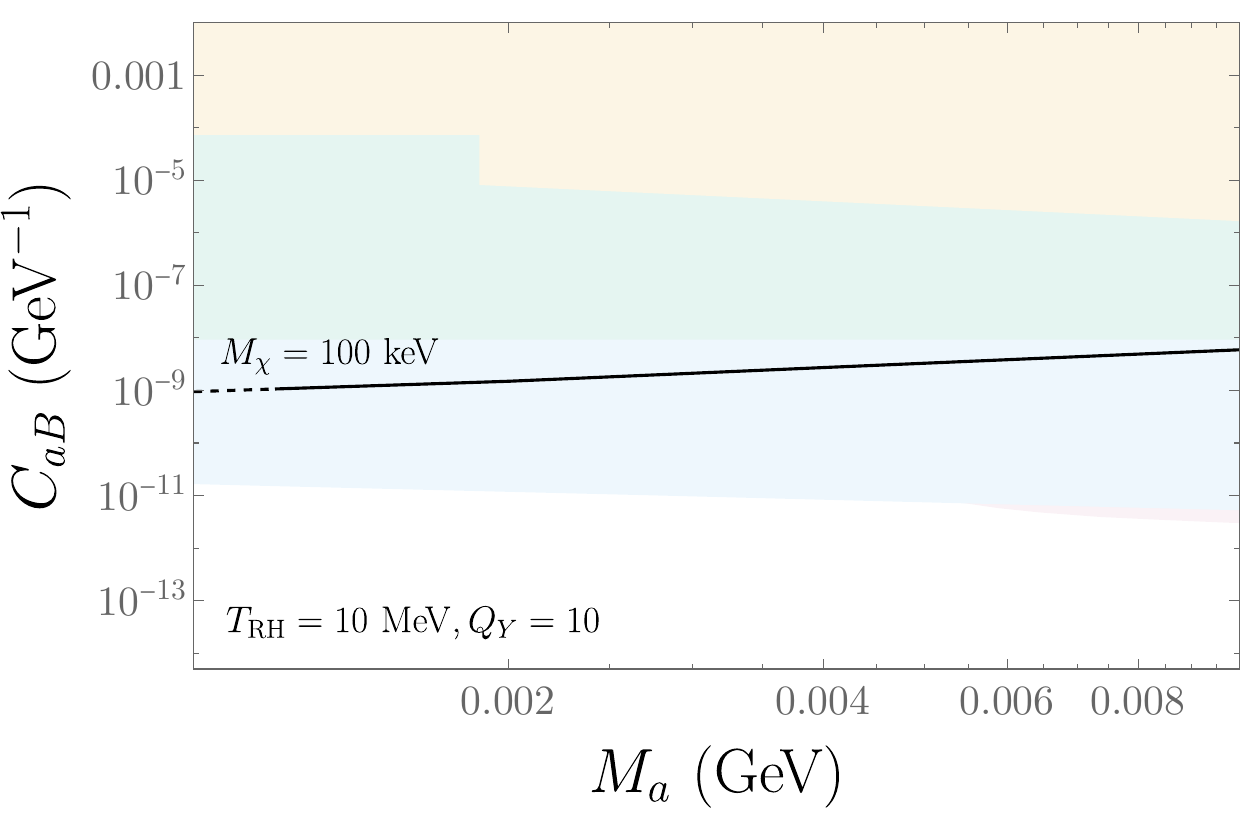}
\caption{
Estimated couplings  for DM freeze-in along with observational constraints. In all plots, $C_{aW}=0$, and we take $Q_Y=1$ ($Q_Y=10$) in the left (right) column. We assume a different reheat temperature in each row:~(top) 1 TeV; (middle) 1 GeV; (bottom) 10 MeV. In this figure, we consider only masses satisfying $M_\chi,M_a\le T_{\rm RH}$. Cosmological constraints are shown in red,  supernova cooling in green, gamma rays and axion-sourced fireballs from supernovae and neutron star mergers in blue, and accelerator/collider constraints in orange. The DM abundance curves are dashed when $\mathrm{BF}(a\to\bar\chi\chi)>50\%$; in these regions, most accelerator/collider, cosmological, and gamma-ray/fireball constraints are relaxed.
}
\label{fig:freezein}
\end{figure*}

For $T_{\rm RH}\gg M_a$, the fact that the ALP reaches equilibrium at the time of reheating does not mean that it \emph{stays} in equilibrium throughout its history since the $2\to2$ scattering rate falls sharply as the universe cools. For sufficiently large couplings, it also maintains equilibrium at $T\sim M_a$ through (inverse) decays to photons. This qualitatively changes the DM prediction:~whereas before we had $Y_{\rm \chi}^{\rm after\,\,decay} \propto Y_a^{\rm before\,\,decay}$, now the ALP abundance is continually replenished from the SM bath as it decays into DM. The result is that the DM abundance depends on    $\Gamma(a\to\bar\chi\chi)$  (\emph{not} the branching fraction) integrated over time. We can approximate this integrated quantity as $\Gamma(a\to\bar\chi\chi)/H(T=M_a)$, which scales as $\frac{1}{f^2 M_a}$. We therefore predict that the coupling $C_{aB}\sim\frac{1}{f}\sim\sqrt{M_a}$ in this regime.\\

\noindent {\bf The ALP mass exceeds the reheat temperature:}~When $M_a > T_{\rm RH}$, the DM production rate from ALP decays becomes Boltzmann suppressed. The result is that DM is produced from photon scattering  via a virtual intermediate ALP with a maximum rate $\Gamma_{\gamma\gamma\to\bar\chi\chi}\sim \frac{C_{aB}^2M_\chi^2T_{\rm RH}^7}{f^2 M_a^4}$. Consequently, arbitrarily large values of $C_{aB}$ are  viable for freeze-in provided $T_{\rm RH}\ll M_a$. At the same time, the ALP EFT might break down in this limit and so care is needed to interpret any results. We do not quantitatively pursue this regime in any further detail, but nevertheless point it out as a phenomenological possibility.

\subsection{Numerical Results}\label{sec:FI_results}
To illustrate the regimes outlined in the previous section and to numerically compare the DM parameter space with current bounds, we select several reheat temperatures and DM masses, and for each combination of $T_{\rm RH}$ and $M_\chi$ we show the DM-motivated coupling as a function of the ALP mass. We consider reheat temperatures of 1 TeV, 1 GeV, and 10 MeV, with the latter approximating the lowest value of $T_{\rm RH}$ consistent with Big-Bang Nucleosynthesis (BBN). We study DM masses ranging from 100 keV (chosen to be consistent with structure-formation constraints on warm freeze-in DM \cite{Zelko:2022tgf}) to 100 GeV, and ALP masses ranging from 1 MeV to 1 TeV. We present results for the hypercharge-coupled scenario with $Q_Y=1$ and 10.

We show our results in Fig.~\ref{fig:freezein}. Collectively, each of the freeze-in regimes from the previous section is present in our results (for example:~the sharp drop in coupling predicted at $M_a=2M_\chi$, the sharp increase in coupling predicted well above $2M_\chi$ where the ALP equilibrates, the various coupling-mass scaling relations, etc.). We see, as expected, that the range of ALP masses for which $\mathrm{BF}(a\to\bar\chi\chi)\approx100\%$ is much smaller for $Q_Y=10$ than for $Q_Y=1$, and consequently the value of $M_a$ at which the ALP comes into equilibrium above the DM threshold is lower at higher $Q_Y$. It is evident that freeze-in DM is consistent with a very wide range of DM masses, ALP masses, and coupling strengths.

We also show constraints on the ALP assuming that it decays visibly, which is true for most ALP masses even above the DM threshold. The accelerator and collider constraints are the same as those derived in  Sec.~\ref{sec:acc}. We impose constraints from BBN on primordial element abundances \cite{ParticleDataGroup:2018ovx} and from Planck  on new relativistic degrees of freedom \cite{Planck:2018vyg}; these were derived  in Ref.~\cite{Depta:2020wmr} for ALP freeze-in with various reheat temperatures. We also include astrophysical constraints from supernova cooling \cite{Ertas:2020xcc}, gamma rays and fireballs from ALP bursts produced in supernovae \cite{Jaeckel:2017tud,Hoof:2022xbe,Muller:2023vjm,Diamond:2023scc}, low-energy supernova \cite{Caputo:2022mah}, and axion-sourced fireballs produced during neutron stars mergers \cite{Diamond:2023cto}. We use the tabulated bounds in \texttt{AxionLimits} when applying astrophysical constraints \cite{AxionLimits}. Note that most of the constraints apart from supernova cooling are relaxed when the ALP decays invisibly. Indirect-detection bounds are not typically strong enough to constrain the freeze-in parameter space.

Taken together, we see that the freeze-in scenario is somewhat constrained for ALP masses below 100 MeV, although there is still viable parameter space depending on the DM mass and reheat temperature, especially when the ALP decays invisibly. For larger ALP masses, there are essentially no constraints on freeze-in:~the predicted couplings are sufficiently small that it will be challenging for even future experiments to achieve sensitivity.

\section{Discussion and Conclusions}\label{sec:conclusion}

We have systematically studied the cosmology and phenomenology of electroweak ALP portal DM. For thermal DM, we find the parameter space already meaningfully constrained by a combination of indirect detection of DM annihilation into photon lines, and direct experimental searches for ALPs. There remains a substantial viable parameter space, much of which will be tested in future experiments. In particular, closing the well known MeV-GeV gap in gamma ray searches could significantly improve sensitivity to DM masses in this range.

Interestingly, over much of the thermal parameter space we find that the validity of the EFT is somewhat marginal, with the PQ-breaking scale typically constrained to $f\lesssim1$ TeV. The radial mode(s) of the PQ-breaking sector and heavy fermions mediating ALP-gauge interactions, all of which we have neglected, could play an appreciable role in the model's cosmology and phenomenology. This is particularly important for LHC searches since deviations from the EFT will first appear at high-energy colliders. A future detailed study is warranted of how ALP phenomenology changes in UV completions of the model, along with the interplay between ALP constraints and direct searches for new heavy states.

We have also investigated the possibility of freeze-in DM, finding that some parameter space, particularly at low ALP masses, is constrained by a combination of astrophysical and cosmological bounds. However, the predicted couplings are sufficiently small that there are few meaningful constraints on freeze-in for ALP masses above 100 MeV. We adopted a simple, instantaneous model of reheating, and it is possible that some of our DM-motivated parameters would change with a more realistic treatment of reheating.

\section*{Acknowledgments}
We are grateful to Benjamin Khoury and Chris Ranlett  for collaboration at early stages of this work. We thank Eder Izaguirre, Tongyan Lin and  Dave Tucker-Smith for helpful conversations. We are grateful to Edoardo Vitagliano for his helpful comments on astrophysical constraints.  This work is supported by Research Corporation for Science Advancement
through Cottrell Scholar Grant \#27632, and by the Campbell Fund of Harvey Mudd College. BS is grateful for the hospitality of Perimeter Institute, where part of this work was carried out. Research at Perimeter Institute is supported in part by the Government of Canada through the Department of Innovation, Science and Economic Development and by the Province of Ontario through the Ministry of Colleges and Universities.

\appendix 

\section{Boltzmann Equations} \label{app:BE}

In this appendix, we present the coupled Boltzmann equations for our model. The cross section and decay rates can be found in App.~\ref{app:CS}.

\subsection{Freeze-Out}

For thermal DM, most of the DM-coupling curve is derived using \texttt{micrOMEGAs}. However, we use the Boltzmann equations presented below to validate results from \texttt{micrOMEGAs}. We also use solutions to our Boltzmann equations in the vicinity of the ALP resonance due to greater numerical stability. We include only ALP-photon couplings, although in Sec.~\ref{sec:resonance} we have described how the cross section on resonance can be modified to take into account decays to weak gauge bosons as well.

For thermal DM, we can assume the ALP is always in equilibrium, while $\chi$ is in kinetic (but not necessarily chemical) equilibrium. We use Maxwell-Boltzmann statistics and neglect quantum statistical factors in the Boltzmann equation. We use $Y_\chi\equiv n_\chi/s$ to track the $\chi$ comoving number density, and $x\equiv M_\chi/T$ as a dimensionless time variable. The Boltzmann equation for $\chi$ is:
\begin{widetext}
 \be\label{eq:BE}
\frac{dY_\chi}{dx} &=& -\frac{s\langle\sigma_{\bar\chi\chi\to\gamma\gamma}v\rangle^{\rm sub} (Y_\chi^{\rm eq})^2}{x H(x)}\left[\frac{Y_\chi^2}{(Y_\chi^{\rm eq})^2}-1\right] -\frac{s\langle\sigma_{\bar\chi\chi\to aa}v\rangle (Y_\chi^{\rm eq})^2}{x H(x)}\left[\frac{Y_\chi^2}{(Y_\chi^{\rm eq})^2}-1\right]\nonumber\\
&&{}-\frac{\langle\Gamma_a\rangle Y_a^{\rm eq}}{x H(x)}\mathrm{BF}(a\to\bar\chi\chi)\left[\frac{Y_\chi^2}{(Y_\chi^{\rm eq})^2}-1\right].
\ee   
\end{widetext}

 In deriving these equations, we have used the real-intermediate-state (RIS) subtraction scheme \cite{Kolb:1979ui,Kolb:1979qa,Fry:1980ph} to remove double-counting from the portion of the $\bar\chi\chi\to\gamma\gamma$ process in which the ALP is on shell,
\begin{align}
\langle\sigma_{\bar\chi\chi\to\gamma\gamma}v\rangle^\text{sub}=& \langle\sigma_{\bar\chi\chi\to\gamma\gamma}v\rangle\nonumber\\
    &{}-\frac{Y_a^{\rm eq}}{s(Y_\chi^{\rm eq})^2}\langle\Gamma_a\rangle\mathrm{BF}(a\to\bar\chi\chi)\mathrm{BF}(a\to\gamma\gamma). 
\end{align}
\subsection{Freeze-In}
For freeze-in, we can no longer assume the ALP is in equilibrium. We now solve a system of Boltzmann equations for $Y_a(x)$ and $Y_\chi(x)$. Because $2\to2$ scattering predominantly occurs in the UV at $T\sim T_{\rm RH}$ while inverse decays into ALPs occur in the IR at $T\sim M_a$, the phase space of these processes does not appreciably overlap and this eliminates the need for the RIS subtraction. The Boltzmann equations are:
\begin{widetext}
 \be\label{eq:BE}
\frac{dY_a}{dx} &=& -\frac{\langle\Gamma_a\rangle Y_a^{\rm eq}}{x H(x)}\mathrm{BF}(a\to\gamma\gamma)\left(\frac{Y_a}{Y_a^{\rm eq}}-1\right)-\frac{\langle\Gamma_a\rangle Y_a^{\rm eq}}{x H(x)}\mathrm{BF}(a\to\bar\chi\chi)\left[\frac{Y_a}{Y_a^{\rm eq}}-\frac{Y_\chi^2}{(Y_\chi^{\rm eq})^2}\right] \\
&&+{}\frac{2s\langle\sigma_{\bar\chi\chi\to aa}v\rangle (Y_\chi^{\rm eq})^2}{x H(x)}\left[\frac{Y_\chi^2}{(Y_\chi^{\rm eq})^2}-\frac{Y_a^2}{(Y_a^{\rm eq})^2}\right]- \frac{s\langle\sigma_{\mathrm{SM}\,a\to\mathrm{SM\,SM}}v\rangle Y_{\rm SM}^{\rm eq}Y_a^{\rm eq}}{x H(x)}\left(\frac{Y_a}{Y_a^{\rm eq}}-1\right),\\
\frac{dY_\chi}{dx} &=& -\frac{s\langle\sigma_{\bar\chi\chi\to\gamma\gamma}v\rangle (Y_\chi^{\rm eq})^2}{x H(x)}\left[\frac{Y_\chi^2}{(Y_\chi^{\rm eq})^2}-1\right] -\frac{s\langle\sigma_{\bar\chi\chi\to aa}v\rangle (Y_\chi^{\rm eq})^2}{x H(x)}\left[\frac{Y_\chi^2}{(Y_\chi^{\rm eq})^2}-\frac{Y_a^2}{(Y_a^{\rm eq})^2}\right]\\
&&{}-\frac{\langle\Gamma_a\rangle Y_a^{\rm eq}}{x H(x)}\mathrm{BF}(a\to\bar\chi\chi)\left[\frac{Y_\chi^2}{(Y_\chi^{\rm eq})^2}-\frac{Y_a}{Y_a^{\rm eq}}\right].
\ee   
\end{widetext}
For the production of ALPs from SM $2\to2$ scattering, we include the processes $f\bar{f}\to\gamma a$, $f\gamma\to fa$, and $\bar f\gamma \to \bar f a$ for all charged SM fermions. These rates are weighted by appropriate squared-charge and color factors, and at temperature $T$ we include all charged fermion species with masses below $T$.

For $T_{\rm RH}=10$ MeV and 1 GeV, we can neglect interactions with the weak gauge bosons. For $T_{\rm RH}=1$ TeV, however, we must include these effects. Since we only consider the hypercharge-coupled scenario for freeze-in, and the $2\to2$ scattering rates above are UV dominated, we can simply replace the photon with a hypercharge boson in the above rates, which is equivalent to taking $g_{a\gamma\gamma}\to C_{aB}$. For (inverse) ALP decays, we include decay rates to all SM boson pairs.

\section{Cross Sections and Decay Rates}\label{app:CS}
\begin{figure}
    \centering
\begin{tikzpicture}
\begin{feynman}
\vertex (ALP); 
\vertex[above right=1cm and 4.7cm of ALP] (achi) {\(\bar\chi\)}; 
\vertex[below right=1cm and 4.7cm of ALP] (chi){\(\chi\)}; 
\vertex[right=2.2 cm of ALP](v1);
\diagram* { {[edges=fermion]
(achi) -- (v1)--(chi)},
(ALP) -- [scalar, edge label=\(a\)] (v1)
};
\end{feynman} 
\end{tikzpicture}\\
\begin{tikzpicture}\begin{feynman}
\vertex (ALP); 
\vertex[above right=1cm and 3.2cm of ALP] (apsi); 
\vertex[below right=1cm and 3.2cm of ALP] (psi); 
\vertex[right=1.5cm of apsi] (aV){\( V\)}; 
\vertex[right=1.5cm of psi] (V){\( V\)}; 
\vertex[right=1.9cm of ALP](v1);
\diagram* { {[edges=fermion]
(apsi)-- [edge label'=\(\bar\Psi\)](v1)--[edge label'=\(\Psi\)](psi)--(apsi)},
(ALP) -- [scalar, edge label=\(a\)] (v1),(apsi) -- [boson] (aV),(psi) -- [boson] (V)
};
\end{feynman} 
\end{tikzpicture}
\caption{\textbf{Top:}~ALP decay to a pair of dark matter and anti-dark matter. \textbf{Bottom:}~ALP decay into a pair of electroweak bosons $ V V\in\{\gamma\gamma,W^+W^-,ZZ,\gamma Z\}$ via a loop of massive charged fermions $\Psi.$ In the effective field theory, the loop can be collapsed into the $a V V$ vertex with coupling strength $g_{aVV}$.}
\label{dig:ALPdecay}
\end{figure}
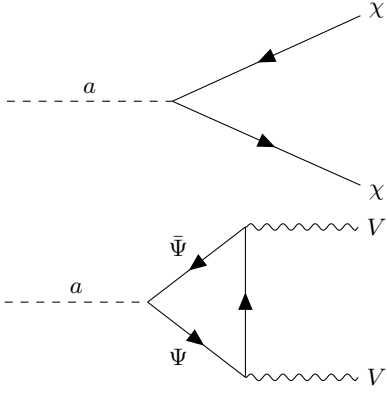
Here, we present various decay rates and cross sections implemented in our  Boltzmann equations.

\begin{figure}[t]
    \centering
\begin{tikzpicture}\begin{feynman}
\vertex (ALP); 
\vertex[above right=1cm and -2.5cm of ALP] (achi) {\(\bar\chi\)}; 
\vertex[below right=1cm and -2.5cm of ALP] (chi){\(\chi\)}; 
\vertex[above right=1cm and 3cm of ALP] (apsi); 
\vertex[below right=1cm and 3cm of ALP] (psi); 
\vertex[right=1.3cm of apsi] (aV){\( V\)}; 
\vertex[right=1.3cm of psi] (V){\( V\)}; 
\vertex[right=1.7cm of ALP](v1);
\diagram* { {[edges=fermion]
(chi) -- (ALP)--(achi)},{[edges=fermion]
(apsi)-- [edge label'=\(\bar\Psi\)](v1)--[edge label'=\(\Psi\)](psi)--(apsi),},
(ALP) -- [scalar, edge label=\(a\)] (v1),(apsi) -- [boson] (aV),(psi) -- [boson] (V)
};
\end{feynman} 
\end{tikzpicture}
\caption{DM annihilates into a pair of electroweak bosons $ V V\in\{\gamma\gamma,W^+W^-,ZZ,\gamma Z\}$ via an on- or off-shell axion.}
\label{dig:DMphoton}
\end{figure}
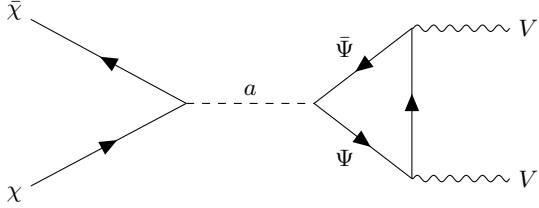

The ALP partial widths are
\be
\Gamma(a\to\bar\chi\chi) &=& \frac{M_a M_\chi^2}{8\pi f^2}\sqrt{1-\frac{4M_\chi^2}{M_a^2}},\\
\Gamma(a\to\gamma\gamma) &=& \frac{g_{a\gamma\gamma}^2M_a^3}{64\pi},\\
\Gamma(a\to\gamma Z) &=& \frac{g_{a\gamma Z}^2M_a^3}{32\pi}\left(1-\frac{M_Z^2}{M_a^2}\right)^3,\\
\Gamma(a\to Z Z) &=& \frac{g_{aZZ}^2M_a^3}{64\pi}\left(1-\frac{4M_Z^2}{M_a^2}\right)^{3/2},\\
\Gamma(a\to W^+ W^-) &=& \frac{g_{aWW}^2M_a^3}{32\pi}\left(1-\frac{4M_W^2}{M_a^2}\right)^{3/2},
\ee
for the decay modes shown in Fig.~\ref{dig:ALPdecay}. All couplings are defined in Sec.~\ref{sec:model}.
The thermally averaged decay width is related to the zero-temperature decay width via
\be
\langle\Gamma\rangle &=& \Gamma\,\frac{K_1(M_a/T)}{K_2(M_a/T)},
\ee
where $K_1(x)$ and $K_2(x)$ are modified Bessel functions of the second kind.

DM can annihilate into a pair of SM bosons via an on- or off-shell axion as shown in Fig. \ref{dig:DMphoton}. Its cross section in the narrow-width approximation is  
\be
\sigma_{\bar\chi\chi\to\gamma\gamma} &=& \frac{g_{a\gamma\gamma}^2 M_\chi^2 }{128\pi f^2}\,\frac{1}{\sqrt{1-4M_\chi^2/s}}\,\frac{s^2}{(s-M_a^2)^2+M_a^2\Gamma_a^2},\nonumber\\
\ee
where $s$ here is the Mandelstam variable, not the entropy density.

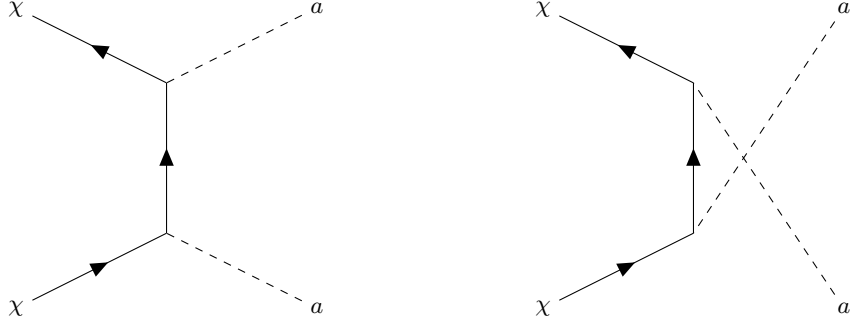
\begin{figure*}
    \centering
  \subfloat{\begin{tikzpicture}
\begin{feynman}
\vertex (achi){$\bar\chi$}; 
\vertex[below right= 1cm and 2cm of achi] (v1); 
\vertex[below = 2cm of v1] (v2); 
\vertex[below = 4cm of achi](chi){$\chi$};
\vertex[right= 4cm of achi](a1){$a$};
\vertex[right= 4cm of chi](a2){$a$};
\diagram* { {[edges=fermion]
(chi) -- (v2)--(v1)--(achi)},
(a1) -- [scalar] (v1), (a2) -- [scalar] (v2)
};
\end{feynman} 
\end{tikzpicture}}
 \qquad\qquad \qquad\qquad
\subfloat{\begin{tikzpicture}
\begin{feynman}
\vertex (achi){$\bar\chi$}; 
\vertex[below right= 1cm and 2cm of achi] (v1); 
\vertex[below = 2cm of v1] (v2); 
\vertex[below = 4cm of achi](chi){$\chi$};
\vertex[right= 4cm of achi](a1){$a$};
\vertex[right= 4cm of chi](a2){$a$};
\diagram* { {[edges=fermion]
(chi) -- (v2)--(v1)--(achi)},
(a2) -- [scalar] (v1), (a1) -- [scalar] (v2)
};
\end{feynman} 
\end{tikzpicture}}
    \caption{Two Feynman diagrams contributing to DM annihilating into a pair of ALPs.}
    \label{dig:DMALP}
\end{figure*} 

The cross section for DM annihilation to ALPs is 
\begin{widetext}
\be
\sigma_{\bar\chi\chi\to a a} &=& \frac{M_\chi^2}{64\pi f^4 s^2}\,\frac{1}{1-4M_\chi^2/s}\left[\frac{M_\chi^2s^2-4M_a^2M_\chi^2s+M_a^4(s-2M_\chi^2)}{M_\chi^2 s-4M_a^2M_\chi^2+M_a^4}\sqrt{s-4M_a^2}\sqrt{s-4M_\chi^2}\right.\\
&& \left.{}-\frac{2M_\chi^2(s^2-4M_a^2s^2+2M_a^4)}{s-2M_a^2}\log\left(\frac{s-2M_a^2+\sqrt{s-4M_a^2}\sqrt{s-4M_\chi^2}}{s-2M_a^2-\sqrt{s-4M_a^2}\sqrt{s-4M_\chi^2}}\right)\right]\nonumber,
\ee
\end{widetext}
with the relevant Feynman diagrams illustrated in Fig.~\ref{dig:DMALP}. The squared matrix elements for some of the processes were calculated with the assistance of \texttt{FeynCalc 9.3} \cite{Shtabovenko:2020gxv}.

In both cases, the thermally averaged cross section is found from the cross section using \cite{Gondolo:1990dk}%
\be
\langle \sigma_{\bar\chi\chi}v\rangle = \frac{1}{8M_\chi^4 T K_2(M_\chi/T)^2}\quad\quad\quad\quad\quad\quad\nonumber\\
\quad\quad\int_{4M_\chi^2}^\infty\,\sigma_{\bar\chi\chi}\sqrt{s}(s-4M_\chi^2)\,K_1(\sqrt{s}/T)\,ds.\label{eq:thermalavg}
\ee 

For freeze-in, we also need to calculate the rates of ALP production. In the limit of approximately massless ALP and fermions, we have
\be
\sigma_{\gamma a \to e^+e^-} &=& \frac{\alpha g_{a\gamma\gamma}^2}{12},\\
\sigma_{e^- a\to e^-\gamma} &=& \frac{\alpha g_{a\gamma\gamma}^2}{2}\log\left(\frac{s^{3/2}}{M_e M_a^2}\right).
\ee
The thermally averaged cross sections for these processes in the massless limit are found using:
\small
\be
\langle \sigma_{\mathrm{SM}\,a}v\rangle &=& \frac{1}{32T^5}\int_0^\infty \,\sigma_{\mathrm{SM}\,a}\,s^{3/2}K_1(\sqrt{s}/T)\,ds.
\ee
\normalsize
The thermal average integrals can be evaluated in closed form, giving
\be
\langle \sigma_{\gamma a \to e^+e^-}v\rangle &=& \frac{\alpha g_{a\gamma\gamma}^2}{12},
\ee
\be
\langle\sigma_{e^- a\to e^-\gamma}v\rangle &\approx& \frac{\alpha g_{a\gamma\gamma}^2}{2}\log\left(\frac{60.2T^3}{M_e M_a^2}\right).
\ee
We include SM fermion thermal masses in the log-divergent term of the cross section when they dominate over tree-level masses:
\be
\frac{M_{Q_{\rm L}}^2}{T^2} &=& \frac{g_3^2}{3}+\frac{3g_2^2}{16} +\frac{g_1^2}{144},\\
\frac{M_{u_{\rm R}}^2}{T^2} &=& \frac{g_3^2}{3} +\frac{g_1^2}{9},\\
\frac{M_{d_{\rm R}}^2}{T^2} &=& \frac{g_3^2}{3} +\frac{g_1^2}{36},\\
\frac{M_{L_{\rm L}}^2}{T^2} &=& \frac{3g_2^2}{16} +\frac{g_1^2}{16},\\
\frac{M_{e_{\rm R}}^2}{T^2} &=& \frac{g_1^2}{4}.
\ee
where $g_3$, $g_2$, and $g_1$ are the $\mathrm{SU}(3)_c$, $\mathrm{SU}(2)_L$, and $U(1)_Y$ gauge couplings, respectively. We neglect Yukawa couplings in these calculations:~they are only relevant for the top quark, which gives only a small contribution to the overall ALP production rate.

\bibliography{biblio}

\begin{thebibliography}{110}%
\makeatletter
\providecommand \@ifxundefined [1]{%
 \@ifx{#1\undefined}
}%
\providecommand \@ifnum [1]{%
 \ifnum #1\expandafter \@firstoftwo
 \else \expandafter \@secondoftwo
 \fi
}%
\providecommand \@ifx [1]{%
 \ifx #1\expandafter \@firstoftwo
 \else \expandafter \@secondoftwo
 \fi
}%
\providecommand \natexlab [1]{#1}%
\providecommand \enquote  [1]{``#1''}%
\providecommand \bibnamefont  [1]{#1}%
\providecommand \bibfnamefont [1]{#1}%
\providecommand \citenamefont [1]{#1}%
\providecommand \href@noop [0]{\@secondoftwo}%
\providecommand \href [0]{\begingroup \@sanitize@url \@href}%
\providecommand \@href[1]{\@@startlink{#1}\@@href}%
\providecommand \@@href[1]{\endgroup#1\@@endlink}%
\providecommand \@sanitize@url [0]{\catcode `\\12\catcode `\$12\catcode `\&12\catcode `\#12\catcode `\^12\catcode `\_12\catcode `\%12\relax}%
\providecommand \@@startlink[1]{}%
\providecommand \@@endlink[0]{}%
\providecommand \url  [0]{\begingroup\@sanitize@url \@url }%
\providecommand \@url [1]{\endgroup\@href {#1}{\urlprefix }}%
\providecommand \urlprefix  [0]{URL }%
\providecommand \Eprint [0]{\href }%
\providecommand \doibase [0]{http://dx.doi.org/}%
\providecommand \selectlanguage [0]{\@gobble}%
\providecommand \bibinfo  [0]{\@secondoftwo}%
\providecommand \bibfield  [0]{\@secondoftwo}%
\providecommand \translation [1]{[#1]}%
\providecommand \BibitemOpen [0]{}%
\providecommand \bibitemStop [0]{}%
\providecommand \bibitemNoStop [0]{.\EOS\space}%
\providecommand \EOS [0]{\spacefactor3000\relax}%
\providecommand \BibitemShut  [1]{\csname bibitem#1\endcsname}%
\let\auto@bib@innerbib\@empty
\bibitem [{\citenamefont {Peccei}\ and\ \citenamefont {Quinn}(1977{\natexlab{a}})}]{Peccei:1977hh}%
  \BibitemOpen
  \bibfield  {author} {\bibinfo {author} {\bibfnamefont {R.~D.}\ \bibnamefont {Peccei}}\ and\ \bibinfo {author} {\bibfnamefont {H.~R.}\ \bibnamefont {Quinn}},\ }\href {\doibase 10.1103/PhysRevLett.38.1440} {\bibfield  {journal} {\bibinfo  {journal} {Phys. Rev. Lett.}\ }\textbf {\bibinfo {volume} {38}},\ \bibinfo {pages} {1440} (\bibinfo {year} {1977}{\natexlab{a}})}\BibitemShut {NoStop}%
\bibitem [{\citenamefont {Peccei}\ and\ \citenamefont {Quinn}(1977{\natexlab{b}})}]{Peccei:1977ur}%
  \BibitemOpen
  \bibfield  {author} {\bibinfo {author} {\bibfnamefont {R.~D.}\ \bibnamefont {Peccei}}\ and\ \bibinfo {author} {\bibfnamefont {H.~R.}\ \bibnamefont {Quinn}},\ }\href {\doibase 10.1103/PhysRevD.16.1791} {\bibfield  {journal} {\bibinfo  {journal} {Phys. Rev. D}\ }\textbf {\bibinfo {volume} {16}},\ \bibinfo {pages} {1791} (\bibinfo {year} {1977}{\natexlab{b}})}\BibitemShut {NoStop}%
\bibitem [{\citenamefont {Weinberg}(1978)}]{Weinberg:1977ma}%
  \BibitemOpen
  \bibfield  {author} {\bibinfo {author} {\bibfnamefont {S.}~\bibnamefont {Weinberg}},\ }\href {\doibase 10.1103/PhysRevLett.40.223} {\bibfield  {journal} {\bibinfo  {journal} {Phys. Rev. Lett.}\ }\textbf {\bibinfo {volume} {40}},\ \bibinfo {pages} {223} (\bibinfo {year} {1978})}\BibitemShut {NoStop}%
\bibitem [{\citenamefont {Wilczek}(1978)}]{Wilczek:1977pj}%
  \BibitemOpen
  \bibfield  {author} {\bibinfo {author} {\bibfnamefont {F.}~\bibnamefont {Wilczek}},\ }\href {\doibase 10.1103/PhysRevLett.40.279} {\bibfield  {journal} {\bibinfo  {journal} {Phys. Rev. Lett.}\ }\textbf {\bibinfo {volume} {40}},\ \bibinfo {pages} {279} (\bibinfo {year} {1978})}\BibitemShut {NoStop}%
\bibitem [{\citenamefont {Witten}(1984)}]{Witten:1984dg}%
  \BibitemOpen
  \bibfield  {author} {\bibinfo {author} {\bibfnamefont {E.}~\bibnamefont {Witten}},\ }\href {\doibase 10.1016/0370-2693(84)90422-2} {\bibfield  {journal} {\bibinfo  {journal} {Phys. Lett. B}\ }\textbf {\bibinfo {volume} {149}},\ \bibinfo {pages} {351} (\bibinfo {year} {1984})}\BibitemShut {NoStop}%
\bibitem [{\citenamefont {Conlon}(2006)}]{Conlon:2006tq}%
  \BibitemOpen
  \bibfield  {author} {\bibinfo {author} {\bibfnamefont {J.~P.}\ \bibnamefont {Conlon}},\ }\href {\doibase 10.1088/1126-6708/2006/05/078} {\bibfield  {journal} {\bibinfo  {journal} {JHEP}\ }\textbf {\bibinfo {volume} {05}},\ \bibinfo {pages} {078} (\bibinfo {year} {2006})},\ \Eprint {http://arxiv.org/abs/hep-th/0602233} {arXiv:hep-th/0602233} \BibitemShut {NoStop}%
\bibitem [{\citenamefont {Svrcek}\ and\ \citenamefont {Witten}(2006)}]{Svrcek:2006yi}%
  \BibitemOpen
  \bibfield  {author} {\bibinfo {author} {\bibfnamefont {P.}~\bibnamefont {Svrcek}}\ and\ \bibinfo {author} {\bibfnamefont {E.}~\bibnamefont {Witten}},\ }\href {\doibase 10.1088/1126-6708/2006/06/051} {\bibfield  {journal} {\bibinfo  {journal} {JHEP}\ }\textbf {\bibinfo {volume} {06}},\ \bibinfo {pages} {051} (\bibinfo {year} {2006})},\ \Eprint {http://arxiv.org/abs/hep-th/0605206} {arXiv:hep-th/0605206} \BibitemShut {NoStop}%
\bibitem [{\citenamefont {Arvanitaki}\ \emph {et~al.}(2010)\citenamefont {Arvanitaki}, \citenamefont {Dimopoulos}, \citenamefont {Dubovsky}, \citenamefont {Kaloper},\ and\ \citenamefont {March-Russell}}]{Arvanitaki:2009fg}%
  \BibitemOpen
  \bibfield  {author} {\bibinfo {author} {\bibfnamefont {A.}~\bibnamefont {Arvanitaki}}, \bibinfo {author} {\bibfnamefont {S.}~\bibnamefont {Dimopoulos}}, \bibinfo {author} {\bibfnamefont {S.}~\bibnamefont {Dubovsky}}, \bibinfo {author} {\bibfnamefont {N.}~\bibnamefont {Kaloper}}, \ and\ \bibinfo {author} {\bibfnamefont {J.}~\bibnamefont {March-Russell}},\ }\href {\doibase 10.1103/PhysRevD.81.123530} {\bibfield  {journal} {\bibinfo  {journal} {Phys. Rev. D}\ }\textbf {\bibinfo {volume} {81}},\ \bibinfo {pages} {123530} (\bibinfo {year} {2010})},\ \Eprint {http://arxiv.org/abs/0905.4720} {arXiv:0905.4720 [hep-th]} \BibitemShut {NoStop}%
\bibitem [{\citenamefont {Frere}\ \emph {et~al.}(1983)\citenamefont {Frere}, \citenamefont {Jones},\ and\ \citenamefont {Raby}}]{Frere:1983ag}%
  \BibitemOpen
  \bibfield  {author} {\bibinfo {author} {\bibfnamefont {J.~M.}\ \bibnamefont {Frere}}, \bibinfo {author} {\bibfnamefont {D.~R.~T.}\ \bibnamefont {Jones}}, \ and\ \bibinfo {author} {\bibfnamefont {S.}~\bibnamefont {Raby}},\ }\href {\doibase 10.1016/0550-3213(83)90606-5} {\bibfield  {journal} {\bibinfo  {journal} {Nucl. Phys. B}\ }\textbf {\bibinfo {volume} {222}},\ \bibinfo {pages} {11} (\bibinfo {year} {1983})}\BibitemShut {NoStop}%
\bibitem [{\citenamefont {Nelson}\ and\ \citenamefont {Seiberg}(1994)}]{Nelson:1993nf}%
  \BibitemOpen
  \bibfield  {author} {\bibinfo {author} {\bibfnamefont {A.~E.}\ \bibnamefont {Nelson}}\ and\ \bibinfo {author} {\bibfnamefont {N.}~\bibnamefont {Seiberg}},\ }\href {\doibase 10.1016/0550-3213(94)90577-0} {\bibfield  {journal} {\bibinfo  {journal} {Nucl. Phys. B}\ }\textbf {\bibinfo {volume} {416}},\ \bibinfo {pages} {46} (\bibinfo {year} {1994})},\ \Eprint {http://arxiv.org/abs/hep-ph/9309299} {arXiv:hep-ph/9309299} \BibitemShut {NoStop}%
\bibitem [{\citenamefont {Bagger}\ \emph {et~al.}(1994)\citenamefont {Bagger}, \citenamefont {Poppitz},\ and\ \citenamefont {Randall}}]{Bagger:1994hh}%
  \BibitemOpen
  \bibfield  {author} {\bibinfo {author} {\bibfnamefont {J.}~\bibnamefont {Bagger}}, \bibinfo {author} {\bibfnamefont {E.}~\bibnamefont {Poppitz}}, \ and\ \bibinfo {author} {\bibfnamefont {L.}~\bibnamefont {Randall}},\ }\href {\doibase 10.1016/0550-3213(94)90123-6} {\bibfield  {journal} {\bibinfo  {journal} {Nucl. Phys. B}\ }\textbf {\bibinfo {volume} {426}},\ \bibinfo {pages} {3} (\bibinfo {year} {1994})},\ \Eprint {http://arxiv.org/abs/hep-ph/9405345} {arXiv:hep-ph/9405345} \BibitemShut {NoStop}%
\bibitem [{\citenamefont {Graham}\ \emph {et~al.}(2015)\citenamefont {Graham}, \citenamefont {Kaplan},\ and\ \citenamefont {Rajendran}}]{Graham:2015cka}%
  \BibitemOpen
  \bibfield  {author} {\bibinfo {author} {\bibfnamefont {P.~W.}\ \bibnamefont {Graham}}, \bibinfo {author} {\bibfnamefont {D.~E.}\ \bibnamefont {Kaplan}}, \ and\ \bibinfo {author} {\bibfnamefont {S.}~\bibnamefont {Rajendran}},\ }\href {\doibase 10.1103/PhysRevLett.115.221801} {\bibfield  {journal} {\bibinfo  {journal} {Phys. Rev. Lett.}\ }\textbf {\bibinfo {volume} {115}},\ \bibinfo {pages} {221801} (\bibinfo {year} {2015})},\ \Eprint {http://arxiv.org/abs/1504.07551} {arXiv:1504.07551 [hep-ph]} \BibitemShut {NoStop}%
\bibitem [{\citenamefont {Nomura}\ and\ \citenamefont {Thaler}(2009)}]{Nomura:2008ru}%
  \BibitemOpen
  \bibfield  {author} {\bibinfo {author} {\bibfnamefont {Y.}~\bibnamefont {Nomura}}\ and\ \bibinfo {author} {\bibfnamefont {J.}~\bibnamefont {Thaler}},\ }\href {\doibase 10.1103/PhysRevD.79.075008} {\bibfield  {journal} {\bibinfo  {journal} {Phys. Rev. D}\ }\textbf {\bibinfo {volume} {79}},\ \bibinfo {pages} {075008} (\bibinfo {year} {2009})},\ \Eprint {http://arxiv.org/abs/0810.5397} {arXiv:0810.5397 [hep-ph]} \BibitemShut {NoStop}%
\bibitem [{\citenamefont {Freytsis}\ and\ \citenamefont {Ligeti}(2011)}]{Freytsis:2010ne}%
  \BibitemOpen
  \bibfield  {author} {\bibinfo {author} {\bibfnamefont {M.}~\bibnamefont {Freytsis}}\ and\ \bibinfo {author} {\bibfnamefont {Z.}~\bibnamefont {Ligeti}},\ }\href {\doibase 10.1103/PhysRevD.83.115009} {\bibfield  {journal} {\bibinfo  {journal} {Phys. Rev. D}\ }\textbf {\bibinfo {volume} {83}},\ \bibinfo {pages} {115009} (\bibinfo {year} {2011})},\ \Eprint {http://arxiv.org/abs/1012.5317} {arXiv:1012.5317 [hep-ph]} \BibitemShut {NoStop}%
\bibitem [{\citenamefont {Dolan}\ \emph {et~al.}(2015)\citenamefont {Dolan}, \citenamefont {Kahlhoefer}, \citenamefont {McCabe},\ and\ \citenamefont {Schmidt-Hoberg}}]{Dolan:2014ska}%
  \BibitemOpen
  \bibfield  {author} {\bibinfo {author} {\bibfnamefont {M.~J.}\ \bibnamefont {Dolan}}, \bibinfo {author} {\bibfnamefont {F.}~\bibnamefont {Kahlhoefer}}, \bibinfo {author} {\bibfnamefont {C.}~\bibnamefont {McCabe}}, \ and\ \bibinfo {author} {\bibfnamefont {K.}~\bibnamefont {Schmidt-Hoberg}},\ }\href {\doibase 10.1007/JHEP03(2015)171} {\bibfield  {journal} {\bibinfo  {journal} {JHEP}\ }\textbf {\bibinfo {volume} {03}},\ \bibinfo {pages} {171} (\bibinfo {year} {2015})},\ \bibinfo {note} {[Erratum: JHEP 07, 103 (2015)]},\ \Eprint {http://arxiv.org/abs/1412.5174} {arXiv:1412.5174 [hep-ph]} \BibitemShut {NoStop}%
\bibitem [{\citenamefont {Kaneta}\ \emph {et~al.}(2017)\citenamefont {Kaneta}, \citenamefont {Lee},\ and\ \citenamefont {Yun}}]{Kaneta:2017wfh}%
  \BibitemOpen
  \bibfield  {author} {\bibinfo {author} {\bibfnamefont {K.}~\bibnamefont {Kaneta}}, \bibinfo {author} {\bibfnamefont {H.-S.}\ \bibnamefont {Lee}}, \ and\ \bibinfo {author} {\bibfnamefont {S.}~\bibnamefont {Yun}},\ }\href {\doibase 10.1103/PhysRevD.95.115032} {\bibfield  {journal} {\bibinfo  {journal} {Phys. Rev. D}\ }\textbf {\bibinfo {volume} {95}},\ \bibinfo {pages} {115032} (\bibinfo {year} {2017})},\ \Eprint {http://arxiv.org/abs/1704.07542} {arXiv:1704.07542 [hep-ph]} \BibitemShut {NoStop}%
\bibitem [{\citenamefont {Kamada}\ \emph {et~al.}(2017)\citenamefont {Kamada}, \citenamefont {Kim},\ and\ \citenamefont {Sekiguchi}}]{Kamada:2017tsq}%
  \BibitemOpen
  \bibfield  {author} {\bibinfo {author} {\bibfnamefont {A.}~\bibnamefont {Kamada}}, \bibinfo {author} {\bibfnamefont {H.}~\bibnamefont {Kim}}, \ and\ \bibinfo {author} {\bibfnamefont {T.}~\bibnamefont {Sekiguchi}},\ }\href {\doibase 10.1103/PhysRevD.96.016007} {\bibfield  {journal} {\bibinfo  {journal} {Phys. Rev. D}\ }\textbf {\bibinfo {volume} {96}},\ \bibinfo {pages} {016007} (\bibinfo {year} {2017})},\ \Eprint {http://arxiv.org/abs/1704.04505} {arXiv:1704.04505 [hep-ph]} \BibitemShut {NoStop}%
\bibitem [{\citenamefont {Hochberg}\ \emph {et~al.}(2018)\citenamefont {Hochberg}, \citenamefont {Kuflik}, \citenamefont {Mcgehee}, \citenamefont {Murayama},\ and\ \citenamefont {Schutz}}]{Hochberg:2018rjs}%
  \BibitemOpen
  \bibfield  {author} {\bibinfo {author} {\bibfnamefont {Y.}~\bibnamefont {Hochberg}}, \bibinfo {author} {\bibfnamefont {E.}~\bibnamefont {Kuflik}}, \bibinfo {author} {\bibfnamefont {R.}~\bibnamefont {Mcgehee}}, \bibinfo {author} {\bibfnamefont {H.}~\bibnamefont {Murayama}}, \ and\ \bibinfo {author} {\bibfnamefont {K.}~\bibnamefont {Schutz}},\ }\href {\doibase 10.1103/PhysRevD.98.115031} {\bibfield  {journal} {\bibinfo  {journal} {Phys. Rev. D}\ }\textbf {\bibinfo {volume} {98}},\ \bibinfo {pages} {115031} (\bibinfo {year} {2018})},\ \Eprint {http://arxiv.org/abs/1806.10139} {arXiv:1806.10139 [hep-ph]} \BibitemShut {NoStop}%
\bibitem [{\citenamefont {Gola}\ \emph {et~al.}(2022)\citenamefont {Gola}, \citenamefont {Mandal},\ and\ \citenamefont {Sinha}}]{Gola:2021abm}%
  \BibitemOpen
  \bibfield  {author} {\bibinfo {author} {\bibfnamefont {S.}~\bibnamefont {Gola}}, \bibinfo {author} {\bibfnamefont {S.}~\bibnamefont {Mandal}}, \ and\ \bibinfo {author} {\bibfnamefont {N.}~\bibnamefont {Sinha}},\ }\href {\doibase 10.1142/S0217751X22501317} {\bibfield  {journal} {\bibinfo  {journal} {Int. J. Mod. Phys. A}\ }\textbf {\bibinfo {volume} {37}},\ \bibinfo {pages} {2250131} (\bibinfo {year} {2022})},\ \Eprint {http://arxiv.org/abs/2106.00547} {arXiv:2106.00547 [hep-ph]} \BibitemShut {NoStop}%
\bibitem [{\citenamefont {Fitzpatrick}\ \emph {et~al.}(2023)\citenamefont {Fitzpatrick}, \citenamefont {Hochberg}, \citenamefont {Kuflik}, \citenamefont {Ovadia},\ and\ \citenamefont {Soreq}}]{Fitzpatrick:2023xks}%
  \BibitemOpen
  \bibfield  {author} {\bibinfo {author} {\bibfnamefont {P.~J.}\ \bibnamefont {Fitzpatrick}}, \bibinfo {author} {\bibfnamefont {Y.}~\bibnamefont {Hochberg}}, \bibinfo {author} {\bibfnamefont {E.}~\bibnamefont {Kuflik}}, \bibinfo {author} {\bibfnamefont {R.}~\bibnamefont {Ovadia}}, \ and\ \bibinfo {author} {\bibfnamefont {Y.}~\bibnamefont {Soreq}},\ }\href {\doibase 10.1103/PhysRevD.108.075003} {\bibfield  {journal} {\bibinfo  {journal} {Phys. Rev. D}\ }\textbf {\bibinfo {volume} {108}},\ \bibinfo {pages} {075003} (\bibinfo {year} {2023})},\ \Eprint {http://arxiv.org/abs/2306.03128} {arXiv:2306.03128 [hep-ph]} \BibitemShut {NoStop}%
\bibitem [{\citenamefont {Dror}\ \emph {et~al.}(2023)\citenamefont {Dror}, \citenamefont {Gori},\ and\ \citenamefont {Munbodh}}]{Dror:2023fyd}%
  \BibitemOpen
  \bibfield  {author} {\bibinfo {author} {\bibfnamefont {J.~A.}\ \bibnamefont {Dror}}, \bibinfo {author} {\bibfnamefont {S.}~\bibnamefont {Gori}}, \ and\ \bibinfo {author} {\bibfnamefont {P.}~\bibnamefont {Munbodh}},\ }\href {\doibase 10.1007/JHEP09(2023)128} {\bibfield  {journal} {\bibinfo  {journal} {JHEP}\ }\textbf {\bibinfo {volume} {09}},\ \bibinfo {pages} {128} (\bibinfo {year} {2023})},\ \Eprint {http://arxiv.org/abs/2306.03145} {arXiv:2306.03145 [hep-ph]} \BibitemShut {NoStop}%
\bibitem [{\citenamefont {Armando}\ \emph {et~al.}(2024)\citenamefont {Armando}, \citenamefont {Panci}, \citenamefont {Weiss},\ and\ \citenamefont {Ziegler}}]{Armando:2023zwz}%
  \BibitemOpen
  \bibfield  {author} {\bibinfo {author} {\bibfnamefont {G.}~\bibnamefont {Armando}}, \bibinfo {author} {\bibfnamefont {P.}~\bibnamefont {Panci}}, \bibinfo {author} {\bibfnamefont {J.}~\bibnamefont {Weiss}}, \ and\ \bibinfo {author} {\bibfnamefont {R.}~\bibnamefont {Ziegler}},\ }\href {\doibase 10.1103/PhysRevD.109.055029} {\bibfield  {journal} {\bibinfo  {journal} {Phys. Rev. D}\ }\textbf {\bibinfo {volume} {109}},\ \bibinfo {pages} {055029} (\bibinfo {year} {2024})},\ \Eprint {http://arxiv.org/abs/2310.05827} {arXiv:2310.05827 [hep-ph]} \BibitemShut {NoStop}%
\bibitem [{\citenamefont {Dolan}\ \emph {et~al.}(2017)\citenamefont {Dolan}, \citenamefont {Ferber}, \citenamefont {Hearty}, \citenamefont {Kahlhoefer},\ and\ \citenamefont {Schmidt-Hoberg}}]{Dolan:2017osp}%
  \BibitemOpen
  \bibfield  {author} {\bibinfo {author} {\bibfnamefont {M.~J.}\ \bibnamefont {Dolan}}, \bibinfo {author} {\bibfnamefont {T.}~\bibnamefont {Ferber}}, \bibinfo {author} {\bibfnamefont {C.}~\bibnamefont {Hearty}}, \bibinfo {author} {\bibfnamefont {F.}~\bibnamefont {Kahlhoefer}}, \ and\ \bibinfo {author} {\bibfnamefont {K.}~\bibnamefont {Schmidt-Hoberg}},\ }\href {\doibase 10.1007/JHEP12(2017)094} {\bibfield  {journal} {\bibinfo  {journal} {JHEP}\ }\textbf {\bibinfo {volume} {12}},\ \bibinfo {pages} {094} (\bibinfo {year} {2017})},\ \bibinfo {note} {[Erratum: JHEP 03, 190 (2021)]},\ \Eprint {http://arxiv.org/abs/1709.00009} {arXiv:1709.00009 [hep-ph]} \BibitemShut {NoStop}%
\bibitem [{\citenamefont {Bjorken}\ \emph {et~al.}(1988)\citenamefont {Bjorken}, \citenamefont {Ecklund}, \citenamefont {Nelson}, \citenamefont {Abashian}, \citenamefont {Church}, \citenamefont {Lu}, \citenamefont {Mo}, \citenamefont {Nunamaker},\ and\ \citenamefont {Rassmann}}]{Bjorken:1988as}%
  \BibitemOpen
  \bibfield  {author} {\bibinfo {author} {\bibfnamefont {J.~D.}\ \bibnamefont {Bjorken}}, \bibinfo {author} {\bibfnamefont {S.}~\bibnamefont {Ecklund}}, \bibinfo {author} {\bibfnamefont {W.~R.}\ \bibnamefont {Nelson}}, \bibinfo {author} {\bibfnamefont {A.}~\bibnamefont {Abashian}}, \bibinfo {author} {\bibfnamefont {C.}~\bibnamefont {Church}}, \bibinfo {author} {\bibfnamefont {B.}~\bibnamefont {Lu}}, \bibinfo {author} {\bibfnamefont {L.~W.}\ \bibnamefont {Mo}}, \bibinfo {author} {\bibfnamefont {T.~A.}\ \bibnamefont {Nunamaker}}, \ and\ \bibinfo {author} {\bibfnamefont {P.}~\bibnamefont {Rassmann}},\ }\href {\doibase 10.1103/PhysRevD.38.3375} {\bibfield  {journal} {\bibinfo  {journal} {Phys. Rev. D}\ }\textbf {\bibinfo {volume} {38}},\ \bibinfo {pages} {3375} (\bibinfo {year} {1988})}\BibitemShut {NoStop}%
\bibitem [{\citenamefont {Blumlein}\ \emph {et~al.}(1991)\citenamefont {Blumlein} \emph {et~al.}}]{Blumlein:1990ay}%
  \BibitemOpen
  \bibfield  {author} {\bibinfo {author} {\bibfnamefont {J.}~\bibnamefont {Blumlein}} \emph {et~al.},\ }\href {\doibase 10.1007/BF01548556} {\bibfield  {journal} {\bibinfo  {journal} {Z. Phys. C}\ }\textbf {\bibinfo {volume} {51}},\ \bibinfo {pages} {341} (\bibinfo {year} {1991})}\BibitemShut {NoStop}%
\bibitem [{\citenamefont {Acciarri}\ \emph {et~al.}(1995{\natexlab{a}})\citenamefont {Acciarri} \emph {et~al.}}]{L3:1994shn}%
  \BibitemOpen
  \bibfield  {author} {\bibinfo {author} {\bibfnamefont {M.}~\bibnamefont {Acciarri}} \emph {et~al.} (\bibinfo {collaboration} {L3}),\ }\href {\doibase 10.1016/0370-2693(95)01612-T} {\bibfield  {journal} {\bibinfo  {journal} {Phys. Lett. B}\ }\textbf {\bibinfo {volume} {345}},\ \bibinfo {pages} {609} (\bibinfo {year} {1995}{\natexlab{a}})}\BibitemShut {NoStop}%
\bibitem [{\citenamefont {Acciarri}\ \emph {et~al.}(1995{\natexlab{b}})\citenamefont {Acciarri} \emph {et~al.}}]{L3:1995nbq}%
  \BibitemOpen
  \bibfield  {author} {\bibinfo {author} {\bibfnamefont {M.}~\bibnamefont {Acciarri}} \emph {et~al.} (\bibinfo {collaboration} {L3}),\ }\href {\doibase 10.1016/0370-2693(95)00527-R} {\bibfield  {journal} {\bibinfo  {journal} {Phys. Lett. B}\ }\textbf {\bibinfo {volume} {353}},\ \bibinfo {pages} {136} (\bibinfo {year} {1995}{\natexlab{b}})}\BibitemShut {NoStop}%
\bibitem [{\citenamefont {Acciarri}\ \emph {et~al.}(1997)\citenamefont {Acciarri} \emph {et~al.}}]{L3:1997exg}%
  \BibitemOpen
  \bibfield  {author} {\bibinfo {author} {\bibfnamefont {M.}~\bibnamefont {Acciarri}} \emph {et~al.} (\bibinfo {collaboration} {L3}),\ }\href {\doibase 10.1016/S0370-2693(97)01003-4} {\bibfield  {journal} {\bibinfo  {journal} {Phys. Lett. B}\ }\textbf {\bibinfo {volume} {412}},\ \bibinfo {pages} {201} (\bibinfo {year} {1997})}\BibitemShut {NoStop}%
\bibitem [{\citenamefont {Artamonov}\ \emph {et~al.}(2005)\citenamefont {Artamonov} \emph {et~al.}}]{E949:2005qiy}%
  \BibitemOpen
  \bibfield  {author} {\bibinfo {author} {\bibfnamefont {A.~V.}\ \bibnamefont {Artamonov}} \emph {et~al.} (\bibinfo {collaboration} {E949}),\ }\href {\doibase 10.1016/j.physletb.2005.07.057} {\bibfield  {journal} {\bibinfo  {journal} {Phys. Lett. B}\ }\textbf {\bibinfo {volume} {623}},\ \bibinfo {pages} {192} (\bibinfo {year} {2005})},\ \Eprint {http://arxiv.org/abs/hep-ex/0505069} {arXiv:hep-ex/0505069} \BibitemShut {NoStop}%
\bibitem [{\citenamefont {Abdallah}\ \emph {et~al.}(2009)\citenamefont {Abdallah} \emph {et~al.}}]{DELPHI:2008uka}%
  \BibitemOpen
  \bibfield  {author} {\bibinfo {author} {\bibfnamefont {J.}~\bibnamefont {Abdallah}} \emph {et~al.} (\bibinfo {collaboration} {DELPHI}),\ }\href {\doibase 10.1140/epjc/s10052-009-0874-9} {\bibfield  {journal} {\bibinfo  {journal} {Eur. Phys. J. C}\ }\textbf {\bibinfo {volume} {60}},\ \bibinfo {pages} {17} (\bibinfo {year} {2009})},\ \Eprint {http://arxiv.org/abs/0901.4486} {arXiv:0901.4486 [hep-ex]} \BibitemShut {NoStop}%
\bibitem [{\citenamefont {Abouzaid}\ \emph {et~al.}(2008)\citenamefont {Abouzaid} \emph {et~al.}}]{KTeV:2008nqz}%
  \BibitemOpen
  \bibfield  {author} {\bibinfo {author} {\bibfnamefont {E.}~\bibnamefont {Abouzaid}} \emph {et~al.} (\bibinfo {collaboration} {KTeV}),\ }\href {\doibase 10.1103/PhysRevD.77.112004} {\bibfield  {journal} {\bibinfo  {journal} {Phys. Rev. D}\ }\textbf {\bibinfo {volume} {77}},\ \bibinfo {pages} {112004} (\bibinfo {year} {2008})},\ \Eprint {http://arxiv.org/abs/0805.0031} {arXiv:0805.0031 [hep-ex]} \BibitemShut {NoStop}%
\bibitem [{\citenamefont {Larin}\ \emph {et~al.}(2011)\citenamefont {Larin} \emph {et~al.}}]{PrimEx:2010fvg}%
  \BibitemOpen
  \bibfield  {author} {\bibinfo {author} {\bibfnamefont {I.}~\bibnamefont {Larin}} \emph {et~al.} (\bibinfo {collaboration} {PrimEx}),\ }\href {\doibase 10.1103/PhysRevLett.106.162303} {\bibfield  {journal} {\bibinfo  {journal} {Phys. Rev. Lett.}\ }\textbf {\bibinfo {volume} {106}},\ \bibinfo {pages} {162303} (\bibinfo {year} {2011})},\ \Eprint {http://arxiv.org/abs/1009.1681} {arXiv:1009.1681 [nucl-ex]} \BibitemShut {NoStop}%
\bibitem [{\citenamefont {Chatrchyan}\ \emph {et~al.}(2012)\citenamefont {Chatrchyan} \emph {et~al.}}]{CMS:2011bsw}%
  \BibitemOpen
  \bibfield  {author} {\bibinfo {author} {\bibfnamefont {S.}~\bibnamefont {Chatrchyan}} \emph {et~al.} (\bibinfo {collaboration} {CMS}),\ }\href {\doibase 10.1103/PhysRevLett.108.111801} {\bibfield  {journal} {\bibinfo  {journal} {Phys. Rev. Lett.}\ }\textbf {\bibinfo {volume} {108}},\ \bibinfo {pages} {111801} (\bibinfo {year} {2012})},\ \Eprint {http://arxiv.org/abs/1112.0688} {arXiv:1112.0688 [hep-ex]} \BibitemShut {NoStop}%
\bibitem [{\citenamefont {Aad}\ \emph {et~al.}(2012)\citenamefont {Aad} \emph {et~al.}}]{ATLAS:2011ab}%
  \BibitemOpen
  \bibfield  {author} {\bibinfo {author} {\bibfnamefont {G.}~\bibnamefont {Aad}} \emph {et~al.} (\bibinfo {collaboration} {ATLAS}),\ }\href {\doibase 10.1016/j.physletb.2012.03.022} {\bibfield  {journal} {\bibinfo  {journal} {Phys. Lett. B}\ }\textbf {\bibinfo {volume} {710}},\ \bibinfo {pages} {538} (\bibinfo {year} {2012})},\ \Eprint {http://arxiv.org/abs/1112.2194} {arXiv:1112.2194 [hep-ex]} \BibitemShut {NoStop}%
\bibitem [{\citenamefont {Aad}\ \emph {et~al.}(2013)\citenamefont {Aad} \emph {et~al.}}]{ATLAS:2012fgo}%
  \BibitemOpen
  \bibfield  {author} {\bibinfo {author} {\bibfnamefont {G.}~\bibnamefont {Aad}} \emph {et~al.} (\bibinfo {collaboration} {ATLAS}),\ }\href {\doibase 10.1007/JHEP01(2013)086} {\bibfield  {journal} {\bibinfo  {journal} {JHEP}\ }\textbf {\bibinfo {volume} {01}},\ \bibinfo {pages} {086} (\bibinfo {year} {2013})},\ \Eprint {http://arxiv.org/abs/1211.1913} {arXiv:1211.1913 [hep-ex]} \BibitemShut {NoStop}%
\bibitem [{\citenamefont {Jaeckel}\ \emph {et~al.}(2013)\citenamefont {Jaeckel}, \citenamefont {Jankowiak},\ and\ \citenamefont {Spannowsky}}]{Jaeckel:2012yz}%
  \BibitemOpen
  \bibfield  {author} {\bibinfo {author} {\bibfnamefont {J.}~\bibnamefont {Jaeckel}}, \bibinfo {author} {\bibfnamefont {M.}~\bibnamefont {Jankowiak}}, \ and\ \bibinfo {author} {\bibfnamefont {M.}~\bibnamefont {Spannowsky}},\ }\href {\doibase 10.1016/j.dark.2013.06.001} {\bibfield  {journal} {\bibinfo  {journal} {Phys. Dark Univ.}\ }\textbf {\bibinfo {volume} {2}},\ \bibinfo {pages} {111} (\bibinfo {year} {2013})},\ \Eprint {http://arxiv.org/abs/1212.3620} {arXiv:1212.3620 [hep-ph]} \BibitemShut {NoStop}%
\bibitem [{\citenamefont {Aaltonen}\ \emph {et~al.}(2014)\citenamefont {Aaltonen} \emph {et~al.}}]{CDF:2013lma}%
  \BibitemOpen
  \bibfield  {author} {\bibinfo {author} {\bibfnamefont {T.~A.}\ \bibnamefont {Aaltonen}} \emph {et~al.} (\bibinfo {collaboration} {CDF}),\ }\href {\doibase 10.1103/PhysRevLett.112.111803} {\bibfield  {journal} {\bibinfo  {journal} {Phys. Rev. Lett.}\ }\textbf {\bibinfo {volume} {112}},\ \bibinfo {pages} {111803} (\bibinfo {year} {2014})},\ \Eprint {http://arxiv.org/abs/1311.3282} {arXiv:1311.3282 [hep-ex]} \BibitemShut {NoStop}%
\bibitem [{\citenamefont {Lazzeroni}\ \emph {et~al.}(2014)\citenamefont {Lazzeroni} \emph {et~al.}}]{NA62:2014ybm}%
  \BibitemOpen
  \bibfield  {author} {\bibinfo {author} {\bibfnamefont {C.}~\bibnamefont {Lazzeroni}} \emph {et~al.} (\bibinfo {collaboration} {NA62}),\ }\href {\doibase 10.1016/j.physletb.2014.03.016} {\bibfield  {journal} {\bibinfo  {journal} {Phys. Lett. B}\ }\textbf {\bibinfo {volume} {732}},\ \bibinfo {pages} {65} (\bibinfo {year} {2014})},\ \Eprint {http://arxiv.org/abs/1402.4334} {arXiv:1402.4334 [hep-ex]} \BibitemShut {NoStop}%
\bibitem [{\citenamefont {Aad}\ \emph {et~al.}(2016)\citenamefont {Aad} \emph {et~al.}}]{ATLAS:2015rsn}%
  \BibitemOpen
  \bibfield  {author} {\bibinfo {author} {\bibfnamefont {G.}~\bibnamefont {Aad}} \emph {et~al.} (\bibinfo {collaboration} {ATLAS}),\ }\href {\doibase 10.1140/epjc/s10052-016-4034-8} {\bibfield  {journal} {\bibinfo  {journal} {Eur. Phys. J. C}\ }\textbf {\bibinfo {volume} {76}},\ \bibinfo {pages} {210} (\bibinfo {year} {2016})},\ \Eprint {http://arxiv.org/abs/1509.05051} {arXiv:1509.05051 [hep-ex]} \BibitemShut {NoStop}%
\bibitem [{\citenamefont {Jaeckel}\ and\ \citenamefont {Spannowsky}(2016)}]{Jaeckel:2015jla}%
  \BibitemOpen
  \bibfield  {author} {\bibinfo {author} {\bibfnamefont {J.}~\bibnamefont {Jaeckel}}\ and\ \bibinfo {author} {\bibfnamefont {M.}~\bibnamefont {Spannowsky}},\ }\href {\doibase 10.1016/j.physletb.2015.12.037} {\bibfield  {journal} {\bibinfo  {journal} {Phys. Lett. B}\ }\textbf {\bibinfo {volume} {753}},\ \bibinfo {pages} {482} (\bibinfo {year} {2016})},\ \Eprint {http://arxiv.org/abs/1509.00476} {arXiv:1509.00476 [hep-ph]} \BibitemShut {NoStop}%
\bibitem [{\citenamefont {Knapen}\ \emph {et~al.}(2017)\citenamefont {Knapen}, \citenamefont {Lin}, \citenamefont {Lou},\ and\ \citenamefont {Melia}}]{Knapen:2016moh}%
  \BibitemOpen
  \bibfield  {author} {\bibinfo {author} {\bibfnamefont {S.}~\bibnamefont {Knapen}}, \bibinfo {author} {\bibfnamefont {T.}~\bibnamefont {Lin}}, \bibinfo {author} {\bibfnamefont {H.~K.}\ \bibnamefont {Lou}}, \ and\ \bibinfo {author} {\bibfnamefont {T.}~\bibnamefont {Melia}},\ }\href {\doibase 10.1103/PhysRevLett.118.171801} {\bibfield  {journal} {\bibinfo  {journal} {Phys. Rev. Lett.}\ }\textbf {\bibinfo {volume} {118}},\ \bibinfo {pages} {171801} (\bibinfo {year} {2017})},\ \Eprint {http://arxiv.org/abs/1607.06083} {arXiv:1607.06083 [hep-ph]} \BibitemShut {NoStop}%
\bibitem [{\citenamefont {Izaguirre}\ \emph {et~al.}(2017)\citenamefont {Izaguirre}, \citenamefont {Lin},\ and\ \citenamefont {Shuve}}]{Izaguirre:2016dfi}%
  \BibitemOpen
  \bibfield  {author} {\bibinfo {author} {\bibfnamefont {E.}~\bibnamefont {Izaguirre}}, \bibinfo {author} {\bibfnamefont {T.}~\bibnamefont {Lin}}, \ and\ \bibinfo {author} {\bibfnamefont {B.}~\bibnamefont {Shuve}},\ }\href {\doibase 10.1103/PhysRevLett.118.111802} {\bibfield  {journal} {\bibinfo  {journal} {Phys. Rev. Lett.}\ }\textbf {\bibinfo {volume} {118}},\ \bibinfo {pages} {111802} (\bibinfo {year} {2017})},\ \Eprint {http://arxiv.org/abs/1611.09355} {arXiv:1611.09355 [hep-ph]} \BibitemShut {NoStop}%
\bibitem [{\citenamefont {Bauer}\ \emph {et~al.}(2017)\citenamefont {Bauer}, \citenamefont {Neubert},\ and\ \citenamefont {Thamm}}]{Bauer:2017ris}%
  \BibitemOpen
  \bibfield  {author} {\bibinfo {author} {\bibfnamefont {M.}~\bibnamefont {Bauer}}, \bibinfo {author} {\bibfnamefont {M.}~\bibnamefont {Neubert}}, \ and\ \bibinfo {author} {\bibfnamefont {A.}~\bibnamefont {Thamm}},\ }\href {\doibase 10.1007/JHEP12(2017)044} {\bibfield  {journal} {\bibinfo  {journal} {JHEP}\ }\textbf {\bibinfo {volume} {12}},\ \bibinfo {pages} {044} (\bibinfo {year} {2017})},\ \Eprint {http://arxiv.org/abs/1708.00443} {arXiv:1708.00443 [hep-ph]} \BibitemShut {NoStop}%
\bibitem [{\citenamefont {Aaboud}\ \emph {et~al.}(2018)\citenamefont {Aaboud} \emph {et~al.}}]{ATLAS:2017jag}%
  \BibitemOpen
  \bibfield  {author} {\bibinfo {author} {\bibfnamefont {M.}~\bibnamefont {Aaboud}} \emph {et~al.} (\bibinfo {collaboration} {ATLAS}),\ }\href {\doibase 10.1007/JHEP03(2018)042} {\bibfield  {journal} {\bibinfo  {journal} {JHEP}\ }\textbf {\bibinfo {volume} {03}},\ \bibinfo {pages} {042} (\bibinfo {year} {2018})},\ \Eprint {http://arxiv.org/abs/1710.07235} {arXiv:1710.07235 [hep-ex]} \BibitemShut {NoStop}%
\bibitem [{\citenamefont {Sirunyan}\ \emph {et~al.}(2019)\citenamefont {Sirunyan} \emph {et~al.}}]{CMS:2018erd}%
  \BibitemOpen
  \bibfield  {author} {\bibinfo {author} {\bibfnamefont {A.~M.}\ \bibnamefont {Sirunyan}} \emph {et~al.} (\bibinfo {collaboration} {CMS}),\ }\href {\doibase 10.1016/j.physletb.2019.134826} {\bibfield  {journal} {\bibinfo  {journal} {Phys. Lett. B}\ }\textbf {\bibinfo {volume} {797}},\ \bibinfo {pages} {134826} (\bibinfo {year} {2019})},\ \Eprint {http://arxiv.org/abs/1810.04602} {arXiv:1810.04602 [hep-ex]} \BibitemShut {NoStop}%
\bibitem [{\citenamefont {Ahn}\ \emph {et~al.}(2019)\citenamefont {Ahn} \emph {et~al.}}]{KOTO:2018dsc}%
  \BibitemOpen
  \bibfield  {author} {\bibinfo {author} {\bibfnamefont {J.~K.}\ \bibnamefont {Ahn}} \emph {et~al.} (\bibinfo {collaboration} {KOTO}),\ }\href {\doibase 10.1103/PhysRevLett.122.021802} {\bibfield  {journal} {\bibinfo  {journal} {Phys. Rev. Lett.}\ }\textbf {\bibinfo {volume} {122}},\ \bibinfo {pages} {021802} (\bibinfo {year} {2019})},\ \Eprint {http://arxiv.org/abs/1810.09655} {arXiv:1810.09655 [hep-ex]} \BibitemShut {NoStop}%
\bibitem [{\citenamefont {Aloni}\ \emph {et~al.}(2019)\citenamefont {Aloni}, \citenamefont {Fanelli}, \citenamefont {Soreq},\ and\ \citenamefont {Williams}}]{Aloni:2019ruo}%
  \BibitemOpen
  \bibfield  {author} {\bibinfo {author} {\bibfnamefont {D.}~\bibnamefont {Aloni}}, \bibinfo {author} {\bibfnamefont {C.}~\bibnamefont {Fanelli}}, \bibinfo {author} {\bibfnamefont {Y.}~\bibnamefont {Soreq}}, \ and\ \bibinfo {author} {\bibfnamefont {M.}~\bibnamefont {Williams}},\ }\href {\doibase 10.1103/PhysRevLett.123.071801} {\bibfield  {journal} {\bibinfo  {journal} {Phys. Rev. Lett.}\ }\textbf {\bibinfo {volume} {123}},\ \bibinfo {pages} {071801} (\bibinfo {year} {2019})},\ \Eprint {http://arxiv.org/abs/1903.03586} {arXiv:1903.03586 [hep-ph]} \BibitemShut {NoStop}%
\bibitem [{\citenamefont {Aad}\ \emph {et~al.}(2021{\natexlab{a}})\citenamefont {Aad} \emph {et~al.}}]{ATLAS:2020hii}%
  \BibitemOpen
  \bibfield  {author} {\bibinfo {author} {\bibfnamefont {G.}~\bibnamefont {Aad}} \emph {et~al.} (\bibinfo {collaboration} {ATLAS}),\ }\href {\doibase 10.1007/JHEP11(2021)050} {\bibfield  {journal} {\bibinfo  {journal} {JHEP}\ }\textbf {\bibinfo {volume} {03}},\ \bibinfo {pages} {243} (\bibinfo {year} {2021}{\natexlab{a}})},\ \bibinfo {note} {[Erratum: JHEP 11, 050 (2021)]},\ \Eprint {http://arxiv.org/abs/2008.05355} {arXiv:2008.05355 [hep-ex]} \BibitemShut {NoStop}%
\bibitem [{\citenamefont {Abudin\'en}\ \emph {et~al.}(2020)\citenamefont {Abudin\'en} \emph {et~al.}}]{Belle-II:2020jti}%
  \BibitemOpen
  \bibfield  {author} {\bibinfo {author} {\bibfnamefont {F.}~\bibnamefont {Abudin\'en}} \emph {et~al.} (\bibinfo {collaboration} {Belle-II}),\ }\href {\doibase 10.1103/PhysRevLett.125.161806} {\bibfield  {journal} {\bibinfo  {journal} {Phys. Rev. Lett.}\ }\textbf {\bibinfo {volume} {125}},\ \bibinfo {pages} {161806} (\bibinfo {year} {2020})},\ \Eprint {http://arxiv.org/abs/2007.13071} {arXiv:2007.13071 [hep-ex]} \BibitemShut {NoStop}%
\bibitem [{\citenamefont {Gori}\ \emph {et~al.}(2020)\citenamefont {Gori}, \citenamefont {Perez},\ and\ \citenamefont {Tobioka}}]{Gori:2020xvq}%
  \BibitemOpen
  \bibfield  {author} {\bibinfo {author} {\bibfnamefont {S.}~\bibnamefont {Gori}}, \bibinfo {author} {\bibfnamefont {G.}~\bibnamefont {Perez}}, \ and\ \bibinfo {author} {\bibfnamefont {K.}~\bibnamefont {Tobioka}},\ }\href {\doibase 10.1007/JHEP08(2020)110} {\bibfield  {journal} {\bibinfo  {journal} {JHEP}\ }\textbf {\bibinfo {volume} {08}},\ \bibinfo {pages} {110} (\bibinfo {year} {2020})},\ \Eprint {http://arxiv.org/abs/2005.05170} {arXiv:2005.05170 [hep-ph]} \BibitemShut {NoStop}%
\bibitem [{\citenamefont {Banerjee}\ \emph {et~al.}(2020)\citenamefont {Banerjee} \emph {et~al.}}]{NA64:2020qwq}%
  \BibitemOpen
  \bibfield  {author} {\bibinfo {author} {\bibfnamefont {D.}~\bibnamefont {Banerjee}} \emph {et~al.} (\bibinfo {collaboration} {NA64}),\ }\href {\doibase 10.1103/PhysRevLett.125.081801} {\bibfield  {journal} {\bibinfo  {journal} {Phys. Rev. Lett.}\ }\textbf {\bibinfo {volume} {125}},\ \bibinfo {pages} {081801} (\bibinfo {year} {2020})},\ \Eprint {http://arxiv.org/abs/2005.02710} {arXiv:2005.02710 [hep-ex]} \BibitemShut {NoStop}%
\bibitem [{\citenamefont {Cortina~Gil}\ \emph {et~al.}(2021)\citenamefont {Cortina~Gil} \emph {et~al.}}]{NA62:2020xlg}%
  \BibitemOpen
  \bibfield  {author} {\bibinfo {author} {\bibfnamefont {E.}~\bibnamefont {Cortina~Gil}} \emph {et~al.} (\bibinfo {collaboration} {NA62}),\ }\href {\doibase 10.1007/JHEP03(2021)058} {\bibfield  {journal} {\bibinfo  {journal} {JHEP}\ }\textbf {\bibinfo {volume} {03}},\ \bibinfo {pages} {058} (\bibinfo {year} {2021})},\ \Eprint {http://arxiv.org/abs/2011.11329} {arXiv:2011.11329 [hep-ex]} \BibitemShut {NoStop}%
\bibitem [{\citenamefont {Lees}\ \emph {et~al.}(2022)\citenamefont {Lees} \emph {et~al.}}]{BaBar:2021ich}%
  \BibitemOpen
  \bibfield  {author} {\bibinfo {author} {\bibfnamefont {J.~P.}\ \bibnamefont {Lees}} \emph {et~al.} (\bibinfo {collaboration} {BaBar}),\ }\href {\doibase 10.1103/PhysRevLett.128.131802} {\bibfield  {journal} {\bibinfo  {journal} {Phys. Rev. Lett.}\ }\textbf {\bibinfo {volume} {128}},\ \bibinfo {pages} {131802} (\bibinfo {year} {2022})},\ \Eprint {http://arxiv.org/abs/2111.01800} {arXiv:2111.01800 [hep-ex]} \BibitemShut {NoStop}%
\bibitem [{\citenamefont {Tumasyan}\ \emph {et~al.}(2021)\citenamefont {Tumasyan} \emph {et~al.}}]{CMS:2021juv}%
  \BibitemOpen
  \bibfield  {author} {\bibinfo {author} {\bibfnamefont {A.}~\bibnamefont {Tumasyan}} \emph {et~al.} (\bibinfo {collaboration} {CMS}),\ }\href {\doibase 10.1103/PhysRevLett.127.261804} {\bibfield  {journal} {\bibinfo  {journal} {Phys. Rev. Lett.}\ }\textbf {\bibinfo {volume} {127}},\ \bibinfo {pages} {261804} (\bibinfo {year} {2021})},\ \Eprint {http://arxiv.org/abs/2107.04838} {arXiv:2107.04838 [hep-ex]} \BibitemShut {NoStop}%
\bibitem [{\citenamefont {Tumasyan}\ \emph {et~al.}(2022)\citenamefont {Tumasyan} \emph {et~al.}}]{CMS:2021klu}%
  \BibitemOpen
  \bibfield  {author} {\bibinfo {author} {\bibfnamefont {A.}~\bibnamefont {Tumasyan}} \emph {et~al.} (\bibinfo {collaboration} {CMS}),\ }\href {\doibase 10.1103/PhysRevD.105.032008} {\bibfield  {journal} {\bibinfo  {journal} {Phys. Rev. D}\ }\textbf {\bibinfo {volume} {105}},\ \bibinfo {pages} {032008} (\bibinfo {year} {2022})},\ \Eprint {http://arxiv.org/abs/2109.06055} {arXiv:2109.06055 [hep-ex]} \BibitemShut {NoStop}%
\bibitem [{\citenamefont {Aad}\ \emph {et~al.}(2021{\natexlab{b}})\citenamefont {Aad} \emph {et~al.}}]{ATLAS:2021uiz}%
  \BibitemOpen
  \bibfield  {author} {\bibinfo {author} {\bibfnamefont {G.}~\bibnamefont {Aad}} \emph {et~al.} (\bibinfo {collaboration} {ATLAS}),\ }\href {\doibase 10.1016/j.physletb.2021.136651} {\bibfield  {journal} {\bibinfo  {journal} {Phys. Lett. B}\ }\textbf {\bibinfo {volume} {822}},\ \bibinfo {pages} {136651} (\bibinfo {year} {2021}{\natexlab{b}})},\ \Eprint {http://arxiv.org/abs/2102.13405} {arXiv:2102.13405 [hep-ex]} \BibitemShut {NoStop}%
\bibitem [{\citenamefont {Ablikim}\ \emph {et~al.}(2023)\citenamefont {Ablikim} \emph {et~al.}}]{BESIII:2022rzz}%
  \BibitemOpen
  \bibfield  {author} {\bibinfo {author} {\bibfnamefont {M.}~\bibnamefont {Ablikim}} \emph {et~al.} (\bibinfo {collaboration} {BESIII}),\ }\href {\doibase 10.1016/j.physletb.2023.137698} {\bibfield  {journal} {\bibinfo  {journal} {Phys. Lett. B}\ }\textbf {\bibinfo {volume} {838}},\ \bibinfo {pages} {137698} (\bibinfo {year} {2023})},\ \Eprint {http://arxiv.org/abs/2211.12699} {arXiv:2211.12699 [hep-ex]} \BibitemShut {NoStop}%
\bibitem [{\citenamefont {Jerhot}\ \emph {et~al.}(2022)\citenamefont {Jerhot}, \citenamefont {D\"obrich}, \citenamefont {Ertas}, \citenamefont {Kahlhoefer},\ and\ \citenamefont {Spadaro}}]{Jerhot:2022chi}%
  \BibitemOpen
  \bibfield  {author} {\bibinfo {author} {\bibfnamefont {J.}~\bibnamefont {Jerhot}}, \bibinfo {author} {\bibfnamefont {B.}~\bibnamefont {D\"obrich}}, \bibinfo {author} {\bibfnamefont {F.}~\bibnamefont {Ertas}}, \bibinfo {author} {\bibfnamefont {F.}~\bibnamefont {Kahlhoefer}}, \ and\ \bibinfo {author} {\bibfnamefont {T.}~\bibnamefont {Spadaro}},\ }\href {\doibase 10.1007/JHEP07(2022)094} {\bibfield  {journal} {\bibinfo  {journal} {JHEP}\ }\textbf {\bibinfo {volume} {07}},\ \bibinfo {pages} {094} (\bibinfo {year} {2022})},\ \Eprint {http://arxiv.org/abs/2201.05170} {arXiv:2201.05170 [hep-ph]} \BibitemShut {NoStop}%
\bibitem [{\citenamefont {Mitridate}\ \emph {et~al.}(2023)\citenamefont {Mitridate}, \citenamefont {Papucci}, \citenamefont {Wang}, \citenamefont {Pe\~na},\ and\ \citenamefont {Xie}}]{Mitridate:2023tbj}%
  \BibitemOpen
  \bibfield  {author} {\bibinfo {author} {\bibfnamefont {A.}~\bibnamefont {Mitridate}}, \bibinfo {author} {\bibfnamefont {M.}~\bibnamefont {Papucci}}, \bibinfo {author} {\bibfnamefont {C.~W.}\ \bibnamefont {Wang}}, \bibinfo {author} {\bibfnamefont {C.}~\bibnamefont {Pe\~na}}, \ and\ \bibinfo {author} {\bibfnamefont {S.}~\bibnamefont {Xie}},\ }\href {\doibase 10.1103/PhysRevD.108.055040} {\bibfield  {journal} {\bibinfo  {journal} {Phys. Rev. D}\ }\textbf {\bibinfo {volume} {108}},\ \bibinfo {pages} {055040} (\bibinfo {year} {2023})},\ \Eprint {http://arxiv.org/abs/2304.06109} {arXiv:2304.06109 [hep-ph]} \BibitemShut {NoStop}%
\bibitem [{\citenamefont {Aad}\ \emph {et~al.}(2023)\citenamefont {Aad} \emph {et~al.}}]{ATLAS:2023zfc}%
  \BibitemOpen
  \bibfield  {author} {\bibinfo {author} {\bibfnamefont {G.}~\bibnamefont {Aad}} \emph {et~al.} (\bibinfo {collaboration} {ATLAS}),\ }\href {\doibase 10.1007/JHEP07(2023)234} {\bibfield  {journal} {\bibinfo  {journal} {JHEP}\ }\textbf {\bibinfo {volume} {07}},\ \bibinfo {pages} {234} (\bibinfo {year} {2023})},\ \Eprint {http://arxiv.org/abs/2304.10953} {arXiv:2304.10953 [hep-ex]} \BibitemShut {NoStop}%
\bibitem [{\citenamefont {Tumasyan}\ \emph {et~al.}(2023)\citenamefont {Tumasyan} \emph {et~al.}}]{CMS:2023jgd}%
  \BibitemOpen
  \bibfield  {author} {\bibinfo {author} {\bibfnamefont {A.}~\bibnamefont {Tumasyan}} \emph {et~al.} (\bibinfo {collaboration} {CMS, TOTEM}),\ }\href@noop {} {\  (\bibinfo {year} {2023})},\ \Eprint {http://arxiv.org/abs/2311.02725} {arXiv:2311.02725 [hep-ex]} \BibitemShut {NoStop}%
\bibitem [{\citenamefont {Bharucha}\ \emph {et~al.}(2023)\citenamefont {Bharucha}, \citenamefont {Br\"ummer}, \citenamefont {Desai},\ and\ \citenamefont {Mutzel}}]{Bharucha:2022lty}%
  \BibitemOpen
  \bibfield  {author} {\bibinfo {author} {\bibfnamefont {A.}~\bibnamefont {Bharucha}}, \bibinfo {author} {\bibfnamefont {F.}~\bibnamefont {Br\"ummer}}, \bibinfo {author} {\bibfnamefont {N.}~\bibnamefont {Desai}}, \ and\ \bibinfo {author} {\bibfnamefont {S.}~\bibnamefont {Mutzel}},\ }\href {\doibase 10.1007/JHEP02(2023)141} {\bibfield  {journal} {\bibinfo  {journal} {JHEP}\ }\textbf {\bibinfo {volume} {02}},\ \bibinfo {pages} {141} (\bibinfo {year} {2023})},\ \Eprint {http://arxiv.org/abs/2209.03932} {arXiv:2209.03932 [hep-ph]} \BibitemShut {NoStop}%
\bibitem [{\citenamefont {Ghosh}\ \emph {et~al.}(2024)\citenamefont {Ghosh}, \citenamefont {Ghoshal},\ and\ \citenamefont {Jeesun}}]{Ghosh:2023tyz}%
  \BibitemOpen
  \bibfield  {author} {\bibinfo {author} {\bibfnamefont {D.~K.}\ \bibnamefont {Ghosh}}, \bibinfo {author} {\bibfnamefont {A.}~\bibnamefont {Ghoshal}}, \ and\ \bibinfo {author} {\bibfnamefont {S.}~\bibnamefont {Jeesun}},\ }\href {\doibase 10.1007/JHEP01(2024)026} {\bibfield  {journal} {\bibinfo  {journal} {JHEP}\ }\textbf {\bibinfo {volume} {01}},\ \bibinfo {pages} {026} (\bibinfo {year} {2024})},\ \Eprint {http://arxiv.org/abs/2305.09188} {arXiv:2305.09188 [hep-ph]} \BibitemShut {NoStop}%
\bibitem [{\citenamefont {Bauer}\ \emph {et~al.}(2021)\citenamefont {Bauer}, \citenamefont {Neubert}, \citenamefont {Renner}, \citenamefont {Schnubel},\ and\ \citenamefont {Thamm}}]{Bauer:2020jbp}%
  \BibitemOpen
  \bibfield  {author} {\bibinfo {author} {\bibfnamefont {M.}~\bibnamefont {Bauer}}, \bibinfo {author} {\bibfnamefont {M.}~\bibnamefont {Neubert}}, \bibinfo {author} {\bibfnamefont {S.}~\bibnamefont {Renner}}, \bibinfo {author} {\bibfnamefont {M.}~\bibnamefont {Schnubel}}, \ and\ \bibinfo {author} {\bibfnamefont {A.}~\bibnamefont {Thamm}},\ }\href {\doibase 10.1007/JHEP04(2021)063} {\bibfield  {journal} {\bibinfo  {journal} {JHEP}\ }\textbf {\bibinfo {volume} {04}},\ \bibinfo {pages} {063} (\bibinfo {year} {2021})},\ \Eprint {http://arxiv.org/abs/2012.12272} {arXiv:2012.12272 [hep-ph]} \BibitemShut {NoStop}%
\bibitem [{\citenamefont {Bauer}\ \emph {et~al.}(2022)\citenamefont {Bauer}, \citenamefont {Neubert}, \citenamefont {Renner}, \citenamefont {Schnubel},\ and\ \citenamefont {Thamm}}]{Bauer:2021mvw}%
  \BibitemOpen
  \bibfield  {author} {\bibinfo {author} {\bibfnamefont {M.}~\bibnamefont {Bauer}}, \bibinfo {author} {\bibfnamefont {M.}~\bibnamefont {Neubert}}, \bibinfo {author} {\bibfnamefont {S.}~\bibnamefont {Renner}}, \bibinfo {author} {\bibfnamefont {M.}~\bibnamefont {Schnubel}}, \ and\ \bibinfo {author} {\bibfnamefont {A.}~\bibnamefont {Thamm}},\ }\href {\doibase 10.1007/JHEP09(2022)056} {\bibfield  {journal} {\bibinfo  {journal} {JHEP}\ }\textbf {\bibinfo {volume} {09}},\ \bibinfo {pages} {056} (\bibinfo {year} {2022})},\ \Eprint {http://arxiv.org/abs/2110.10698} {arXiv:2110.10698 [hep-ph]} \BibitemShut {NoStop}%
\bibitem [{\citenamefont {Belyaev}\ \emph {et~al.}(2013)\citenamefont {Belyaev}, \citenamefont {Christensen},\ and\ \citenamefont {Pukhov}}]{Belyaev:2012qa}%
  \BibitemOpen
  \bibfield  {author} {\bibinfo {author} {\bibfnamefont {A.}~\bibnamefont {Belyaev}}, \bibinfo {author} {\bibfnamefont {N.~D.}\ \bibnamefont {Christensen}}, \ and\ \bibinfo {author} {\bibfnamefont {A.}~\bibnamefont {Pukhov}},\ }\href {\doibase 10.1016/j.cpc.2013.01.014} {\bibfield  {journal} {\bibinfo  {journal} {Comput. Phys. Commun.}\ }\textbf {\bibinfo {volume} {184}},\ \bibinfo {pages} {1729} (\bibinfo {year} {2013})},\ \Eprint {http://arxiv.org/abs/1207.6082} {arXiv:1207.6082 [hep-ph]} \BibitemShut {NoStop}%
\bibitem [{\citenamefont {Alloul}\ \emph {et~al.}(2014)\citenamefont {Alloul}, \citenamefont {Christensen}, \citenamefont {Degrande}, \citenamefont {Duhr},\ and\ \citenamefont {Fuks}}]{Alloul:2013bka}%
  \BibitemOpen
  \bibfield  {author} {\bibinfo {author} {\bibfnamefont {A.}~\bibnamefont {Alloul}}, \bibinfo {author} {\bibfnamefont {N.~D.}\ \bibnamefont {Christensen}}, \bibinfo {author} {\bibfnamefont {C.}~\bibnamefont {Degrande}}, \bibinfo {author} {\bibfnamefont {C.}~\bibnamefont {Duhr}}, \ and\ \bibinfo {author} {\bibfnamefont {B.}~\bibnamefont {Fuks}},\ }\href {\doibase 10.1016/j.cpc.2014.04.012} {\bibfield  {journal} {\bibinfo  {journal} {Comput. Phys. Commun.}\ }\textbf {\bibinfo {volume} {185}},\ \bibinfo {pages} {2250} (\bibinfo {year} {2014})},\ \Eprint {http://arxiv.org/abs/1310.1921} {arXiv:1310.1921 [hep-ph]} \BibitemShut {NoStop}%
\bibitem [{\citenamefont {B\'elanger}\ \emph {et~al.}(2018)\citenamefont {B\'elanger}, \citenamefont {Boudjema}, \citenamefont {Goudelis}, \citenamefont {Pukhov},\ and\ \citenamefont {Zaldivar}}]{Belanger:2018ccd}%
  \BibitemOpen
  \bibfield  {author} {\bibinfo {author} {\bibfnamefont {G.}~\bibnamefont {B\'elanger}}, \bibinfo {author} {\bibfnamefont {F.}~\bibnamefont {Boudjema}}, \bibinfo {author} {\bibfnamefont {A.}~\bibnamefont {Goudelis}}, \bibinfo {author} {\bibfnamefont {A.}~\bibnamefont {Pukhov}}, \ and\ \bibinfo {author} {\bibfnamefont {B.}~\bibnamefont {Zaldivar}},\ }\href {\doibase 10.1016/j.cpc.2018.04.027} {\bibfield  {journal} {\bibinfo  {journal} {Comput. Phys. Commun.}\ }\textbf {\bibinfo {volume} {231}},\ \bibinfo {pages} {173} (\bibinfo {year} {2018})},\ \Eprint {http://arxiv.org/abs/1801.03509} {arXiv:1801.03509 [hep-ph]} \BibitemShut {NoStop}%
\bibitem [{\citenamefont {Albert}\ \emph {et~al.}(2014)\citenamefont {Albert}, \citenamefont {Gomez-Vargas}, \citenamefont {Grefe}, \citenamefont {Munoz}, \citenamefont {Weniger}, \citenamefont {Bloom}, \citenamefont {Charles}, \citenamefont {Mazziotta},\ and\ \citenamefont {Morselli}}]{Albert:2014hwa}%
  \BibitemOpen
  \bibfield  {author} {\bibinfo {author} {\bibfnamefont {A.}~\bibnamefont {Albert}}, \bibinfo {author} {\bibfnamefont {G.~A.}\ \bibnamefont {Gomez-Vargas}}, \bibinfo {author} {\bibfnamefont {M.}~\bibnamefont {Grefe}}, \bibinfo {author} {\bibfnamefont {C.}~\bibnamefont {Munoz}}, \bibinfo {author} {\bibfnamefont {C.}~\bibnamefont {Weniger}}, \bibinfo {author} {\bibfnamefont {E.~D.}\ \bibnamefont {Bloom}}, \bibinfo {author} {\bibfnamefont {E.}~\bibnamefont {Charles}}, \bibinfo {author} {\bibfnamefont {M.~N.}\ \bibnamefont {Mazziotta}}, \ and\ \bibinfo {author} {\bibfnamefont {A.}~\bibnamefont {Morselli}} (\bibinfo {collaboration} {Fermi-LAT}),\ }\href {\doibase 10.1088/1475-7516/2014/10/023} {\bibfield  {journal} {\bibinfo  {journal} {JCAP}\ }\textbf {\bibinfo {volume} {10}},\ \bibinfo {pages} {023} (\bibinfo {year} {2014})},\ \Eprint {http://arxiv.org/abs/1406.3430} {arXiv:1406.3430 [astro-ph.HE]} \BibitemShut {NoStop}%
\bibitem [{\citenamefont {Foster}\ \emph {et~al.}(2023)\citenamefont {Foster}, \citenamefont {Park}, \citenamefont {Safdi}, \citenamefont {Soreq},\ and\ \citenamefont {Xu}}]{Foster:2022nva}%
  \BibitemOpen
  \bibfield  {author} {\bibinfo {author} {\bibfnamefont {J.~W.}\ \bibnamefont {Foster}}, \bibinfo {author} {\bibfnamefont {Y.}~\bibnamefont {Park}}, \bibinfo {author} {\bibfnamefont {B.~R.}\ \bibnamefont {Safdi}}, \bibinfo {author} {\bibfnamefont {Y.}~\bibnamefont {Soreq}}, \ and\ \bibinfo {author} {\bibfnamefont {W.~L.}\ \bibnamefont {Xu}},\ }\href {\doibase 10.1103/PhysRevD.107.103047} {\bibfield  {journal} {\bibinfo  {journal} {Phys. Rev. D}\ }\textbf {\bibinfo {volume} {107}},\ \bibinfo {pages} {103047} (\bibinfo {year} {2023})},\ \Eprint {http://arxiv.org/abs/2212.07435} {arXiv:2212.07435 [hep-ph]} \BibitemShut {NoStop}%
\bibitem [{\citenamefont {Abe}\ \emph {et~al.}(2023)\citenamefont {Abe} \emph {et~al.}}]{MAGIC:2022acl}%
  \BibitemOpen
  \bibfield  {author} {\bibinfo {author} {\bibfnamefont {H.}~\bibnamefont {Abe}} \emph {et~al.} (\bibinfo {collaboration} {MAGIC}),\ }\href {\doibase 10.1103/PhysRevLett.130.061002} {\bibfield  {journal} {\bibinfo  {journal} {Phys. Rev. Lett.}\ }\textbf {\bibinfo {volume} {130}},\ \bibinfo {pages} {061002} (\bibinfo {year} {2023})},\ \Eprint {http://arxiv.org/abs/2212.10527} {arXiv:2212.10527 [astro-ph.HE]} \BibitemShut {NoStop}%
\bibitem [{\citenamefont {Abdallah}\ \emph {et~al.}(2018)\citenamefont {Abdallah} \emph {et~al.}}]{HESS:2018cbt}%
  \BibitemOpen
  \bibfield  {author} {\bibinfo {author} {\bibfnamefont {H.}~\bibnamefont {Abdallah}} \emph {et~al.} (\bibinfo {collaboration} {HESS}),\ }\href {\doibase 10.1103/PhysRevLett.120.201101} {\bibfield  {journal} {\bibinfo  {journal} {Phys. Rev. Lett.}\ }\textbf {\bibinfo {volume} {120}},\ \bibinfo {pages} {201101} (\bibinfo {year} {2018})},\ \Eprint {http://arxiv.org/abs/1805.05741} {arXiv:1805.05741 [astro-ph.HE]} \BibitemShut {NoStop}%
\bibitem [{\citenamefont {Pullen}\ \emph {et~al.}(2007)\citenamefont {Pullen}, \citenamefont {Chary},\ and\ \citenamefont {Kamionkowski}}]{Pullen:2006sy}%
  \BibitemOpen
  \bibfield  {author} {\bibinfo {author} {\bibfnamefont {A.~R.}\ \bibnamefont {Pullen}}, \bibinfo {author} {\bibfnamefont {R.-R.}\ \bibnamefont {Chary}}, \ and\ \bibinfo {author} {\bibfnamefont {M.}~\bibnamefont {Kamionkowski}},\ }\href {\doibase 10.1103/PhysRevD.76.063006} {\bibfield  {journal} {\bibinfo  {journal} {Phys. Rev. D}\ }\textbf {\bibinfo {volume} {76}},\ \bibinfo {pages} {063006} (\bibinfo {year} {2007})},\ \bibinfo {note} {[Erratum: Phys.Rev.D 83, 029904 (2011)]},\ \Eprint {http://arxiv.org/abs/astro-ph/0610295} {arXiv:astro-ph/0610295} \BibitemShut {NoStop}%
\bibitem [{\citenamefont {Boddy}\ and\ \citenamefont {Kumar}(2015)}]{Boddy:2015efa}%
  \BibitemOpen
  \bibfield  {author} {\bibinfo {author} {\bibfnamefont {K.~K.}\ \bibnamefont {Boddy}}\ and\ \bibinfo {author} {\bibfnamefont {J.}~\bibnamefont {Kumar}},\ }\href {\doibase 10.1103/PhysRevD.92.023533} {\bibfield  {journal} {\bibinfo  {journal} {Phys. Rev. D}\ }\textbf {\bibinfo {volume} {92}},\ \bibinfo {pages} {023533} (\bibinfo {year} {2015})},\ \Eprint {http://arxiv.org/abs/1504.04024} {arXiv:1504.04024 [astro-ph.CO]} \BibitemShut {NoStop}%
\bibitem [{\citenamefont {De~Angelis}\ \emph {et~al.}(2017)\citenamefont {De~Angelis} \emph {et~al.}}]{e-ASTROGAM:2016bph}%
  \BibitemOpen
  \bibfield  {author} {\bibinfo {author} {\bibfnamefont {A.}~\bibnamefont {De~Angelis}} \emph {et~al.} (\bibinfo {collaboration} {e-ASTROGAM}),\ }\href {\doibase 10.1007/s10686-017-9533-6} {\bibfield  {journal} {\bibinfo  {journal} {Exper. Astron.}\ }\textbf {\bibinfo {volume} {44}},\ \bibinfo {pages} {25} (\bibinfo {year} {2017})},\ \Eprint {http://arxiv.org/abs/1611.02232} {arXiv:1611.02232 [astro-ph.HE]} \BibitemShut {NoStop}%
\bibitem [{\citenamefont {Navarro}\ \emph {et~al.}(1996)\citenamefont {Navarro}, \citenamefont {Frenk},\ and\ \citenamefont {White}}]{Navarro:1995iw}%
  \BibitemOpen
  \bibfield  {author} {\bibinfo {author} {\bibfnamefont {J.~F.}\ \bibnamefont {Navarro}}, \bibinfo {author} {\bibfnamefont {C.~S.}\ \bibnamefont {Frenk}}, \ and\ \bibinfo {author} {\bibfnamefont {S.~D.~M.}\ \bibnamefont {White}},\ }\href {\doibase 10.1086/177173} {\bibfield  {journal} {\bibinfo  {journal} {Astrophys. J.}\ }\textbf {\bibinfo {volume} {462}},\ \bibinfo {pages} {563} (\bibinfo {year} {1996})},\ \Eprint {http://arxiv.org/abs/astro-ph/9508025} {arXiv:astro-ph/9508025} \BibitemShut {NoStop}%
\bibitem [{\citenamefont {Cirelli}\ \emph {et~al.}(2011)\citenamefont {Cirelli}, \citenamefont {Corcella}, \citenamefont {Hektor}, \citenamefont {Hutsi}, \citenamefont {Kadastik}, \citenamefont {Panci}, \citenamefont {Raidal}, \citenamefont {Sala},\ and\ \citenamefont {Strumia}}]{Cirelli:2010xx}%
  \BibitemOpen
  \bibfield  {author} {\bibinfo {author} {\bibfnamefont {M.}~\bibnamefont {Cirelli}}, \bibinfo {author} {\bibfnamefont {G.}~\bibnamefont {Corcella}}, \bibinfo {author} {\bibfnamefont {A.}~\bibnamefont {Hektor}}, \bibinfo {author} {\bibfnamefont {G.}~\bibnamefont {Hutsi}}, \bibinfo {author} {\bibfnamefont {M.}~\bibnamefont {Kadastik}}, \bibinfo {author} {\bibfnamefont {P.}~\bibnamefont {Panci}}, \bibinfo {author} {\bibfnamefont {M.}~\bibnamefont {Raidal}}, \bibinfo {author} {\bibfnamefont {F.}~\bibnamefont {Sala}}, \ and\ \bibinfo {author} {\bibfnamefont {A.}~\bibnamefont {Strumia}},\ }\href {\doibase 10.1088/1475-7516/2012/10/E01} {\bibfield  {journal} {\bibinfo  {journal} {JCAP}\ }\textbf {\bibinfo {volume} {03}},\ \bibinfo {pages} {051} (\bibinfo {year} {2011})},\ \bibinfo {note} {[Erratum: JCAP 10, E01 (2012)]},\ \Eprint {http://arxiv.org/abs/1012.4515} {arXiv:1012.4515 [hep-ph]} \BibitemShut {NoStop}%
\bibitem [{\citenamefont {Aghanim}\ \emph {et~al.}(2020)\citenamefont {Aghanim} \emph {et~al.}}]{Planck:2018vyg}%
  \BibitemOpen
  \bibfield  {author} {\bibinfo {author} {\bibfnamefont {N.}~\bibnamefont {Aghanim}} \emph {et~al.} (\bibinfo {collaboration} {Planck}),\ }\href {\doibase 10.1051/0004-6361/201833910} {\bibfield  {journal} {\bibinfo  {journal} {Astron. Astrophys.}\ }\textbf {\bibinfo {volume} {641}},\ \bibinfo {pages} {A6} (\bibinfo {year} {2020})},\ \bibinfo {note} {[Erratum: Astron.Astrophys. 652, C4 (2021)]},\ \Eprint {http://arxiv.org/abs/1807.06209} {arXiv:1807.06209 [astro-ph.CO]} \BibitemShut {NoStop}%
\bibitem [{\citenamefont {Slatyer}(2016)}]{Slatyer:2015jla}%
  \BibitemOpen
  \bibfield  {author} {\bibinfo {author} {\bibfnamefont {T.~R.}\ \bibnamefont {Slatyer}},\ }\href {\doibase 10.1103/PhysRevD.93.023527} {\bibfield  {journal} {\bibinfo  {journal} {Phys. Rev. D}\ }\textbf {\bibinfo {volume} {93}},\ \bibinfo {pages} {023527} (\bibinfo {year} {2016})},\ \Eprint {http://arxiv.org/abs/1506.03811} {arXiv:1506.03811 [hep-ph]} \BibitemShut {NoStop}%
\bibitem [{\citenamefont {Egana-Ugrinovic}\ \emph {et~al.}(2018)\citenamefont {Egana-Ugrinovic}, \citenamefont {Low},\ and\ \citenamefont {Ruderman}}]{Egana-Ugrinovic:2018roi}%
  \BibitemOpen
  \bibfield  {author} {\bibinfo {author} {\bibfnamefont {D.}~\bibnamefont {Egana-Ugrinovic}}, \bibinfo {author} {\bibfnamefont {M.}~\bibnamefont {Low}}, \ and\ \bibinfo {author} {\bibfnamefont {J.~T.}\ \bibnamefont {Ruderman}},\ }\href {\doibase 10.1007/JHEP05(2018)012} {\bibfield  {journal} {\bibinfo  {journal} {JHEP}\ }\textbf {\bibinfo {volume} {05}},\ \bibinfo {pages} {012} (\bibinfo {year} {2018})},\ \Eprint {http://arxiv.org/abs/1801.05432} {arXiv:1801.05432 [hep-ph]} \BibitemShut {NoStop}%
\bibitem [{\citenamefont {Altakach}\ \emph {et~al.}(2022)\citenamefont {Altakach}, \citenamefont {Lamba}, \citenamefont {Mase\l{}ek}, \citenamefont {Mitsou},\ and\ \citenamefont {Sakurai}}]{Altakach:2022hgn}%
  \BibitemOpen
  \bibfield  {author} {\bibinfo {author} {\bibfnamefont {M.~M.}\ \bibnamefont {Altakach}}, \bibinfo {author} {\bibfnamefont {P.}~\bibnamefont {Lamba}}, \bibinfo {author} {\bibfnamefont {R.}~\bibnamefont {Mase\l{}ek}}, \bibinfo {author} {\bibfnamefont {V.~A.}\ \bibnamefont {Mitsou}}, \ and\ \bibinfo {author} {\bibfnamefont {K.}~\bibnamefont {Sakurai}},\ }\href {\doibase 10.1140/epjc/s10052-022-10805-z} {\bibfield  {journal} {\bibinfo  {journal} {Eur. Phys. J. C}\ }\textbf {\bibinfo {volume} {82}},\ \bibinfo {pages} {848} (\bibinfo {year} {2022})},\ \Eprint {http://arxiv.org/abs/2204.03667} {arXiv:2204.03667 [hep-ph]} \BibitemShut {NoStop}%
\bibitem [{\citenamefont {Hayrapetyan}\ \emph {et~al.}(2024)\citenamefont {Hayrapetyan} \emph {et~al.}}]{CMS:2024eyx}%
  \BibitemOpen
  \bibfield  {author} {\bibinfo {author} {\bibfnamefont {A.}~\bibnamefont {Hayrapetyan}} \emph {et~al.} (\bibinfo {collaboration} {CMS}),\ }\href@noop {} {\  (\bibinfo {year} {2024})},\ \Eprint {http://arxiv.org/abs/2402.09932} {arXiv:2402.09932 [hep-ex]} \BibitemShut {NoStop}%
\bibitem [{\citenamefont {Frandsen}\ \emph {et~al.}(2012)\citenamefont {Frandsen}, \citenamefont {Haisch}, \citenamefont {Kahlhoefer}, \citenamefont {Mertsch},\ and\ \citenamefont {Schmidt-Hoberg}}]{Frandsen:2012db}%
  \BibitemOpen
  \bibfield  {author} {\bibinfo {author} {\bibfnamefont {M.~T.}\ \bibnamefont {Frandsen}}, \bibinfo {author} {\bibfnamefont {U.}~\bibnamefont {Haisch}}, \bibinfo {author} {\bibfnamefont {F.}~\bibnamefont {Kahlhoefer}}, \bibinfo {author} {\bibfnamefont {P.}~\bibnamefont {Mertsch}}, \ and\ \bibinfo {author} {\bibfnamefont {K.}~\bibnamefont {Schmidt-Hoberg}},\ }\href {\doibase 10.1088/1475-7516/2012/10/033} {\bibfield  {journal} {\bibinfo  {journal} {JCAP}\ }\textbf {\bibinfo {volume} {10}},\ \bibinfo {pages} {033} (\bibinfo {year} {2012})},\ \Eprint {http://arxiv.org/abs/1207.3971} {arXiv:1207.3971 [hep-ph]} \BibitemShut {NoStop}%
\bibitem [{\citenamefont {Darm\'e}\ \emph {et~al.}(2021)\citenamefont {Darm\'e}, \citenamefont {Giacchino}, \citenamefont {Nardi},\ and\ \citenamefont {Raggi}}]{Darme:2020sjf}%
  \BibitemOpen
  \bibfield  {author} {\bibinfo {author} {\bibfnamefont {L.}~\bibnamefont {Darm\'e}}, \bibinfo {author} {\bibfnamefont {F.}~\bibnamefont {Giacchino}}, \bibinfo {author} {\bibfnamefont {E.}~\bibnamefont {Nardi}}, \ and\ \bibinfo {author} {\bibfnamefont {M.}~\bibnamefont {Raggi}},\ }\href {\doibase 10.1007/JHEP06(2021)009} {\bibfield  {journal} {\bibinfo  {journal} {JHEP}\ }\textbf {\bibinfo {volume} {06}},\ \bibinfo {pages} {009} (\bibinfo {year} {2021})},\ \Eprint {http://arxiv.org/abs/2012.07894} {arXiv:2012.07894 [hep-ph]} \BibitemShut {NoStop}%
\bibitem [{\citenamefont {Alwall}\ \emph {et~al.}(2014)\citenamefont {Alwall}, \citenamefont {Frederix}, \citenamefont {Frixione}, \citenamefont {Hirschi}, \citenamefont {Maltoni}, \citenamefont {Mattelaer}, \citenamefont {Shao}, \citenamefont {Stelzer}, \citenamefont {Torrielli},\ and\ \citenamefont {Zaro}}]{Alwall:2014hca}%
  \BibitemOpen
  \bibfield  {author} {\bibinfo {author} {\bibfnamefont {J.}~\bibnamefont {Alwall}}, \bibinfo {author} {\bibfnamefont {R.}~\bibnamefont {Frederix}}, \bibinfo {author} {\bibfnamefont {S.}~\bibnamefont {Frixione}}, \bibinfo {author} {\bibfnamefont {V.}~\bibnamefont {Hirschi}}, \bibinfo {author} {\bibfnamefont {F.}~\bibnamefont {Maltoni}}, \bibinfo {author} {\bibfnamefont {O.}~\bibnamefont {Mattelaer}}, \bibinfo {author} {\bibfnamefont {H.~S.}\ \bibnamefont {Shao}}, \bibinfo {author} {\bibfnamefont {T.}~\bibnamefont {Stelzer}}, \bibinfo {author} {\bibfnamefont {P.}~\bibnamefont {Torrielli}}, \ and\ \bibinfo {author} {\bibfnamefont {M.}~\bibnamefont {Zaro}},\ }\href {\doibase 10.1007/JHEP07(2014)079} {\bibfield  {journal} {\bibinfo  {journal} {JHEP}\ }\textbf {\bibinfo {volume} {07}},\ \bibinfo {pages} {079} (\bibinfo {year} {2014})},\ \Eprint {http://arxiv.org/abs/1405.0301} {arXiv:1405.0301 [hep-ph]} \BibitemShut {NoStop}%
\bibitem [{\citenamefont {Darm\'e}\ \emph {et~al.}(2023)\citenamefont {Darm\'e} \emph {et~al.}}]{Darme:2023jdn}%
  \BibitemOpen
  \bibfield  {author} {\bibinfo {author} {\bibfnamefont {L.}~\bibnamefont {Darm\'e}} \emph {et~al.},\ }\href {\doibase 10.1140/epjc/s10052-023-11780-9} {\bibfield  {journal} {\bibinfo  {journal} {Eur. Phys. J. C}\ }\textbf {\bibinfo {volume} {83}},\ \bibinfo {pages} {631} (\bibinfo {year} {2023})},\ \Eprint {http://arxiv.org/abs/2304.09883} {arXiv:2304.09883 [hep-ph]} \BibitemShut {NoStop}%
\bibitem [{\citenamefont {Aaboud}\ \emph {et~al.}(2017)\citenamefont {Aaboud} \emph {et~al.}}]{ATLAS:2017zdf}%
  \BibitemOpen
  \bibfield  {author} {\bibinfo {author} {\bibfnamefont {M.}~\bibnamefont {Aaboud}} \emph {et~al.} (\bibinfo {collaboration} {ATLAS}),\ }\href {\doibase 10.1007/JHEP10(2017)112} {\bibfield  {journal} {\bibinfo  {journal} {JHEP}\ }\textbf {\bibinfo {volume} {10}},\ \bibinfo {pages} {112} (\bibinfo {year} {2017})},\ \Eprint {http://arxiv.org/abs/1708.00212} {arXiv:1708.00212 [hep-ex]} \BibitemShut {NoStop}%
\bibitem [{\citenamefont {Ablikim}\ \emph {et~al.}(2024)\citenamefont {Ablikim} \emph {et~al.}}]{BESIII:2024hdv}%
  \BibitemOpen
  \bibfield  {author} {\bibinfo {author} {\bibfnamefont {M.}~\bibnamefont {Ablikim}} \emph {et~al.} (\bibinfo {collaboration} {BESIII}),\ }\href@noop {} {\  (\bibinfo {year} {2024})},\ \Eprint {http://arxiv.org/abs/2404.04640} {arXiv:2404.04640 [hep-ex]} \BibitemShut {NoStop}%
\bibitem [{\citenamefont {Adler}\ \emph {et~al.}(2004)\citenamefont {Adler} \emph {et~al.}}]{E787:2004ovg}%
  \BibitemOpen
  \bibfield  {author} {\bibinfo {author} {\bibfnamefont {S.}~\bibnamefont {Adler}} \emph {et~al.} (\bibinfo {collaboration} {E787}),\ }\href {\doibase 10.1103/PhysRevD.70.037102} {\bibfield  {journal} {\bibinfo  {journal} {Phys. Rev. D}\ }\textbf {\bibinfo {volume} {70}},\ \bibinfo {pages} {037102} (\bibinfo {year} {2004})},\ \Eprint {http://arxiv.org/abs/hep-ex/0403034} {arXiv:hep-ex/0403034} \BibitemShut {NoStop}%
\bibitem [{\citenamefont {Artamonov}\ \emph {et~al.}(2009)\citenamefont {Artamonov} \emph {et~al.}}]{BNL-E949:2009dza}%
  \BibitemOpen
  \bibfield  {author} {\bibinfo {author} {\bibfnamefont {A.~V.}\ \bibnamefont {Artamonov}} \emph {et~al.} (\bibinfo {collaboration} {BNL-E949}),\ }\href {\doibase 10.1103/PhysRevD.79.092004} {\bibfield  {journal} {\bibinfo  {journal} {Phys. Rev. D}\ }\textbf {\bibinfo {volume} {79}},\ \bibinfo {pages} {092004} (\bibinfo {year} {2009})},\ \Eprint {http://arxiv.org/abs/0903.0030} {arXiv:0903.0030 [hep-ex]} \BibitemShut {NoStop}%
\bibitem [{\citenamefont {Ertas}\ and\ \citenamefont {Kahlhoefer}(2020)}]{Ertas:2020xcc}%
  \BibitemOpen
  \bibfield  {author} {\bibinfo {author} {\bibfnamefont {F.}~\bibnamefont {Ertas}}\ and\ \bibinfo {author} {\bibfnamefont {F.}~\bibnamefont {Kahlhoefer}},\ }\href {\doibase 10.1007/JHEP07(2020)050} {\bibfield  {journal} {\bibinfo  {journal} {JHEP}\ }\textbf {\bibinfo {volume} {07}},\ \bibinfo {pages} {050} (\bibinfo {year} {2020})},\ \Eprint {http://arxiv.org/abs/2004.01193} {arXiv:2004.01193 [hep-ph]} \BibitemShut {NoStop}%
\bibitem [{\citenamefont {Aaij}\ \emph {et~al.}(2015)\citenamefont {Aaij} \emph {et~al.}}]{LHCb:2015nkv}%
  \BibitemOpen
  \bibfield  {author} {\bibinfo {author} {\bibfnamefont {R.}~\bibnamefont {Aaij}} \emph {et~al.} (\bibinfo {collaboration} {LHCb}),\ }\href {\doibase 10.1103/PhysRevLett.115.161802} {\bibfield  {journal} {\bibinfo  {journal} {Phys. Rev. Lett.}\ }\textbf {\bibinfo {volume} {115}},\ \bibinfo {pages} {161802} (\bibinfo {year} {2015})},\ \Eprint {http://arxiv.org/abs/1508.04094} {arXiv:1508.04094 [hep-ex]} \BibitemShut {NoStop}%
\bibitem [{\citenamefont {Aaij}\ \emph {et~al.}(2017)\citenamefont {Aaij} \emph {et~al.}}]{LHCb:2016awg}%
  \BibitemOpen
  \bibfield  {author} {\bibinfo {author} {\bibfnamefont {R.}~\bibnamefont {Aaij}} \emph {et~al.} (\bibinfo {collaboration} {LHCb}),\ }\href {\doibase 10.1103/PhysRevD.95.071101} {\bibfield  {journal} {\bibinfo  {journal} {Phys. Rev. D}\ }\textbf {\bibinfo {volume} {95}},\ \bibinfo {pages} {071101} (\bibinfo {year} {2017})},\ \Eprint {http://arxiv.org/abs/1612.07818} {arXiv:1612.07818 [hep-ex]} \BibitemShut {NoStop}%
\bibitem [{\citenamefont {Elahi}\ \emph {et~al.}(2015)\citenamefont {Elahi}, \citenamefont {Kolda},\ and\ \citenamefont {Unwin}}]{Elahi:2014fsa}%
  \BibitemOpen
  \bibfield  {author} {\bibinfo {author} {\bibfnamefont {F.}~\bibnamefont {Elahi}}, \bibinfo {author} {\bibfnamefont {C.}~\bibnamefont {Kolda}}, \ and\ \bibinfo {author} {\bibfnamefont {J.}~\bibnamefont {Unwin}},\ }\href {\doibase 10.1007/JHEP03(2015)048} {\bibfield  {journal} {\bibinfo  {journal} {JHEP}\ }\textbf {\bibinfo {volume} {03}},\ \bibinfo {pages} {048} (\bibinfo {year} {2015})},\ \Eprint {http://arxiv.org/abs/1410.6157} {arXiv:1410.6157 [hep-ph]} \BibitemShut {NoStop}%
\bibitem [{\citenamefont {Garcia}\ \emph {et~al.}(2017)\citenamefont {Garcia}, \citenamefont {Mambrini}, \citenamefont {Olive},\ and\ \citenamefont {Peloso}}]{Garcia:2017tuj}%
  \BibitemOpen
  \bibfield  {author} {\bibinfo {author} {\bibfnamefont {M.~A.~G.}\ \bibnamefont {Garcia}}, \bibinfo {author} {\bibfnamefont {Y.}~\bibnamefont {Mambrini}}, \bibinfo {author} {\bibfnamefont {K.~A.}\ \bibnamefont {Olive}}, \ and\ \bibinfo {author} {\bibfnamefont {M.}~\bibnamefont {Peloso}},\ }\href {\doibase 10.1103/PhysRevD.96.103510} {\bibfield  {journal} {\bibinfo  {journal} {Phys. Rev. D}\ }\textbf {\bibinfo {volume} {96}},\ \bibinfo {pages} {103510} (\bibinfo {year} {2017})},\ \Eprint {http://arxiv.org/abs/1709.01549} {arXiv:1709.01549 [hep-ph]} \BibitemShut {NoStop}%
\bibitem [{\citenamefont {Zelko}\ \emph {et~al.}(2022)\citenamefont {Zelko}, \citenamefont {Treu}, \citenamefont {Abazajian}, \citenamefont {Gilman}, \citenamefont {Benson}, \citenamefont {Birrer}, \citenamefont {Nierenberg},\ and\ \citenamefont {Kusenko}}]{Zelko:2022tgf}%
  \BibitemOpen
  \bibfield  {author} {\bibinfo {author} {\bibfnamefont {I.~A.}\ \bibnamefont {Zelko}}, \bibinfo {author} {\bibfnamefont {T.}~\bibnamefont {Treu}}, \bibinfo {author} {\bibfnamefont {K.~N.}\ \bibnamefont {Abazajian}}, \bibinfo {author} {\bibfnamefont {D.}~\bibnamefont {Gilman}}, \bibinfo {author} {\bibfnamefont {A.~J.}\ \bibnamefont {Benson}}, \bibinfo {author} {\bibfnamefont {S.}~\bibnamefont {Birrer}}, \bibinfo {author} {\bibfnamefont {A.~M.}\ \bibnamefont {Nierenberg}}, \ and\ \bibinfo {author} {\bibfnamefont {A.}~\bibnamefont {Kusenko}},\ }\href {\doibase 10.1103/PhysRevLett.129.191301} {\bibfield  {journal} {\bibinfo  {journal} {Phys. Rev. Lett.}\ }\textbf {\bibinfo {volume} {129}},\ \bibinfo {pages} {191301} (\bibinfo {year} {2022})},\ \Eprint {http://arxiv.org/abs/2205.09777} {arXiv:2205.09777 [hep-ph]} \BibitemShut {NoStop}%
\bibitem [{\citenamefont {Tanabashi}\ \emph {et~al.}(2018)\citenamefont {Tanabashi} \emph {et~al.}}]{ParticleDataGroup:2018ovx}%
  \BibitemOpen
  \bibfield  {author} {\bibinfo {author} {\bibfnamefont {M.}~\bibnamefont {Tanabashi}} \emph {et~al.} (\bibinfo {collaboration} {Particle Data Group}),\ }\href {\doibase 10.1103/PhysRevD.98.030001} {\bibfield  {journal} {\bibinfo  {journal} {Phys. Rev. D}\ }\textbf {\bibinfo {volume} {98}},\ \bibinfo {pages} {030001} (\bibinfo {year} {2018})}\BibitemShut {NoStop}%
\bibitem [{\citenamefont {Depta}\ \emph {et~al.}(2020)\citenamefont {Depta}, \citenamefont {Hufnagel},\ and\ \citenamefont {Schmidt-Hoberg}}]{Depta:2020wmr}%
  \BibitemOpen
  \bibfield  {author} {\bibinfo {author} {\bibfnamefont {P.~F.}\ \bibnamefont {Depta}}, \bibinfo {author} {\bibfnamefont {M.}~\bibnamefont {Hufnagel}}, \ and\ \bibinfo {author} {\bibfnamefont {K.}~\bibnamefont {Schmidt-Hoberg}},\ }\href {\doibase 10.1088/1475-7516/2020/05/009} {\bibfield  {journal} {\bibinfo  {journal} {JCAP}\ }\textbf {\bibinfo {volume} {05}},\ \bibinfo {pages} {009} (\bibinfo {year} {2020})},\ \Eprint {http://arxiv.org/abs/2002.08370} {arXiv:2002.08370 [hep-ph]} \BibitemShut {NoStop}%
\bibitem [{\citenamefont {Jaeckel}\ \emph {et~al.}(2018)\citenamefont {Jaeckel}, \citenamefont {Malta},\ and\ \citenamefont {Redondo}}]{Jaeckel:2017tud}%
  \BibitemOpen
  \bibfield  {author} {\bibinfo {author} {\bibfnamefont {J.}~\bibnamefont {Jaeckel}}, \bibinfo {author} {\bibfnamefont {P.~C.}\ \bibnamefont {Malta}}, \ and\ \bibinfo {author} {\bibfnamefont {J.}~\bibnamefont {Redondo}},\ }\href {\doibase 10.1103/PhysRevD.98.055032} {\bibfield  {journal} {\bibinfo  {journal} {Phys. Rev. D}\ }\textbf {\bibinfo {volume} {98}},\ \bibinfo {pages} {055032} (\bibinfo {year} {2018})},\ \Eprint {http://arxiv.org/abs/1702.02964} {arXiv:1702.02964 [hep-ph]} \BibitemShut {NoStop}%
\bibitem [{\citenamefont {Hoof}\ and\ \citenamefont {Schulz}(2023)}]{Hoof:2022xbe}%
  \BibitemOpen
  \bibfield  {author} {\bibinfo {author} {\bibfnamefont {S.}~\bibnamefont {Hoof}}\ and\ \bibinfo {author} {\bibfnamefont {L.}~\bibnamefont {Schulz}},\ }\href {\doibase 10.1088/1475-7516/2023/03/054} {\bibfield  {journal} {\bibinfo  {journal} {JCAP}\ }\textbf {\bibinfo {volume} {03}},\ \bibinfo {pages} {054} (\bibinfo {year} {2023})},\ \Eprint {http://arxiv.org/abs/2212.09764} {arXiv:2212.09764 [hep-ph]} \BibitemShut {NoStop}%
\bibitem [{\citenamefont {M\"uller}\ \emph {et~al.}(2023)\citenamefont {M\"uller}, \citenamefont {Calore}, \citenamefont {Carenza}, \citenamefont {Eckner},\ and\ \citenamefont {Marsh}}]{Muller:2023vjm}%
  \BibitemOpen
  \bibfield  {author} {\bibinfo {author} {\bibfnamefont {E.}~\bibnamefont {M\"uller}}, \bibinfo {author} {\bibfnamefont {F.}~\bibnamefont {Calore}}, \bibinfo {author} {\bibfnamefont {P.}~\bibnamefont {Carenza}}, \bibinfo {author} {\bibfnamefont {C.}~\bibnamefont {Eckner}}, \ and\ \bibinfo {author} {\bibfnamefont {M.~C.~D.}\ \bibnamefont {Marsh}},\ }\href {\doibase 10.1088/1475-7516/2023/07/056} {\bibfield  {journal} {\bibinfo  {journal} {JCAP}\ }\textbf {\bibinfo {volume} {07}},\ \bibinfo {pages} {056} (\bibinfo {year} {2023})},\ \Eprint {http://arxiv.org/abs/2304.01060} {arXiv:2304.01060 [astro-ph.HE]} \BibitemShut {NoStop}%
\bibitem [{\citenamefont {Diamond}\ \emph {et~al.}(2023)\citenamefont {Diamond}, \citenamefont {Fiorillo}, \citenamefont {Marques-Tavares},\ and\ \citenamefont {Vitagliano}}]{Diamond:2023scc}%
  \BibitemOpen
  \bibfield  {author} {\bibinfo {author} {\bibfnamefont {M.}~\bibnamefont {Diamond}}, \bibinfo {author} {\bibfnamefont {D.~F.~G.}\ \bibnamefont {Fiorillo}}, \bibinfo {author} {\bibfnamefont {G.}~\bibnamefont {Marques-Tavares}}, \ and\ \bibinfo {author} {\bibfnamefont {E.}~\bibnamefont {Vitagliano}},\ }\href {\doibase 10.1103/PhysRevD.107.103029} {\bibfield  {journal} {\bibinfo  {journal} {Phys. Rev. D}\ }\textbf {\bibinfo {volume} {107}},\ \bibinfo {pages} {103029} (\bibinfo {year} {2023})},\ \bibinfo {note} {[Erratum: Phys.Rev.D 108, 049902 (2023)]},\ \Eprint {http://arxiv.org/abs/2303.11395} {arXiv:2303.11395 [hep-ph]} \BibitemShut {NoStop}%
\bibitem [{\citenamefont {Caputo}\ \emph {et~al.}(2022)\citenamefont {Caputo}, \citenamefont {Janka}, \citenamefont {Raffelt},\ and\ \citenamefont {Vitagliano}}]{Caputo:2022mah}%
  \BibitemOpen
  \bibfield  {author} {\bibinfo {author} {\bibfnamefont {A.}~\bibnamefont {Caputo}}, \bibinfo {author} {\bibfnamefont {H.-T.}\ \bibnamefont {Janka}}, \bibinfo {author} {\bibfnamefont {G.}~\bibnamefont {Raffelt}}, \ and\ \bibinfo {author} {\bibfnamefont {E.}~\bibnamefont {Vitagliano}},\ }\href {\doibase 10.1103/PhysRevLett.128.221103} {\bibfield  {journal} {\bibinfo  {journal} {Phys. Rev. Lett.}\ }\textbf {\bibinfo {volume} {128}},\ \bibinfo {pages} {221103} (\bibinfo {year} {2022})},\ \Eprint {http://arxiv.org/abs/2201.09890} {arXiv:2201.09890 [astro-ph.HE]} \BibitemShut {NoStop}%
\bibitem [{\citenamefont {Diamond}\ \emph {et~al.}(2024)\citenamefont {Diamond}, \citenamefont {Fiorillo}, \citenamefont {Marques-Tavares}, \citenamefont {Tamborra},\ and\ \citenamefont {Vitagliano}}]{Diamond:2023cto}%
  \BibitemOpen
  \bibfield  {author} {\bibinfo {author} {\bibfnamefont {M.}~\bibnamefont {Diamond}}, \bibinfo {author} {\bibfnamefont {D.~F.~G.}\ \bibnamefont {Fiorillo}}, \bibinfo {author} {\bibfnamefont {G.}~\bibnamefont {Marques-Tavares}}, \bibinfo {author} {\bibfnamefont {I.}~\bibnamefont {Tamborra}}, \ and\ \bibinfo {author} {\bibfnamefont {E.}~\bibnamefont {Vitagliano}},\ }\href {\doibase 10.1103/PhysRevLett.132.101004} {\bibfield  {journal} {\bibinfo  {journal} {Phys. Rev. Lett.}\ }\textbf {\bibinfo {volume} {132}},\ \bibinfo {pages} {101004} (\bibinfo {year} {2024})},\ \Eprint {http://arxiv.org/abs/2305.10327} {arXiv:2305.10327 [hep-ph]} \BibitemShut {NoStop}%
\bibitem [{\citenamefont {O'Hare}(2020)}]{AxionLimits}%
  \BibitemOpen
  \bibfield  {author} {\bibinfo {author} {\bibfnamefont {C.}~\bibnamefont {O'Hare}},\ }\href {\doibase 10.5281/zenodo.3932430} {\enquote {\bibinfo {title} {cajohare/axionlimits: Axionlimits},}\ }\bibinfo {howpublished} {\url{https://cajohare.github.io/AxionLimits/}} (\bibinfo {year} {2020})\BibitemShut {NoStop}%
\bibitem [{\citenamefont {Kolb}\ and\ \citenamefont {Wolfram}(1980{\natexlab{a}})}]{Kolb:1979ui}%
  \BibitemOpen
  \bibfield  {author} {\bibinfo {author} {\bibfnamefont {E.~W.}\ \bibnamefont {Kolb}}\ and\ \bibinfo {author} {\bibfnamefont {S.}~\bibnamefont {Wolfram}},\ }\href {\doibase 10.1016/0370-2693(80)90435-9} {\bibfield  {journal} {\bibinfo  {journal} {Phys. Lett. B}\ }\textbf {\bibinfo {volume} {91}},\ \bibinfo {pages} {217} (\bibinfo {year} {1980}{\natexlab{a}})}\BibitemShut {NoStop}%
\bibitem [{\citenamefont {Kolb}\ and\ \citenamefont {Wolfram}(1980{\natexlab{b}})}]{Kolb:1979qa}%
  \BibitemOpen
  \bibfield  {author} {\bibinfo {author} {\bibfnamefont {E.~W.}\ \bibnamefont {Kolb}}\ and\ \bibinfo {author} {\bibfnamefont {S.}~\bibnamefont {Wolfram}},\ }\href {\doibase 10.1016/0550-3213(82)90012-8} {\bibfield  {journal} {\bibinfo  {journal} {Nucl. Phys. B}\ }\textbf {\bibinfo {volume} {172}},\ \bibinfo {pages} {224} (\bibinfo {year} {1980}{\natexlab{b}})},\ \bibinfo {note} {[Erratum: Nucl.Phys.B 195, 542 (1982)]}\BibitemShut {NoStop}%
\bibitem [{\citenamefont {Fry}\ \emph {et~al.}(1980)\citenamefont {Fry}, \citenamefont {Olive},\ and\ \citenamefont {Turner}}]{Fry:1980ph}%
  \BibitemOpen
  \bibfield  {author} {\bibinfo {author} {\bibfnamefont {J.~N.}\ \bibnamefont {Fry}}, \bibinfo {author} {\bibfnamefont {K.~A.}\ \bibnamefont {Olive}}, \ and\ \bibinfo {author} {\bibfnamefont {M.~S.}\ \bibnamefont {Turner}},\ }\href {\doibase 10.1103/PhysRevD.22.2953} {\bibfield  {journal} {\bibinfo  {journal} {Phys. Rev. D}\ }\textbf {\bibinfo {volume} {22}},\ \bibinfo {pages} {2953} (\bibinfo {year} {1980})}\BibitemShut {NoStop}%
\bibitem [{\citenamefont {Shtabovenko}\ \emph {et~al.}(2020)\citenamefont {Shtabovenko}, \citenamefont {Mertig},\ and\ \citenamefont {Orellana}}]{Shtabovenko:2020gxv}%
  \BibitemOpen
  \bibfield  {author} {\bibinfo {author} {\bibfnamefont {V.}~\bibnamefont {Shtabovenko}}, \bibinfo {author} {\bibfnamefont {R.}~\bibnamefont {Mertig}}, \ and\ \bibinfo {author} {\bibfnamefont {F.}~\bibnamefont {Orellana}},\ }\href {\doibase 10.1016/j.cpc.2020.107478} {\bibfield  {journal} {\bibinfo  {journal} {Comput. Phys. Commun.}\ }\textbf {\bibinfo {volume} {256}},\ \bibinfo {pages} {107478} (\bibinfo {year} {2020})},\ \Eprint {http://arxiv.org/abs/2001.04407} {arXiv:2001.04407 [hep-ph]} \BibitemShut {NoStop}%
\bibitem [{\citenamefont {Gondolo}\ and\ \citenamefont {Gelmini}(1991)}]{Gondolo:1990dk}%
  \BibitemOpen
  \bibfield  {author} {\bibinfo {author} {\bibfnamefont {P.}~\bibnamefont {Gondolo}}\ and\ \bibinfo {author} {\bibfnamefont {G.}~\bibnamefont {Gelmini}},\ }\href {\doibase 10.1016/0550-3213(91)90438-4} {\bibfield  {journal} {\bibinfo  {journal} {Nucl. Phys. B}\ }\textbf {\bibinfo {volume} {360}},\ \bibinfo {pages} {145} (\bibinfo {year} {1991})}\BibitemShut {NoStop}%
\end{thebibliography}%

\end{document}